\documentclass[a4paper,11pt]{article}
\pdfoutput=1 
\usepackage{grffile}
\usepackage{aas_macros,braket}
\usepackage{jcappub,natbib} 
\usepackage[T1]{fontenc} 
\usepackage{color}
\usepackage{comment}
\usepackage{multirow}
\usepackage{mathrsfs}
\usepackage{mathtools}
\usepackage{amsmath}
\allowdisplaybreaks
\usepackage{float}

\usepackage{amssymb}
\usepackage{soul}
\def\be{\begin{equation}}
\def\ee{\end{equation}}
\def\bea{\begin{eqnarray}}
\def\eea{\end{eqnarray}}

\def\ba#1\ea{\begin{align}#1\end{align}}

\def\up{\;\raise1.0pt\hbox{$'$}\hskip-6pt\partial\;}
\def\down{\;\overline{\raise1.0pt\hbox{$'$}\hskip-6pt
		\partial}\;}

\def\apjs{Astrophys.\ J. Supp.}

\renewcommand{\v}[1]{\mathbf{#1}}

\newcommand{\vx}{\v{x}}
\newcommand{\vk}{\v{k}}
\newcommand{\vq}{\v{q}}
\newcommand{\vp}{\v{p}}
\newcommand{\bq}{\mathbf{q}}
\newcommand{\bk}{\mathbf{k}}
\newcommand{\bx}{\mathbf{x}}

\newcommand{\Deg}{^{\circ}}

\newcommand{\nhat}{\hat{n}}

\renewcommand{\Re}{{\rm Re}\,}

\def\nhat{\hat{n}}

\newcommand{\nell}{N_{\ell}^{\rm GW}}
\newcommand{\Omegagw}{{\Omega_{{\rm{{GW}}}}}}
\newcommand{\deltagw}{\ensuremath{\delta^{\rm{{GW}}}}}
\newcommand{\deltaT}{\ensuremath{\delta^{\rm{{T}}}}}

\newcommand{\tj}[6]{\ensuremath{\begin{pmatrix}
			#1 & #2 & #3 \\
			#4 &#5 &#6
\end{pmatrix}}}
\newcommand{\fnls}{\ensuremath{F^{\text{tts}}_{\rm NL}}}
\newcommand{\fnlt}{\ensuremath{F^{\text{ttt}}_{\rm NL}}}
\newcommand{\tlfnls}{\ensuremath{\tilde{F}^{\text{tts}}_{\rm NL}}}

\newcommand{\GGS}{\ensuremath{C^{\rm{  GW,tts}}_{\ell}}}
\newcommand{\GGT}{\ensuremath{C^{\rm{ GW,ttt}}_{\ell}}}
\newcommand{\GTS}{\ensuremath{C^{\rm{  GW-T,tts}}_{\ell}}}
\newcommand{\GTT}{\ensuremath{C^{\rm{ GW-T,ttt}}_{\ell}}}
\newcommand{\GT}{\ensuremath{C^{\rm{GW-T}}_{\ell}}}
\newcommand{\GG}{\ensuremath{C^{\rm{ GW}}_{\ell}}}
\newcommand{\TT}{\ensuremath{C^{\rm{ TT}}_{\ell}}}
\newcommand{\NGW}{\ensuremath{N^{\rm{ GW}}_{\ell}}}
\newcommand{\GGind}{\ensuremath{C_{\ell}^{\rm{GW,ind}}}}
\newcommand{\GTind}{\ensuremath{C_{\ell}^{\rm{GW-T,ind}}}}

\newcommand{\SNRx}{\ensuremath{\rm{SNR}^{\times}}}

\definecolor{green2}{cmyk}{1, 0, 1, 0.1}

\title{Testing the Early Universe with Anisotropies of the Gravitational Wave Background} 

\author[a,b]{Ema Dimastrogiovanni,}\author[c,d]{Matteo Fasiello,}\author[b]{Ameek Malhotra,}\author[a]{P.~Daniel Meerburg}\author[a]{and Giorgio Orlando} 

\affiliation[a]{Van Swinderen Institute for Particle Physics and Gravity,
University of Groningen, Nijenborgh 4, 9747 AG Groningen, The Netherlands}
\affiliation[b]{Sydney Consortium for Particle Physics and Cosmology, School of Physics, The University of New South Wales, Sydney NSW 2052, Australia}
\affiliation[c]{Instituto de F\'{i}sica T\'{e}orica UAM/CSIC, calle Nicol\'{a}s Cabrera 13-15, Cantoblanco, 28049, Madrid, Spain}
\affiliation[d]{Institute of Cosmology \& Gravitation, University of Portsmouth, PO1 3FX, UK}

\emailAdd{e.dimastrogiovanni@rug.nl}
\emailAdd{matteo.fasiello@csic.es}
\emailAdd{ameek.malhotra@unsw.edu.au}
\emailAdd{p.d.meerburg@rug.nl}
\emailAdd{g.orlando@rug.nl}

\abstract{In this work we analyse in detail the possibility of using small and intermediate-scale gravitational wave anisotropies to constrain the inflationary particle content. First, we develop a phenomenological approach focusing on anisotropies generated by primordial tensor-tensor-scalar and purely gravitational non-Gaussianities. We highlight the quantities that play a key role in determining the detectability of the signal. To amplify the power of anisotropies as a probe of early universe physics, we consider cross-correlations with CMB temperature anisotropies. We assess the size of the signal from inflationary interactions against so-called induced anisotropies. In order to arrive at realistic estimates, we obtain the projected constraints on the non-linear primordial parameter $F_{\rm NL}$ for several upcoming gravitational wave probes in the presence of the astrophysical gravitational wave background. We further illustrate our findings by considering a concrete inflationary realisation and use it to underscore a few subtleties in the phenomenological analysis.}

\begin{document}

\maketitle
\flushbottom

\section{Introduction}
\label{intro}
The advent of laser interferometers has opened up a new era for gravitational wave (GW) astronomy. From the very first direct  detection \cite{LIGOScientific:2016aoc} of a GW event in 2015  we have learned precious lessons in stellar evolution and astrophysics. The growing number of GW events detected also provides the ideal testing ground for general relativity. In this sense, the GW170817 event and its optical counterpart have been especially consequential \cite{Baker:2017hug,Creminelli:2017sry,Ezquiaga:2017ekz}. Operational and upcoming GW detectors hold the potential to bring about transformative changes also  in cosmology and the particle physics of the early universe. The increasing number and sensitivity of laser interferometers (LIGO/Virgo/Kagra, LISA, Taiji, Einstein Telescope(ET), Cosmic Explorer (CE) to mention but a few), together with the possibility of detecting GWs at intermediate scales via pulsar timing arrays (PTA) puts us in the enviable position to access key information on cosmological GW sources and, possibly, distinguish their signal from the astrophysical GW background.

Mechanism for GW production are abound in the early universe \cite{Caprini:2015tfa}. Gravitational waves are a universal prediction of inflation and may also result from pre-heating dynamics, the energy loss of cosmic strings via gravitational radiation, and first order phase transitions (this typically requires beyond-the-Standard-Model physics). In this work we will focus on the stochastic gravitational wave background (SGWB) from inflation. We shall be interested in probing inflationary (self)interactions, and thus the  particle content of the very early universe, through GW probes at intermediate and small scales (from the $10^{-9}$ Hz of PTAs to the 10kHz of, for example, LIGO). \\
\indent A GW signal at PTA scales or at the frequency range accessible via laser interferometers is typically associated to a multi-field or multi-clock inflationary realisation\footnote{The proposed Big Bang Observer (BBO) is an exception in that it might be able to detect even a signal from single-field slow roll models provided that the tensor-to-scalar ratio $r$ is close to the current combined Planck + BICEP2/Keck Array BK15 upper bound of $r<0.056$ \cite{Planck:2018jri}. There are interesting proposals for single-field models generating a GW signal detectable at small scales. One such mechanism \cite{Mylova:2018yap} postulates the existence of a non-attractor phase followed by an attractor solution \cite{Ozsoy:2019slf}.}. Detection of a primordial SGWB of inflationary origin would then by itself be strongly suggestive of a rich inflationary field content. In this manuscript we take the analysis one step further and detail on how one may use GW probes to test inflation beyond the ``standard'' power spectrum (i.e. beyond the quadratic Lagrangian).  Anisotropies of the GW spectrum are the key observables here, and in particular those engendered by primordial tensor-tensor-scalar (TTS) and purely GW (TTT) non-Gaussianities.\\
\indent Primordial non-Gaussianities are well constrained on large scales by cosmic microwave background (CMB) observations \cite{Planck:2019kim,Shiraishi:2019yux}. The physics of the intermediate and small scales constraints is also very interesting. As show in \cite{Bartolo:2018evs,Bartolo:2018rku}, propagation effects suppress e.g. the primordial TTT signal in most of the momentum configurations. Let us offer an intuitive understanding of such suppression. Direct access to the bispectrum at intermediate/small scales requires that the modes have all entered the horizon during radiation domination and thus have accumulated a history of propagation through structure. Given that the momenta are to be different in view of the overall momentum conservation, different modes have different propagation history, so much so that this ``washes out'' any initial correlation, i.e. any correlation due to the  initial conditions set by inflation. \\
\indent A proven way to get around this suppression is to consider specific momenta configurations, such as the ultra-squeezed one \cite{Dimastrogiovanni:2019bfl}. In this configuration there is one long (up to horizon-size) mode and two short ones\footnote{Another momenta configuration that can be probed is the folded one, see for example \cite{Powell:2019kid}.}. This implies that  only small modes can be accessed directly, but it also guarantees a much milder suppression: a very long mode does not suffer suppression effects and the two short ones may have a much more similar propagation history, thereby avoiding the wash out effect.  The anisotropies we shall study in this work probe precisely this momentum configuration and  therefore provide a precious handle on non-Gaussianities at intermediate and small scales\footnote{For alternative examples of probes of primordial non-Gaussianities, see e.g. \cite{Dalal:2007cu,Matarrese:2008nc,Jeong:2012df,Dimastrogiovanni:2014ina,Munoz:2015eqa,Pajer:2012vz,Emami:2015xqa,Shiraishi:2015lma,Ota:2016mqd,Ravenni:2017lgw,Cabass:2018jgj,Orlando:2021nkv}.}.\\
\indent We will explore the constraining power of anisotropies on primordial physics in two steps. First, we will consider anisotropies due to non-trivial squeezed primordial TTS and TTT bispectra. Second, we will study the cross-correlation of such GW anisotropies with temperature anisotropies of the CMB. As we shall see, the latter observable can take us a long way towards verifying the primordial nature of possible GW anisotropies even in the presence of a large stochastic GW background of astrophysical origin.  \\
\indent We find it worthwhile in this manuscript to first provide a phenomenological, agnostic, approach to anisotropies from primordial long-short mode couplings, without  committing to any specific inflationary model. It will allow us to show what is typically expected in terms of observables and single out the key quantities that determine the likelihood for a given signal to be observed. Besides the usual suspects, i.e. the non-linear parameters $F^{\rm tts}_{\rm NL}$, $F^{\rm ttt}_{\rm NL}$ and the tensor-to-scalar ratio $r$, we will see how e.g. cross-correlations are also rather sensitive to the angular dependence of primordial bispectra.\\
\indent Our analysis would be incomplete without a detailed example of an inflationary mechanism that supports a GW signal on small scales as well as a sufficiently large primordial non-Gaussian amplitude to give the leading contribution to GW anisotropies, beyond  
the ever-present \textit{induced} component \cite{Contaldi:2016koz,Bartolo:2019oiq,Bartolo:2019yeu,Domcke:2020xmn}. To this aim, we employ an EFT formulation of the inflationary Lagrangian comprising an extra\footnote{That is, beyond   the massless spin-2  field of general relativity.} spin-2 field non-minimally coupled to the inflaton \cite{Bordin:2018pca,Iacconi:2019vgc}. Such coupling is necessary to weaken unitarity bounds on the spin-2 particle mass range (see e.g. \cite{Higuchi:1986py,Fasiello:2013woa}), thus allowing it to have a small (compared to the Hubble rate $H$) mass. \vspace{0.25cm}\\
\indent This paper is organised as follows. In \textit{Section} \ref{sec:self_corr} we briefly review induced anisotropies and compare them with those generated by primordial non-Gaussianities. We give a rule of thumb criterion to estimate the size of each contribution. We study the case of both monopolar and quadrupolar-type contributions to GW anisotropies from a TTS correlator, providing both the general result and its simplified analytical expression in the scale-invariant case. The contribution to anisotropies due to a primordial TTT correlation is also scrutinised.\\ \indent \textit{Section} \ref{sec:cross} is devoted to cross-correlations with the CMB temperature anisotropies. After reviewing the induced contribution, we focus on cross-correlating the TTS term with the CMB and derive the projected constraints on $F^{\rm tts}_{\rm NL}$ from a range of GW probes: PTAs, LISA\&Taiji, ET\&CE, and BBO. We also account for the presence of an astrophysical gravitational wave background (AGWB), which allows us to highlight the power of cross-correlations: a primordial signal may be detected even if the AGWB dominates over the primordial SGWB. The TTT contribution to cross-correlations is also analysed.\\
\indent In \textit{Section} \ref{sec:model_spin} we consider a specific inflationary realisation and show how, interestingly, in this specific case it is the TTT primordial correlation that provides the leading contribution both at the level of the auto- and the cross-correlation. This is true despite the factor of $r$ suppression any TTT correlator inherits with respect to its TTS counterpart. We find that an instrument such as BBO can deliver a percent level relative error on $F^{\rm ttt}_{\rm NL}$ in this model. 
\indent We discuss our findings and comment on future work in \textit{Section} \ref{conclusions}.

\section{SGWB anisotropies} \label{sec:self_corr}

Cosmological backgrounds of gravitational waves (CGWB) are typically characterised in terms of the their normalised energy density \cite{Maggiore:1999vm},
\begin{align}
\Omegagw(k) \equiv \frac{1}{\rho_{\rm{cr}}}\frac{d \rho_{\rm GW}}{d \ln k} \, ,
\end{align}
where $\rho_{\rm GW}$ is the energy density of GW and $\rho_{\rm cr}$ is the critical energy density. For the primordial GW background from inflation, the energy density observed today at time $\eta_0$ is related to the primordial tensor power spectrum as
\begin{align}
    \Omegagw(k,\eta_0) = \frac{k^2}{12 a_0^2 H_0^2}\mathcal{T}^2(k,\eta_0)\cdot\frac{1}{4\pi}\int d^2\nhat\, \mathcal{P}_{\gamma}(\vk,\v{d},\eta_{\rm in})\,,
\end{align}
where  $\mathcal{T}(k,\eta_0)$ is the tensor transfer function \cite{Caprini:2018mtu} and $\vk = k\hat{n}$. The primordial power spectrum is evaluated at the point $\v{d}=-d\hat n$ with $d=\eta_0-\eta_{\rm in}$ being the conformal time elapsed from horizon-entry of the mode $k$ to the present. Allowing for anisotropies in the energy density we can write the above quanitity as
\begin{align}
  \Omegagw(f) =  \overline{\Omega}_{\rm GW}(f)\Big[1+\frac{1}{4\pi}\int d^2\hat{n}\, \delta_{\rm GW}(f,\hat{n})\Big]\,,
\end{align}
where $\delta_{\rm GW}$ and $\overline{\Omega}_{\text{\tiny GW}}(f)$ denote the anisotropic and isotropic component of the energy density respectively. For a generic cosmological gravitational wave background, these anisotropies can arise from (i) propagation in the perturbed universe,  and (ii) an inhomogeneous production mechanism. We shall refer to these anisotropies as `induced' and `intrinsic', respectively.

\subsection{Induced anisotropies}

The induced anisotropies arise from the propagation of GW through the large scale scalar perturbations of the universe\footnotemark. Note that such anisotropies are universal in nature, in the sense that they are rather model-independent. These anisotropies have been studied using the standard Boltzmann formalism in \cite{Contaldi:2016koz,Bartolo:2019oiq,Bartolo:2019yeu,DallArmi:2020dar}. Similarly to what happens with the CMB, one finds that the SGWB is affected by both the Sachs-Wolfe (SW) and Integrated Sachs-Wolfe (ISW) effects. On large angular scales, the dominant contribution is given by the SW term which can be written as \cite{Bartolo:2019oiq,Bartolo:2019yeu}
\begin{align}
    \deltagw_{\rm ind} \simeq \left[4-\frac{\partial \ln{\Omegagw(k)}}{\partial \ln{k}}\right]\int \frac{d^3q}{(2\pi)^3}e^{-id\hat{n}\cdot\vq}\cdot \frac{2}{3}\zeta(\vec{q})\, .
    \label{eq:deltagw_ind}
\end{align}
\footnotetext{There is also a contribution from the large scale tensor peturbations but similar to the CMB, this is subdominant compared to the contribution from the scalar perturbations \cite{Bartolo:2019oiq,Bartolo:2019yeu}.}
\subsection{Intrinsic anisotropies from primordial non-Gaussianity}
The instrinsic anisotropies of interest here are those arising from large primordial  non-Gaussianities in the squeezed limit, i.e. primordial bispectra of the form $\langle \gamma_{\vk_1}\gamma_{\vk_2}\,X_{\vq\to 0}\rangle$. Here $X_\vq$ denotes a long wavelength mode of either a scalar or tensor perturbation while $\gamma_{\vk_{1,2}}$ are the short wavelength tensor modes which we take to be at interferometer scales (these modes re-enter the horizon during radiation domination). 
Let us begin with the case where the long mode $X_{\vq}$ corresponds to the long wavelength mode of the primordial curvature perturbation $\zeta$. The existence of this coupling between the long modes $\zeta_{\vq}$ and the short modes $\gamma_{\vk}$ modulates the primordial power spectrum as \cite{Jeong:2012df,Dai:2013kra,Brahma:2013rua,Dimastrogiovanni:2014ina,Dimastrogiovanni:2015pla,Adshead:2020bji},
\begin{align}
\label{eq:Pmod_scalar_lambda}
 \mathcal{P}^{\rm mod}_{\gamma}(\vk,\vx) = \sum_{\lambda}\mathcal{P}^\lambda_{\gamma}(k)\left[1+\int_{q \ll k}\frac{d^3q}{(2\pi)^3}\,e^{i\vq\cdot\vx}F_{\rm NL}^{\lambda,\rm tts}(\vk,\vq)\zeta(\vq)\right]\,,
\end{align}
where
\begin{align}
\label{eq:fnltts_def}
   F_{\rm NL}^{\lambda,\rm tts}(\vk,\vq) = \frac{B^{\lambda}_
   {\rm tts}(\vk-\vq/2,-\vk-\vq/2,\vq)}{P_{\zeta}(q)P^{\lambda}_{\gamma}(k)},
\end{align}
and the primordial bispectrum $B_{\rm tts}(\vk_1,\vk_2,\vk_3)$ in the squeezed limit is defined as 
\begin{align}
    \langle \gamma_{\vk_1}^{\lambda}\gamma_{\vk_2}^{\lambda'}\zeta_{\vk_3\to 0}\rangle' =\delta^{\lambda \lambda'} B^{\lambda}_{\rm TTS}(\vk_1,\vk_2,\vk_3)\,.
\end{align}
The prime here denotes the fact that we have omitted the factor of $(2\pi)^3\,\delta^{(3)}(\vk_1+\vk_2+\vk_3)$ that ensures momentum conservation. Note that while in principle it is possible to have polarisation-dependent power spectra and bispectra and hence $F_{\rm NL}^{\lambda_1,\rm tts}\neq F_{\rm NL}^{\lambda_2,\rm tts}$ for the different helicities $\lambda_1\neq \lambda_2$, this shall not be the case for the inflationary models we will consider here. Hence, in what follows further we will write $F_{\rm NL}^{\lambda,\rm tts}\equiv \fnls$ and Eq.~\eqref{eq:Pmod_scalar_lambda} becomes,
\begin{align}
\label{eq:Pmod_scalar}
\mathcal{P}^{\rm mod}_{\gamma}(\vk,\vx) = \mathcal{P}_{\gamma}(k)\left[1+\int_{q \ll k}\frac{d^3q}{(2\pi)^3}\,e^{i\vq\cdot\vx}\fnls(\vk,\vq)\zeta(\vq)\right]\,.
\end{align}
The isotropic and anisotropic\footnote{In the context of the Boltzmann formalism for GW \cite{Contaldi:2016koz,Bartolo:2019oiq,Bartolo:2019yeu}, these anisotropies are captured by the initial condition term. } components of the energy density can then be expressed in terms of this primordial tensor power spectrum as
\begin{align}
\overline{\Omega}_{\text{\tiny GW}}(k,\eta_0) = \frac{k^2}{12 a_0^2 H_0^2}\mathcal{T}^2(k,\eta_0)\mathcal{P}_{\gamma}(k)
\end{align}
and
\begin{align}\label{eq:deltagw_tts}
\deltagw_{\rm tts}(k,\hat{n}) = \int_{q_{}\ll k}\frac{d^3q}{(2\pi)^3}\,e^{-id\hat{n}\cdot\vq}\fnls(\vk,\vq)\zeta(\vec{q})\,.
\end{align}
We also consider here the anisotropies generated from the modulation of the power spectrum by a long wavelength tensor mode \cite{Dimastrogiovanni:2019bfl},
\begin{align}
\label{eq:Pmod_tensor}
\mathcal{P}^{\rm mod}_{\gamma}(\vk,\vx) = \mathcal{P}_{\gamma}(k)\left[1+\int_{q \ll k}\frac{d^3q}{(2\pi)^3}\,e^{i\vq\cdot\vx}F^{ttt}_{\rm NL}(\vk,\vq)\sum_{\lambda}\gamma^{\lambda}(\vq)\epsilon^{\lambda}_{ij}(\hat q)\hat{n}^i\hat{n}^j\right]\,,
\end{align}
where (once again we have dropped the polarisation dependence in $\fnlt$),
\begin{align}
\label{eq:fnlttt_def}
   \fnlt(\vk,\vq) = \frac{B_
   {\rm ttt}(\vk-\vq/2,-\vk+\vq/2,\vq)}{P^{\lambda_1}_{\gamma}(q)P^{\lambda_2}_{\gamma}(k)} \, ,
\end{align}
and $B_{\rm ttt}(\vk_1,\vk_2,\vk_3)$ is the squeezed-limit primordial bispectrum defined as
\begin{align}
    \langle \gamma^{\lambda_1}_{\vk_1}\gamma^{\lambda_1}_{\vk_2}\gamma^{\lambda_3}_{\vk_3\to 0}\rangle' = -\delta^{\lambda_2\lambda_3}\,\epsilon_{ij}^{\lambda_3}(k_3)\,k_2^i k_1^j \,B_{\rm TTT}(\vk_1,\vk_2,\vk_3) \,.
\end{align}
Thus, the anisotropies in this case are given by,
\begin{align}
\label{eq:deltagw_ttt}
    \deltagw_{\rm ttt}(k,\nhat) = -\int_{q\ll k}\frac{d^3q}{\left(2\pi\right)^3}\,e^{-i  d\nhat\cdot\vq}\fnlt(\vk,\vq)\sum_{s}\gamma^s(\vq)\epsilon_{ij}^{s}(\hat q)\hat{n}^i \hat{n}^j \,.
\end{align}
From the above discussion, one can expect the typical amplitude of these anisotropies to be
\begin{align}
 \label{eq:deltagw_estimate}
    \deltagw_{\rm ind}&\sim \sqrt{A_S}\nonumber\\
    \deltagw_{\rm tts}&\sim \fnls \sqrt{A_S}\\
    \deltagw_{\rm ttt}&\sim \fnlt \sqrt{r A_S}\nonumber \, ,
\end{align}
where $A_S$ denotes the amplitude of the scalar power spectra on CMB scales and $r$ is the tensor-to-scalar ratio. Thus, if the primordial bispectrum in the squeezed limit is large enough $(F_{\rm NL}\gg 1)$, the intrinsic anisotropies can dominate over the induced anisotropies. Furthermore, when $\fnls\sim\fnlt$ (e.g. as is the case for the solid inflation model of \cite{Endlich:2012pz,Endlich:2013jia}) we will have $\deltagw_{\rm tts}\gg\deltagw_{\rm ttt}$. We will explicitly evaluate these two contributions to the anisotropies in Sec.~\ref{sec:clgw} where we will also elaborate on  the effects of the angular dependence of $\fnls(\vk,\vq)$. Such a  dependence will turn out to be particularly important in calculating the cross-correlation of these GW anisotropies with the CMB temperature anisotropies in Section~\ref{sec:cross_TTS}.   
\subsection{Angular power spectra of SGWB anisotropies}
\label{sec:clgw}
To compute the angular power spectrum of the SGWB anisotropies we first expand them in spherical harmonics,
\begin{align}
    \deltagw_{\ell m} = \int d^2\nhat\, \deltagw(\nhat)Y_{\ell m}^{*}(\nhat) \, . 
\end{align}
The rotationally invariant\footnote{See \cite{Malhotra:2020ket} for a discussion of intrinsic anisotropies of the CGWB in a statistically anisotropic background.} angular power spectra are then defined as
\begin{align}
     \langle \deltagw_{\ell m}\delta^{\rm GW *}_{\ell' m'} \rangle \equiv \delta_{\ell \ell'}\delta_{m m'} \GG\,.
\end{align}
\subsubsection*{Induced anisotropies}
For the induced anisotropies, Eq.~\eqref{eq:deltagw_ind}, one finds \cite{Bartolo:2019oiq,Bartolo:2019yeu}
\begin{align}
\GGind = \left[4-\frac{\partial \ln{\Omegagw(k)}}{\partial \ln{k}}\right]^2\,\frac{2}{\pi}\int q^2dq\,j_{\ell}(q d)^2\cdot\frac{4}{9}P_{\zeta}(q) \,,
\label{eq:cl_ind1}
\end{align}
where $j_\ell$ are the spherical Bessel functions of the first kind and $P_{\zeta}$ denotes the primordial curvature power spectrum. The factor of $4/9$ is a consequence of the relation between the curvature perturbation $\zeta$ and the scalar potential $\Phi$ on super-horizon scales during the radiation dominated era. To evaluate Eq.~\eqref{eq:cl_ind1} analytically, we assume a scale invariant power spectrum for the curvature perturbation $P_{\zeta}(q)=(2\pi^2/q^3)A_S$, as well as a scale invariant spectrum of GW (i.e. $\partial \ln \Omegagw(k)/\partial \ln k=0$). We can then use the identity for the spherical Bessel functions,
\begin{align}
    \int_{0}^{\infty} \frac{dx}{x}\,j^2_{\ell}(x) = \frac{1}{2\ell(\ell+1)}\,,
    \label{eq:bessel_id1}
\end{align}
to get
\begin{align}
    \GGind \simeq \frac{128\pi A_S}{9\ell(\ell+1)}\,.
    \label{eq:cl_ind2}
\end{align}
We will now move on to the calculation of the angular power spectrum for the intrinsic anisotropies of the CGWB.

\subsubsection*{Anisotropies from \texorpdfstring{$\langle  \gamma \gamma \zeta\rangle$}{TTS} bispectrum}
 Here, we shall consider two cases. First we consider the case where the TTS bispectrum is independent of $\hat q\cdot \hat k$, i.e. the angle between the long wavelength scalar mode $\zeta_{q\to 0}$ and the short wavelength GW, $\gamma_k$. The second case of interest is a scenario where the bispectrum has a quadrupolar angular dependence in $\hat q \cdot \hat k$. This kind of angular dependence can arise in the inflationary scenarios presented in \cite{Endlich:2012pz,Endlich:2013jia,Ricciardone:2016lym} as well as in the model we consider in Sec.~\ref{sec:model_spin}.
\subsubsection*{Monopolar TTS}
Let us begin with the case where the parameter $\fnls$ defined in Eq.~\eqref{eq:fnltts_def} has no angular dependence in $\hat q\cdot \hat k$, i.e. we write
\begin{align}
    \fnls(\vk,\vq) = \tlfnls(k,q)\,.
    \label{eq:tts_mono}
\end{align}
The GW anisotropies of this form of the bispectrum have been previously considered in Ref. \cite{Adshead:2020bji} and the result, starting from Eq.~\eqref{eq:deltagw_tts}, can be written as
\begin{align}
    \GGS = \frac{2}{\pi}\int_{q\ll k} q^2dq\,j_{\ell}(q d)^2\tlfnls(q,k)^2P_{\zeta}(q)\,.
\end{align}
This can be calculated analytically assuming for simplicity a scale-independent $\tlfnls$ and a scale invariant $P_{\zeta}$ as before. 
We find
\begin{align}
    \GGS \simeq \left(\tlfnls\right)^2 \frac{2\pi A_S}{\ell(\ell+1)}\,.
\end{align}
As expected, we find the same scaling with $\ell$ as for the induced anisotropies,
\begin{align}
    \GGS = \left(\frac{3}{8}\tlfnls\right)^2\GGind \, .
\end{align}
Thus, for $\tlfnls\gg 1$, we find that the intrinsic anisotropies dominate, as anticipated in Eq.~\eqref{eq:deltagw_estimate}. 
\subsubsection*{Quadrupolar TTS}
\noindent Next, we consider the case where $\fnls$ has a quadrupolar angular dependence in $\hat q\cdot \hat k$. In this case we parametrise $\fnls$ as
\begin{align}
    \fnls(\vk,\vq) = \tlfnls(q,k)\,\left[\frac{ 4\pi}{5}\sum_{M} Y_{2 M}(\hat k)Y^*_{2 M}(\hat q)\right]=\tlfnls\,\mathcal{P}_2(\hat q \cdot \hat k)\,,
    \label{eq:tts_quad}
\end{align}
where $\mathcal{P}_2$ is the second Legendre polynomial. The angular power spectrum of the anisotropies for such an angular dependence is given by,
\begin{align}
\GGS={16\pi^2}\sum_{L_1,L_2}i^{L_1-L_2}h_{\ell L_1 2}^2 h_{{ \ell} L_2 2}^2\frac{H_{L_1 L_2}}{(2\ell+1)^2}, 
\label{eq:quadgw_tts}
\end{align}
where the sum is over $L_1,\,L_2 = \ell-2,\ell,\ell+2$, the quantity $h_{\ell_1 \ell_2 \ell_3}$ is defined in terms of the Wigner $3j$ symbols as
\begin{align}
h_{\ell_1 \ell_2 \ell_3 } \equiv \sqrt{\frac{(2\ell_1+1)(2\ell_2+1)(2\ell_3+1)}{4\pi}}\tj{\ell_1}{\ell_2}{\ell_3}{0}{0}{0}\,,
\label{eq:hldef}
\end{align}
and
\begin{align}
H_{L_1 L_2} &\equiv  \frac{2}{25\pi}\int_{q\ll k} q^2dq\,j_{L_1}(q d)j_{L_2}(q {d} )\tlfnls(k,q)^2P_{\zeta}(q)\,.
\label{eq:Hl_tts}
\end{align}
The result up to Eq.~\eqref{eq:quadgw_tts} was previously derived in \cite{Malhotra:2020ket}. As before, we now obtain an analytic expression for this auto-correlation: for $\ell>2$, this can be estimated using the identity \cite{Gradshteyn7ed},
\begin{align}
    \int_0^\infty dx\, J_\nu(ax)J_\mu(ax)x^{-\lambda} = &\frac{a^\lambda\,\Gamma(\lambda)\Gamma(\frac{\mu+\nu-\lambda+1}{2})}{2^\lambda\, \Gamma(\frac{\mu-\nu+\lambda+1}{2})\Gamma(\frac{-\mu+\nu+\lambda+1}{2})\Gamma(\frac{\mu+\nu+\lambda+1}{2})}\,,\\
    &\left[\text{for } \Re{(\mu+\nu+1)}>\Re{\lambda}>0,\;a>0 \right]\nonumber 
    \label{eq:bessel_id_2}
\end{align}
where $J_{n}$ are the Bessel functions of the first kind whose relation to the spherical Bessel functions is given by
\begin{align}
    j_n(x) = \sqrt{\frac{\pi}{2 x}}J_{n+1/2}(x) \,. 
\end{align}
We finally get
\begin{align}
   \GGS \simeq \frac{2\pi }{5} \frac{(\tlfnls)^2A_S}{(\ell-2)(\ell+3)} \,.
\end{align}

\subsubsection*{Anisotropies from \texorpdfstring{$\langle \gamma \gamma \gamma \rangle$}{TTT} bispectrum}
Let us now compute the intrinsic anisotropies from the TTT bispectrum. Starting from Eq.~\eqref{eq:deltagw_ttt} and assuming that $\fnlt(\vq,\vk)=\fnlt(q,k)$, we can obtain the spherical harmonic coefficients for $\deltagw_{\rm ttt}$ using the following relation \cite{Hu:1997hp,Challinor_2009},
\begin{align}
    \gamma_{ij}^{R/L}(q)\,n^i n^j\,e^{- i d\nhat\cdot \vq} = - (2\pi) \,\gamma^{R/L}(q)\sum_{LM}(-i)^L \sqrt{\frac{(L+2)!}{(L-2)!}}\frac{j_L(q d)}{(qd)^2}{}_{\mp 2}Y_{L M}^*(\hat q)Y_{LM}(\nhat)\, . 
    \label{eq:tensor_id}
\end{align}
This gives
\begin{align}
    \deltagw_{\ell m} = (2 \pi) (-i)^\ell\sqrt{\frac{(\ell+2)!}{(\ell-2)!}}\sum_{s=\pm 2}\int \frac{d^3q}{(2\pi)^3}\,\fnlt(k,q)\gamma_{\vq}^{s}\,\frac{j_\ell(q d)}{(qd)^2}{}_{- s}Y_{\ell m}^*(\hat q)\,.
\end{align}
Thus, the angular power spectrum of these anisotropies is given by
\begin{align}
    \GGT = 
    \frac{(\ell-1)\ell(\ell+1)(\ell+2)}{2\pi}\sum_{s=\pm 2}\int_{q\ll k} \,q^2dq\,\fnlt(k,q)^2 P_{\gamma}^s(q)\frac{j_{\ell}(qd)^2}{(qd)^4}\,.
\end{align}
Once again, an analytic form (for $\ell>2$) can be obtained by assuming a scale independent $\fnlt$ and a scale invariant $P_{\gamma} = (2\pi^2/q^3)rA_S$,
\begin{align}
    \GGT \simeq \frac{4\pi\, (\fnlt)^2 \,r A_S}{15(\ell-2)(\ell+3)}\,,
\end{align}
where we have used the identity
\begin{align}
    \int_0^\infty dx \,\frac{j_\ell^2(x)}{x^5} = \frac{4}{15}\frac{(\ell-2)!}{(\ell+2)!(\ell+3)(\ell-2)}\,.
    \label{eq:bessel_id_3}
\end{align}

\begin{figure}
    \centering
    \includegraphics[width=0.75\linewidth]{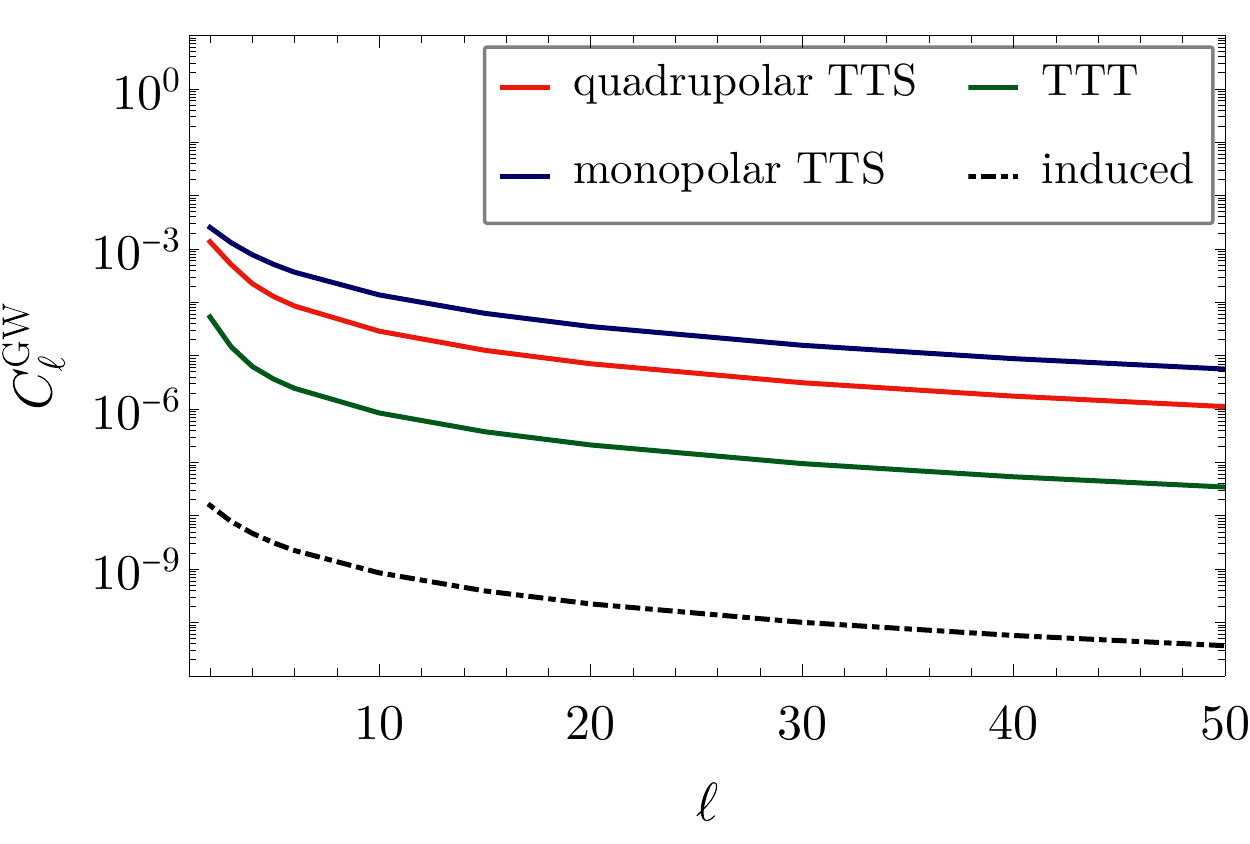}
    \caption{The auto-correlation of the SGWB anisotropies as a function of $\ell$ plotted for $|\tlfnls|=|\fnlt|=10^3$ and the tensor-scalar ratio $r=0.05$.}
    \label{fig:auto_model_ind}
\end{figure}

\noindent In Fig.~\ref{fig:auto_model_ind} we plot the angular power spectra calculated in this section for a representative value $|\tlfnls|=|\fnlt|=10^3$  and taking $r=0.05$. As anticipated in Eq.~\eqref{eq:deltagw_estimate}, for $|F_{\rm NL}|\gg 1$, we see that the angular power spectra for the intrinsic anisotropies are larger than the induced ones by roughly a factor $(\tlfnls)^2$ in the TTS case and $r(\fnlt)^2$ in the TTT case. Thus, for inflationary models with a significant enhancement of squeezed primordial non-Gaussianity, these anisotropies will be dominant. 

\section{CMB-GW Cross-correlation}
\label{sec:cross}
The GW anisotropies considered in the previous section arise from the modulation of the primordial tensor power spectrum by the long wavelength scalar/tensor modes and will be correlated with the scalar/tensor contributions to the CMB temperature anisotropies $\deltaT$ \footnote{One could also consider correlating the SGWB anisotropies with the E-mode polarisation of the CMB. Although this turns out to be smaller than the induced GW-T cross-correlation by at least one order of magnitude \cite{Braglia:2021fxn}, including it can help put tighter constraints on $\tlfnls$. As for the cross-correlation with CMB B-mode polarisation, this will be non-zero only when there is parity violation, similar to the CMB case where  $\langle \rm TB \rangle=0$ if parity is preserved in the theory.}. Here, we calculate this cross-correlation and for the TTS case we comment briefly on the dependence of this cross-correlation on the angular structure of the primordial bispectrum. As in the previous section, we provide analytic estimates wherever possible by assuming a scale independent $F_{\rm NL}$ and $P_{\zeta}(q)=(2\pi^2/q^3)A_S$.
\subsection{Cross-correlations with induced anisotropies}
Since the induced anisotropies are sourced by the large scale curvature perturbation, they are also correlated with the CMB temperature anisotropies. Their cross-correlation with the CMB is given by 
\begin{align}
\langle \deltagw_{\ell m}\delta^{\rm T*}_{\ell' m'} \rangle = \delta_{\ell \ell'}\delta_{m m'}\GTind
\end{align}
with 
\begin{align}
\GTind \simeq \frac{16}{15\pi}\int q^2dq\,j_{\ell}(q d)j_{\ell}(qr_{\rm lss})P_{\zeta}(q) \,,
\label{eq:clgwt_ind}
\end{align}
assuming a flat spectrum for $\Omegagw$. To get to Eq.~\eqref{eq:clgwt_ind}, we have assumed that the temperature anisotropies are given by the SW term which is a good approximation on large angular scales. Its spherical harmonic coefficients are given by \cite{Dodelson:2003ft},
\begin{align}
\deltaT_{\ell m} = \frac{4\pi}{5}(-i)^\ell\int \frac{d^3p}{(2\pi)^3}Y^{*}_{\ell m}(\hat{p})j_{\ell}(pr_{\rm lss})\zeta(\vp) \,, 
\label{eq:SW_T}
\end{align}
where $r_{\rm lss}$ denotes the comoving distance to the last scattering surface. Detailed numerical analyses of this cross-correlation highlighting the relative contribution of the various terms (SW, early and late ISW, Doppler etc.) and the effects of pre-recombination physics have been recently carried out in Refs.~\cite{Ricciardone:2021kel} and \cite{Braglia:2021fxn} respectively. For our purposes, it suffices to take only the SW term Eq.~\eqref{eq:SW_T} since this is the main contribution to the cross-correlation (intrinsic as well as induced) on large angular scales.
We can now evaluate this correlation analytically as 
\begin{align}
    \GT = \frac{32\pi}{15} A_S \int_{q\ll k} \frac{dq}{q}\,j_{\ell}(q d)j_{\ell}(qr_{\rm lss})\,.
    \label{eq:cross_ind_1}
\end{align}
For the low multipole range $(\ell<20)$, a good approximation can be obtained by letting $d=r_{\rm lss}$ and using the identity Eq.~\eqref{eq:bessel_id1} to get
\begin{align}
    \GT \simeq \frac{32\pi}{15}\frac{A_S}{2\ell(\ell+1)}\,.
    \label{eq:cross_ind_analytic1}
\end{align}
If instead one does not make this approximation, one can derive a more accurate but complicated expression in terms of the Gamma functions and the Hypergeometric function ${}_2 F_1$ using the identity \cite{Gradshteyn7ed},
\begin{align}\label{eq:bessel_id_4}
    \int_0^\infty dx\, J_\nu(ax)J_\mu(bx)x^{-\lambda} =& \frac{a^\nu\,\Gamma(\frac{\mu+\nu-\lambda+1}{2})}{2^\lambda\, b^{\nu-\lambda+1}\Gamma(\frac{\mu-\nu+\lambda+1}{2})\Gamma(\nu+1)}\nonumber\\
    &\times{}_2 F_1\left(\frac{\mu+\nu-\lambda+1}{2},\frac{\nu-\mu-\lambda+1}{2};\nu+1;\frac{a^2}{b^2}\right)\,,\\
    &\left[\text{for } \Re{(\mu+\nu-\lambda+1)}>0,\;\Re{\lambda}>-1,\;0<a<b \right]\nonumber
\end{align}
With this, from Eq.~\eqref{eq:cross_ind_1} we find
\begin{align}
    \GT =   \frac{8\pi^{3/2}}{15}A_S  \left(\frac{r_{\rm lss}}{d}\right)^\ell\frac{\Gamma(\ell)}{\Gamma(\ell+\frac{3}{2})} \; {}_2{F}_1\left(-\frac{1}{2},\ell,\ell+\frac{3}{2},\frac{r_{\rm lss}^2}{d^2}\right)\,.
    \label{eq:cross_ind_analytic2}
\end{align}
Thus we see that this cross-correlation decays sharply with $\ell$ due to the factor $(r_{\rm lss}/d)^\ell$, as also pointed out in Refs.~\cite{Adshead:2020bji,Ricciardone:2021kel}. This suppression arises from the fact that the time when the gravitons begin their free streaming $\eta_{\rm in}$ is different from that of the CMB photons $\eta_{\rm lss}$, thus $r_{\rm lss}/d<1$ (recall that $d = \eta_0-\eta_{\rm in}$ for the short mode $k$ so $\eta_{\rm in}$ corresponds to a conformal time deep within radiation domination, whereas $r_{\rm lss} = \eta_0 -\eta_{\rm lss}$). As a result, these anisotropies become uncorrelated on small scales (large $\ell$). Since $d\approx r_{\rm lss}$, this also explains why the estimate of Eq.~\eqref{eq:cross_ind_analytic1} works well for small $\ell$ but fails for $\ell\gtrsim 20$ (see Figure~\ref{fig:cross_ratio}). To see the explicit scaling with $\ell$ we use the following numerical fit,
\begin{align}
    \frac{\Gamma(\ell)}{\Gamma(\ell+\frac{3}{2})} \, {}_2{F}_1\left(-\frac{1}{2},\ell;\ell+\frac{3}{2};\frac{r_{\rm lss}^2}{d^2}\right) \simeq \frac{0.8}{\ell  (\ell +0.44)^{0.78}}\,
\end{align}
which gives
\begin{align}
    \GTS \sim \left(\frac{r_{\rm lss}}{d}\right)^\ell\frac{1}{\ell^{2}}\,.
\end{align}
A similar behaviour will also be present for the TTS cross-correlation, as we shall see below. 

\subsection{Cross-correlations with anisotropies from \texorpdfstring{$\langle  \gamma \gamma \zeta\rangle$}{TTS} bispectrum}
\label{sec:cross_TTS}
\subsubsection*{Monopolar TTS}
We consider here the CMB-GW cross-correlation for the intrinsic CGWB anisotropies where the long-wavelength mode is a scalar. 
For the monopolar $\tlfnls$ we have
\begin{align}
    \GT = \frac{2}{5\pi}\int_{q\ll k} q^2dq\,j_{\ell}(q d)j_{\ell}(qr_{\rm lss})\tlfnls(k,q)P_{\zeta}(q)\,,
    \label{eq:cross_mono_1}
\end{align}
which was previously derived in \cite{Adshead:2020bji}. Similar to the previous section, we can now analytically evaluate this cross-correlation,
\begin{align}
    \GT = \frac{4\pi}{5}\tlfnls A_S \int_{q\ll k} \frac{dq}{q}\,j_{\ell}(q d)j_{\ell}(qr_{\rm lss})\,.
    \label{eq:cross_mono_2}
\end{align}
The approximation with $d=r_{\rm lss}$ gives
\begin{align}
    \GT \simeq \frac{4\pi}{5}\tlfnls \frac{A_S}{2\ell(\ell+1)}\,.
    \label{eq:cross_mono_analytic1}
\end{align}
The full result with $d\neq r_{\rm lss}$ and using Eq.~\eqref{eq:bessel_id_4}
is instead
\begin{align}
    \GT =  \frac{\pi^{3/2}}{5} \tlfnls A_S \left(\frac{r_{\rm lss}}{d}\right)^\ell\frac{\Gamma(\ell)}{\Gamma(\ell+\frac{3}{2})} \, {}_2{F}_1\left(-\frac{1}{2},\ell;\ell+\frac{3}{2};\frac{r_{\rm lss}^2}{d^2}\right)\,.
    \label{eq:cross_mono_analytic2}
\end{align}
We see that the cross-correlation for the monopolar TTS scales with $\ell$ in exactly the same manner as the cross-correlation for the induced anisotropies,
\begin{align}
    \GTS = \left(\frac{3}{8}\tlfnls\right)\GTind\,.
\end{align}
\subsubsection*{Quadrupolar TTS}
\noindent For the quadrupolar case we have,
\begin{align}
\GTS = 4\pi\sum_{L}i^{L- \ell}h_{2 L \ell}^{2}\frac{G_{L \ell}}{2\ell+1} \,,
\label{eq:cross_quad_result}
\end{align}
where the sum is over $L=\ell-2,\ell,\ell+2$ and the function $G_{\ell_1 \ell_2}$ is defined as
\begin{align}
G_{\ell_1 \ell_2} &=  \frac{2}{25\pi}\int_{q\ll k} q^2dq\,j_{\ell_1}(q d)j_{\ell_2}(qr_{\rm lss})\tlfnls(k,q)P_{\zeta}(q),
\label{eq:cross_quad_integral}
\end{align}
and $h_{\ell_1 \ell_2 \ell_3}$ was defined in Eq.~\eqref{eq:hldef}. The result Eq.~\eqref{eq:cross_quad_result}, previously derived in \cite{Malhotra:2020ket}, can now be analytically estimated using Eq.~\eqref{eq:bessel_id_4} to get 
\begin{align}
    \GTS = \frac{\pi^{3/2} \tlfnls A_S}{20}\left(\frac{r_{\rm lss}}{d}\right)^\ell\left(\frac{r_{\rm lss}}{d}-1\right)\left(\frac{r_{\rm lss}}{d}+1\right)\Delta F_{\ell}\,,
    \label{eq:cross_quad_analytic}
\end{align}
where $\Delta F_{\ell}$ is given by
\begin{align}
    \Delta F_{\ell} \equiv \frac{\Gamma(\ell)}{\Gamma(\ell+3/2)}\left[(\ell +1) \, {}_2{F}_1\left(-\frac{1}{2},\ell ;\ell +\frac{3}{2};\frac{r_{\text{lss}}^2}{d^2}\right)-\,
   {}_2{F}_1\left(\frac{1}{2},\ell ;\ell +\frac{3}{2};\frac{r_{\text{lss}}^2}{d^2}\right)\right]\,.
\end{align}
The function $\Delta F_{\ell}$ is well fit by
\begin{align}
    \Delta F_{\ell} \simeq \frac{0.2}{(\ell+2)^{0.45}}\,.
\end{align}
Thus, the $\GTS$ for the quadrupolar case scales with $\ell$ as
\begin{align}
    \GTS \sim \left(\frac{r_{\rm lss}}{d}\right)^\ell\frac{1}{\ell^{1/2}}\,.
\end{align}
From the above results we notice the familiar $(r_{\rm lss}/d)^\ell$ suppression that we have seen previously in Eq.~\eqref{eq:cross_ind_analytic2} and Eq.~\eqref{eq:cross_mono_analytic2} for the induced and monopolar TTS anisotropies, thus the cross-correlation again decreases as we go towards the smaller scales. However from Eq.~\eqref{eq:cross_quad_analytic}, we also see that for the quadrupolar case when $d\to r_{\rm lss}$, this cross-correlation does not increase, instead it drops to zero. Physically, this can be understood from the fact that the source term for the GW anisotropy is locally a quadrupole, whereas the source term for the CMB is a monopole. Therefore, if the sources operate at the same point in space, their cross-correlation will be zero (orthogonality of the Legendre polynomials $\mathcal P_\ell(\hat k\cdot \hat q)$ and $\mathcal P_{\ell'}(\hat k\cdot \hat q)$ for $\ell\neq \ell'$). A further point of difference is that this cross-correlation decays more slowly compared to the monopolar one owing to the term proportional to $(\ell+1)$ in $\Delta F_\ell$.
\begin{figure}
    \centering
    \includegraphics[width=0.7\linewidth]{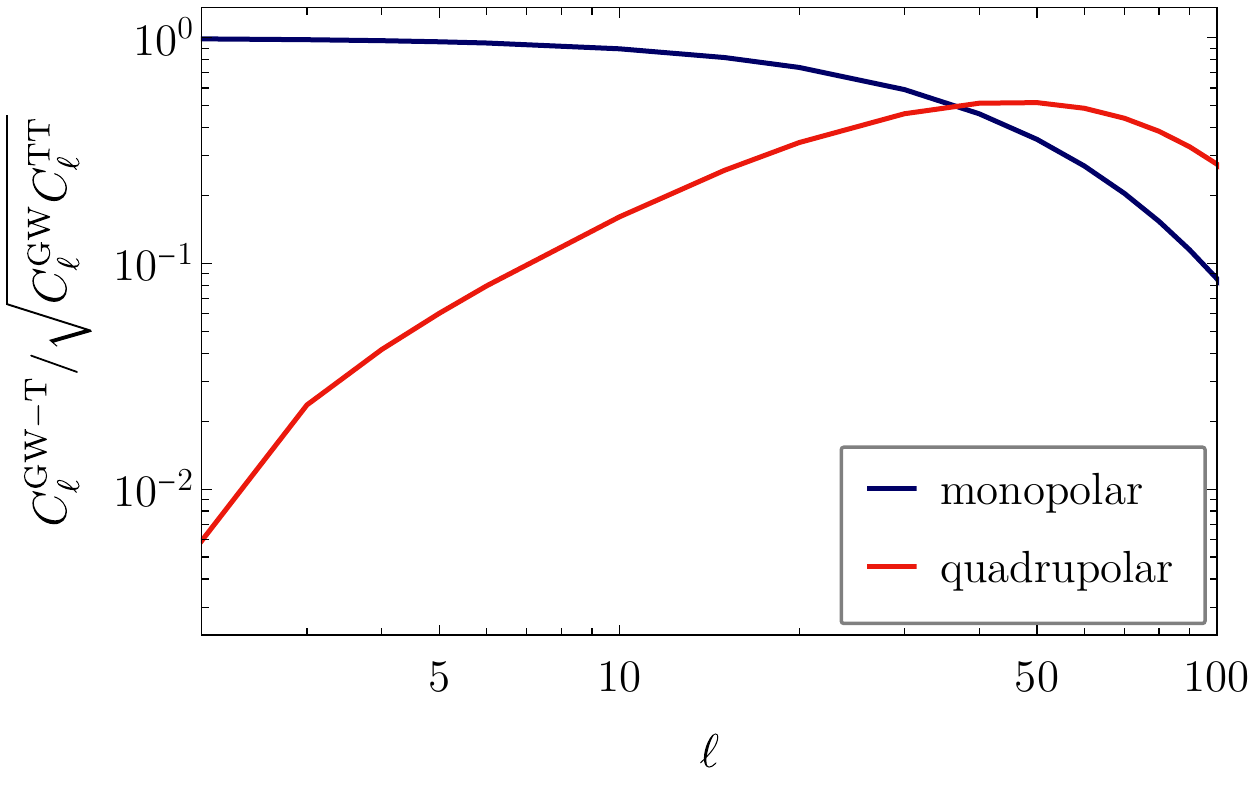}
    \caption{The ratio $\GT/\sqrt{\GG\TT}$ plotted for the monopolar and the quadrupolar TTS cross-correlation. The corresponding ratio for the induced anisotropies follows the same curve as the one for the monopolar TTS.}
    \label{fig:cross_ratio}
\end{figure}

Overall, from the results of this section one can expect that the cross-correlation for the quadrupolar TTS will be smaller compared to the one for the monopolar TTS which will have important consequences for cross-correlation-based observations/constraints on CGWB anisotropies. We can understand why this is to be expected by looking at the expression for the relative error in estimating the individual $\GT$ (e.g. see \cite{Afshordi:2003xu}),
\begin{align}
    \delta \GT \propto \left[{\frac{\TT \GG}{(2\ell+1)(\GT)^2}}\right]^{1/2}\,.
\end{align} 
The ratio $\GT/\sqrt{\GG\TT}$ is plotted in Fig.~\ref{fig:cross_ratio} for the monopolar and the quadrupolar cross-correlations. In the low $\ell$ range which is the relevant range for GW detectors\footnote{Even though the ratio $\GT/\sqrt{\GG\TT}$ increases initially for the quadrupolar case, higher multipoles do not offer any improvement on the constraints. This is due to the fact that the angular resolution of GW detectors is quite poor and we typically have $\GG \simeq \NGW$, i.e. a noise dominated map with the $N_{\ell}^{\text{\tiny GW}}$ increasing quite rapidly with $\ell$ whereas the signal typically decreases. Thus, in practice one is limited to $\ell_{\rm max} \sim 15\text{--}30$ (e.g. see \cite{Contaldi:2020rht,Alonso:2020rar} for ground-based networks and LISA, and see Fig.~\ref{fig:Nell_Omega} for other examples).}, we have $(\GT)^2\simeq \GG \TT$ for the monopolar TTS, whereas for the quadrupolar case we have $(\GT)^2\ll \GG \TT$. Thus, based on the above considerations, we expect cross-correlations to be more effective in the case of the monopolar TTS as compared to the quadrupolar TTS. We  confirm  this in the following section.

\subsubsection{\texorpdfstring{Projected constraints on $\tlfnls$}{Projected constraints on Fnl TTS}}
\label{sec:tts_error}
\begin{figure}
    \centering
    \includegraphics[width=0.95\textwidth]{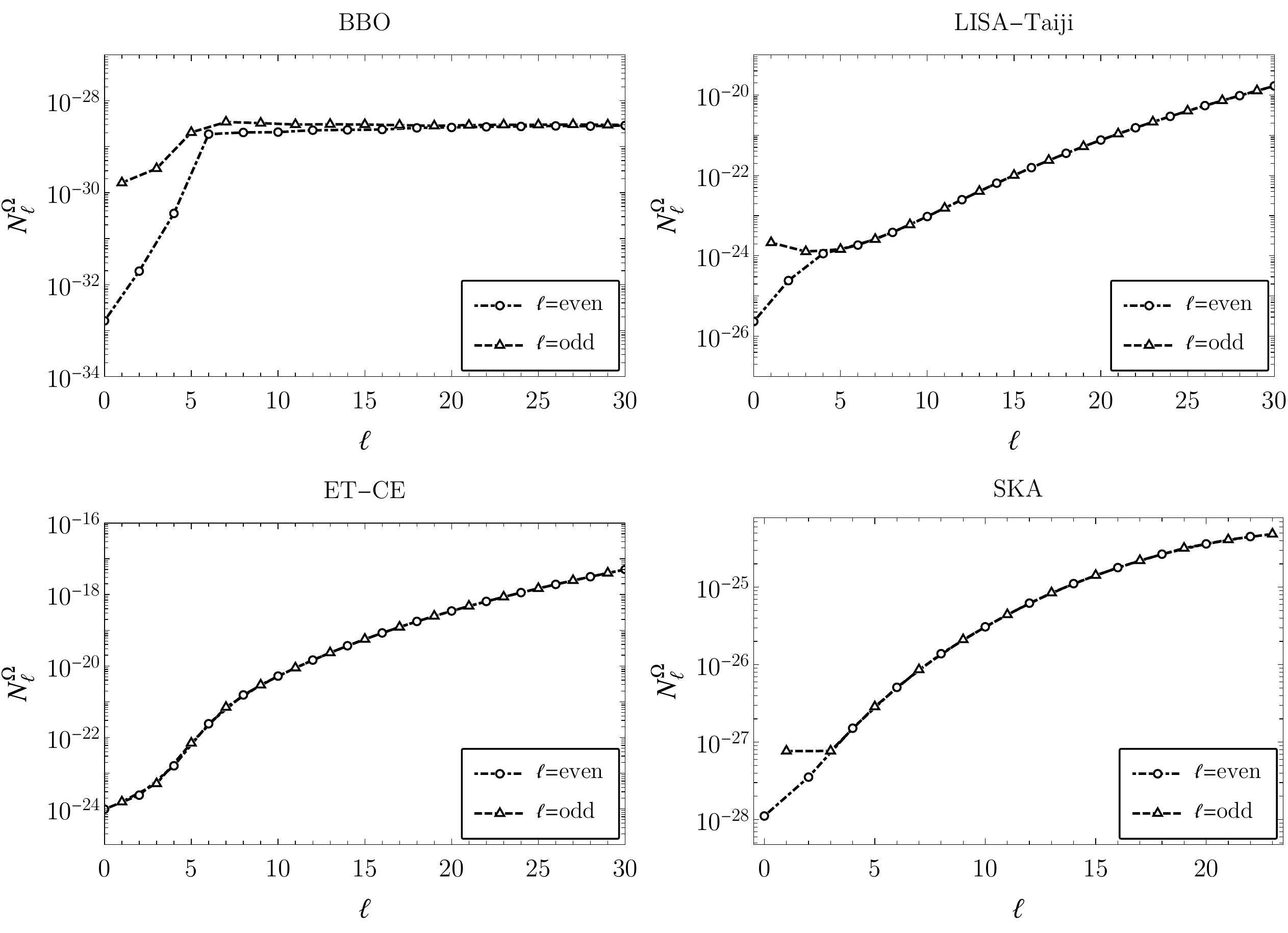}
    \caption{$N_{\ell}^{\Omega}$ plotted for BBO, LISA-Taiji, ET-CE and SKA at $f_{\rm ref}^{\rm BBO}=0.1$ Hz, $f_{\rm ref}^{\rm LISA-Taiji}=0.01$ Hz, $f_{\rm ref}^{\rm ET-CE}=63$ Hz and $f_{\rm ref}^{\rm SKA}=1\, {\rm year}^{-1}$ Hz. The quantity $\nell$ is defined as $\nell \equiv N_{\ell}^{\Omega}/\overline{\Omega}^2_{\rm GW}$}
    \label{fig:Nell_Omega}
\end{figure}

We now estimate the error in the measurement of $\tlfnls$ using a joint auto- and cross-correlation measurement. The Fisher matrix in this case is given by \cite{Verde_2010,Malhotra:2020ket},
\begin{align}
F_{ ij} = \sum_{X Y}\sum_{{\ell =}\ell_{\rm min}}^{\ell_{\rm max}}\frac{\partial C^{X}_{\ell}}{\partial \theta_i}\left(\mathscr{C}_{\ell}^{XY}\right)^{-1} \frac{\partial C^{Y}_{\ell }}{\partial \theta_j}\,,
\label{eq:Fisherdef}
\end{align}
where $X,Y = \{{\text{TT,GW,GW-T}}\}$ and $\vec{\theta}_i$ are the parameters being measured. The elements of the matrix $\mathscr{C}_{\ell}$ are
\begin{align}
\mathscr{C}_{\ell} = \frac{2}{2\ell+1}\begin{bmatrix}
(\TT)^2 & (\GT)^2 & \TT \GT \\
(\GT)^2 & (\GG)^2 & \GG\GT \\
\TT\GT & \GG\GT & \frac{1}{2}(\GT)^2 + \frac{1}{2}\TT\GG
\end{bmatrix}\,.
\label{eq:clmatrix}
\end{align}
The error is then estimated as $\Delta \theta_i = \sqrt{(F^{-1})_{ii}}$ with
\begin{align}
    \TT &\simeq \frac{2\pi A_S}{25\ell(\ell+1)},\nonumber\\
    \GG &= \GGS+\GGind+\NGW,\\
    \GT &= \GTS+\GTind,\nonumber
\end{align}
with the $N_{\ell}^{\text{\tiny GW}}$ being the noise angular power spectra of the detector network being used for the measurement (Fig.~\ref{fig:Nell_Omega}).
The calculation of the $\nell$ is based on the formalism of \cite{Alonso:2020rar} and employs the associated code schNell\footnote{\url{https://github.com/damonge/schNell}}. The $\nell$ for ET-CE were already calculated in \cite{Alonso:2020rar} and following \cite{Malhotra:2020ket} we have adapted this code to calculate the $\nell$ for BBO, LISA-Taiji and SKA. The details of their detector configurations and noise curves are described below.
\subsubsection*{BBO}
We consider the full BBO configuration with 4 LISA-like constellations, 2 of which will be arranged as a six-pointed star. In addition there will be 2 outer constellations trailing and leading the star constellation by $120\Deg$ in an earth-like orbit around the sun \cite{Crowder:2005nr}. The full BBO configuration improves upon the star configuration by reducing the noise at the $\ell > 4$ multipoles (compare to Fig.~5 of \cite{Malhotra:2020ket}). 
The noise curve for BBO is given in Ref.~\cite{Smith:2016jqs} and the total time of observation is taken to be $T_{\rm obs}=4$ years. 
\subsubsection*{LISA-Taiji}
The total time of observation is taken to be $T_{\rm obs}=4$ years and the noise curves for LISA and Taiji are obtained from \cite{Smith:2019wny} and \cite{Ruan:2020smc} respectively. Both LISA and Taiji will be in an earth-like orbit around the sun with an angular separation of $40\Deg$ \cite{Ruan:2020smc}.
\subsubsection*{ET-CE}
The total time of observation is taken to be $T_{\rm obs}=4$ years and the noise curves for ET\footnote{\url{http://www.et-gw.eu/index.php}} and CE\footnote{\url{https://dcc.ligo.org/LIGO-T1500293/public}
} are also available online. The locations for ET and CE are taken to be the same as those assumed in \cite{Alonso:2020rar}.
\subsubsection*{PTAs}
We consider a futuristic PTA experiment like SKA with a network of $N_{\rm psr}$ identical pulsars distributed isotropically across the sky whose timing noise is of the form \cite{Hazboun:2019vhv},
\begin{align}
	N_f = 2\sigma_t^2 \Delta T\,.
\end{align} 
Here $1/\Delta T$ is the cadence of the observations and $\sigma_t$ is the rms error of the timing residuals. For SKA we assume the following set of values $N_{\rm psr}=50$, $\Delta T=2$ weeks, $\sigma_t = 30$ ns and a total time of observation $T_{\rm obs}=20$ years. The choice of values is similar to that considered in \cite{Moore:2014lga}. This estimate for the $\nell$ could be made more realistic by dropping the assumption of identical pulsars and including additional sources of noise, e.g. a red timing noise as well as correlated noise sources arising from clock or solar system ephemeris errors \cite{NANOGrav:2020bcs}. One could also consider including more pulsars ($N_{\rm psr}\sim 100\operatorname{--}1000$), as expected for SKA2 \cite{Weltman:2018zrl}. We leave this for future work.\\

\begin{figure}
    \begin{tabular}{c}
        \includegraphics[width=0.98\linewidth]{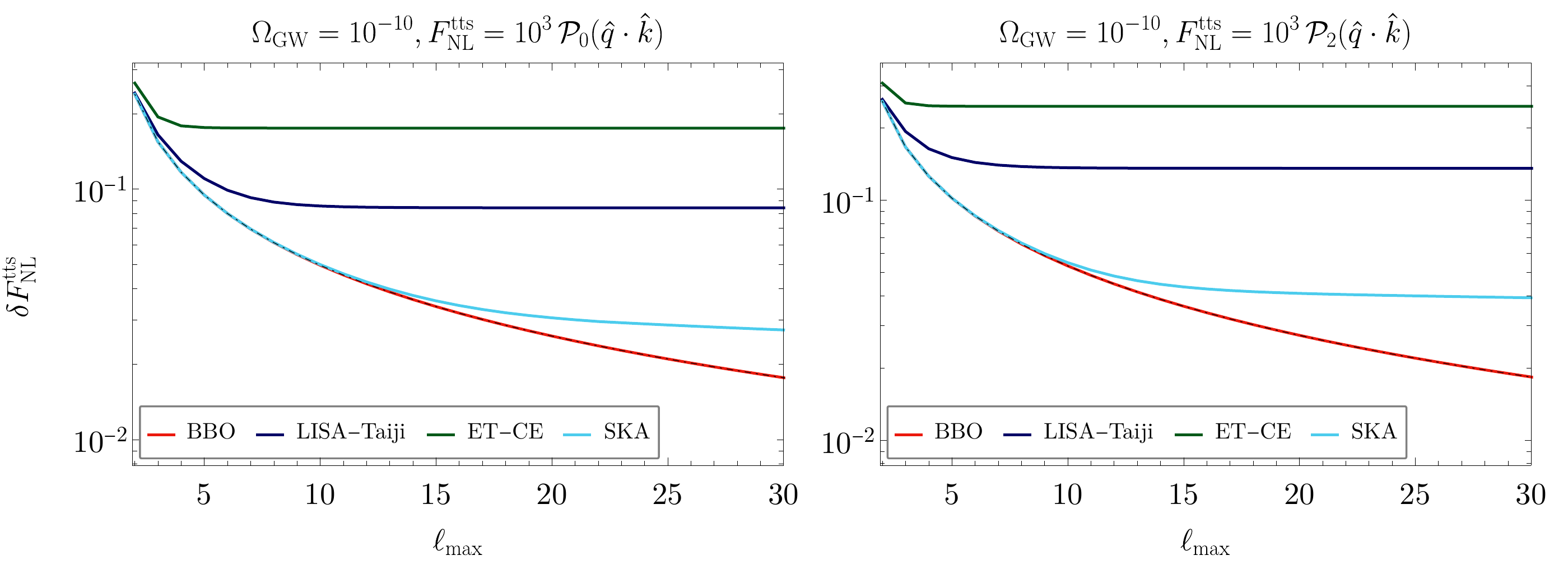} \\
        \includegraphics[width=0.98\linewidth]{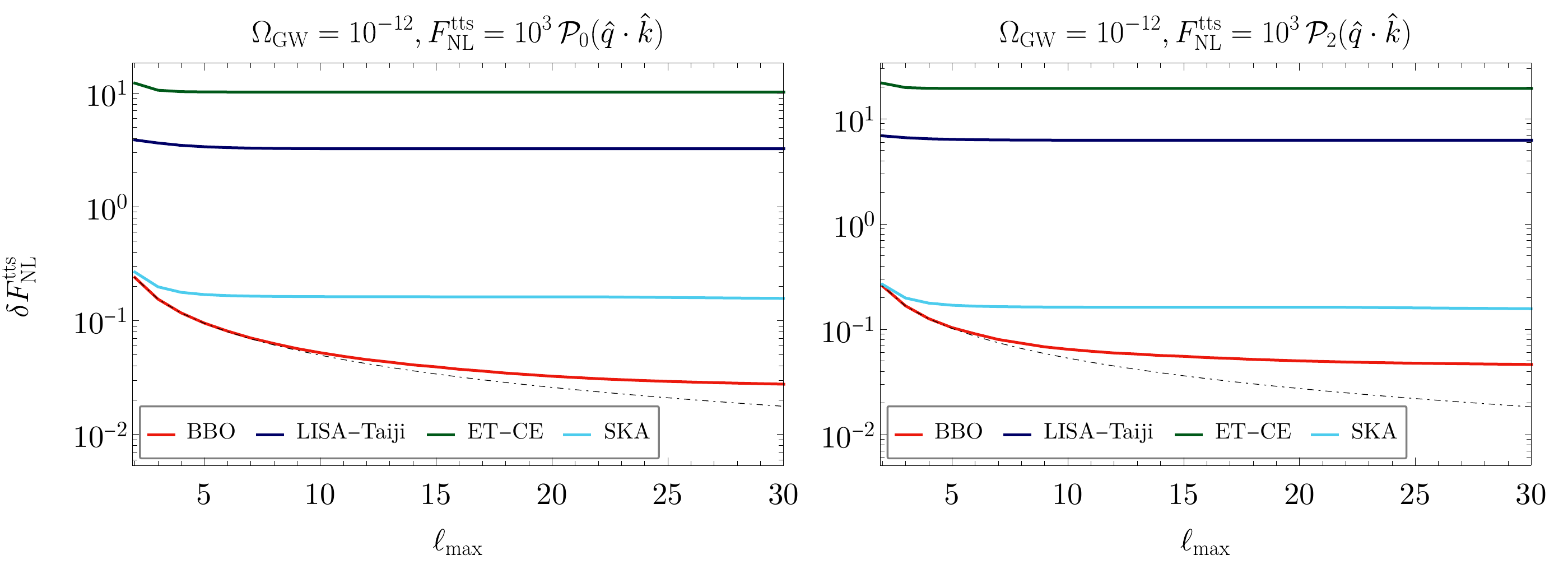} \\
    \end{tabular}
    \caption{The relative error in the measurement of $\tlfnls$ as a function of $\ell_{\rm max}$ for BBO, SKA, LISA-Taiji and ET-CE. The dashed curves show the errors for an idealised, cosmic variance limited measurement.}
    \label{fig:error_omega_tts}
\end{figure}

\noindent In Fig.~\ref{fig:error_omega_tts} we plot the relative error in the measurement of $\tlfnls$ defined as
\begin{align}
    \delta\tlfnls\equiv {\Delta \tlfnls}/{\tlfnls}\,,
\end{align} for different values of $\Omegagw$ and taking a scale independent $\tlfnls=10^3$. 
In calculating the error we have also assumed for simplicity a spectrum for $\Omegagw$ that is flat on small scales, in the frequency range relevant to the particular GW detector. For reference, note that we previously defined
\begin{align}
    \fnls(\vk,\vq) = \tlfnls\,\mathcal{P}_\ell(\hat q\cdot \hat k)
\end{align}
in terms of the Legendre polynomials with $\ell=0,2$ for the monopolar and quadrupolar cases respectively. 
We see that for a large value of the CGWB monopole $\Omegagw=10^{-10}$ we can achieve a relative error $\delta\tilde{F}_{\rm NL}^{\rm tts}\simeq 10^{-2}$ with BBO for both the monopolar and the quadrupolar $\fnls$ and a slightly larger error with SKA. With the ET-CE and LISA-Taiji networks the relative error in this case is of the order $10^{-1}$ and saturates quickly around $\ell_{\rm max}\sim 10$ due to their lower sensitivity compared to BBO. For a smaller value of $\Omegagw=10^{-12}$ only BBO and SKA are able to detect $\fnls$, reaching a relative error of the order $10^{-1}$. For both values of $\Omegagw$ BBO is cosmic variance limited, especially for the monopolar TTS. We also see that the error is smaller in the case of the monopolar TTS as compared to the quadrupolar one, which is to be expected from the discussion of the previous section.

\subsubsection*{Astrophysical foregrounds}
\label{sec:astro_fg}
The analysis carried out so far implicitly assumes that only the cosmological background contributes to the SGWB. However, in addition to this background, one also expects a background of gravitational waves arising from unresolved astrophysical sources to contribute to the SGWB and its anisotropies \cite{Regimbau:2011rp,Cusin:2017fwz,Cusin:2018ump,Cusin:2018rsq,Cusin:2019jhg,Cusin:2019jpv,Jenkins:2018kxc,Jenkins:2018uac,Jenkins:2019nks,Jenkins:2019uzp,Bertacca:2019fnt,Pitrou:2019rjz,Capurri:2021zli}. Detecting the cosmological background in the presence of this astrophysical foreground will be a major challenge and various methods to separate the monopoles of these backgrounds have been proposed in \cite{Regimbau:2016ike,Pan:2019uyn,Sharma:2020btq,Pieroni:2020rob,Biscoveanu:2020gds,Martinovic:2020hru,Poletti:2021ytu,Boileau:2020rpg,Boileau:2021sni}. Importantly, these techniques exploit the fact that astrophysical and cosmological backgrounds have different properties w.r.t their frequency range, spectral dependence, yearly modulation (e.g. in the case of galactic binaries), and can thus be distinguished from each other. In particular for the inflationary background, there is the possibility of detection on vastly different scales ranging from the CMB up to interferometers whereas the various different astrophysical backgrounds are each expected to be limited to a much smaller frequency window. \\ \indent Moving beyond the monopole, one can also study the prospects of detecting the anisotropies of such backgrounds, both cosmological and astrophysical. Since the properties of the astrophysical background depend strongly on the distribution of the large scale structure, it will have cross-correlations with probes like galaxy clustering and weak lensing \cite{Cusin:2017fwz,Cusin:2018rsq,Canas-Herrera:2019npr,Alonso:2020mva,Yang:2020usq,Banagiri:2020kqd}. As for cross-correlation of the AGWB anisotropies with the CMB, this is a direction which is being actively investigated \cite{Ricciardone:2021kel}. Interestingly, the findings of \cite{Ricciardone:2021kel} suggest that cross-correlating the CMB with the cosmological SGWB provides a stronger signal\footnote{A GW signal of cosmological origin that is detectable at e.g. BBO frequencies is assumed here.}. One may then  exploit these cross-correlations to help distinguish between astrophysical and primordial anisotropies.

To highlight the effectiveness of this approach, we compute here the signal to noise ratio of the cross-correlation of the primordial GW anisotropies with the CMB in the presence of an astrophysical background which acts as a foreground to the primordial signal.  The signal to noise ratio (SNR) of this cross-correlation is defined as
\begin{align}
\SNRx = \left[\sum_{\ell_{\rm min}}^{\ell_{\rm max}}(2\ell+1)\frac{\left(C_{\ell}^{\rm{ GW-T,signal}}\right)^2}{\left(C_{\ell}^{\rm{ GW-T,total}}\right)^2 + C_{\ell}^{\rm{ GW,total}}\TT}\right]^{1/2} \, ,  
\label{eq:snr_def}
\end{align}
where
\begin{align}
    &C_{\ell}^{\rm{ GW-T,signal}} = \GTS\,\nonumber\\ 
    &C_{\ell}^{\rm{ GW-T,total}} =  C_{\ell}^{\rm{ GW-T,signal}}+ C_{\ell}^{\rm{ GW-T,induced}}\,,\\
    &C_{\ell}^{\rm{ GW,total}} = C_{\ell}^{\rm{ GW,tts}}+C_{\ell}^{\rm{ GW,induced}}+C_{\ell}^{\rm{GW, astro}}+\nell\,.\nonumber
\end{align}
We assume an astrophysical background of the form $(\ell+1/2)C_{\ell}^{\rm{ GW, astro}}\approx A_{\rm GWB} $, which is based on the astrophysical models of \cite{Cusin:2018rsq,Cusin:2019jhg,Cusin:2019jpv}. The upper limit $A_{\rm GWB}=10^{-25}$ roughly corresponds to the expected magnitude of the background around $f=63$ Hz, while the lower limit $A_{\rm GWB}=10^{-30}$ corresponds to the magnitude around $f=0.01$ Hz \cite{Cusin:2019jhg,Cusin:2019jpv}. Since our aim here is to estimate the SNR for the cross-correlation at the frequency range relevant to BBO, we assume for simplicity that the quantity $A_{\rm GWB}$ takes on values between these two limits.
\begin{figure}
    \centering
    \includegraphics[width=0.7\textwidth]{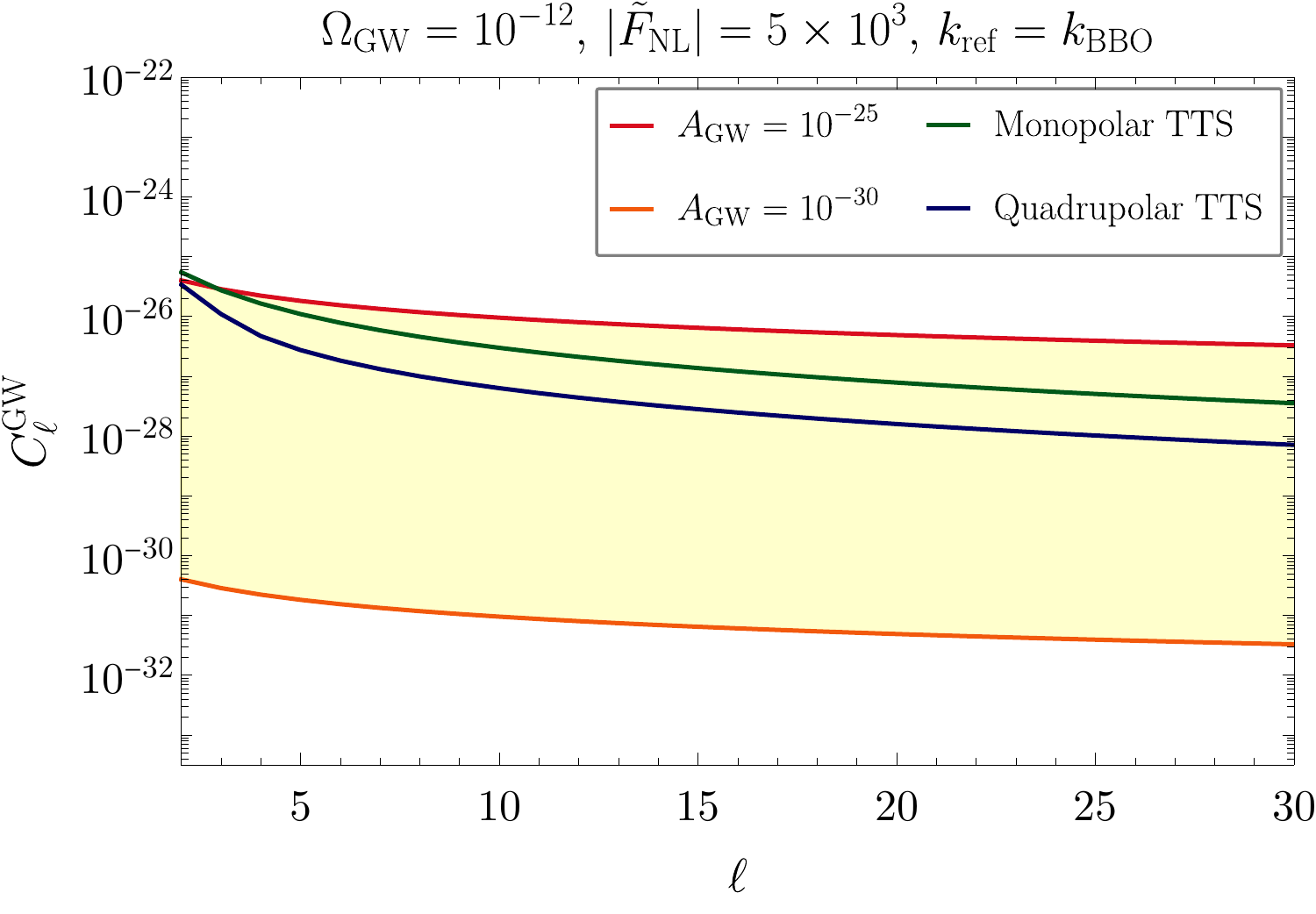}
    \caption{The $\GG$ for the astrophysical background (yellow shaded region) and for the CGWB with $\Omegagw=10^{-12},|\tlfnls|=5\times 10^3$. 
    We have assumed an astrophysical background of the form $(\ell+1/2)C_{\ell}=A_{\rm GWB}$. For comparison the $\GG$ of the CGWB have been rescaled as $\GG\to\overline{\Omega}_{\text{\tiny GW}}^2 \GG$.}
    \label{fig:astro_gw_1}
\end{figure}
The resulting SNR is plotted in Fig.~\ref{fig:snr_cross}. As discussed earlier in this section, the SNR for the monopolar TTS cross-correlation is much larger than that of the quadrupolar TTS. The primordial signal can be detected at a statistically significant level even for the upper limit of the astrophysical background, $A_{\rm GWB}=10^{-25}$, despite the fact this upper limit is larger than the primordial $\GG$ whenever $\Omegagw<10^{-12}$ and $|\tlfnls|<5\times 10^3$ (see Fig.~\ref{fig:astro_gw_1}). Thus, even if a direct observation of the primordial anisotropies is made difficult by the presence of the astrophysical foreground, cross-correlations with the CMB can still prove to be useful. We also see that for the quadrupolar TTS, a similar SNR is possible for a much weaker astrophysical signal, with $A_{\rm GWB}=10^{-27}$ or lower.
 \begin{figure}
 \begin{minipage}{0.49\textwidth}
    \centering
\includegraphics[width=0.99\linewidth]{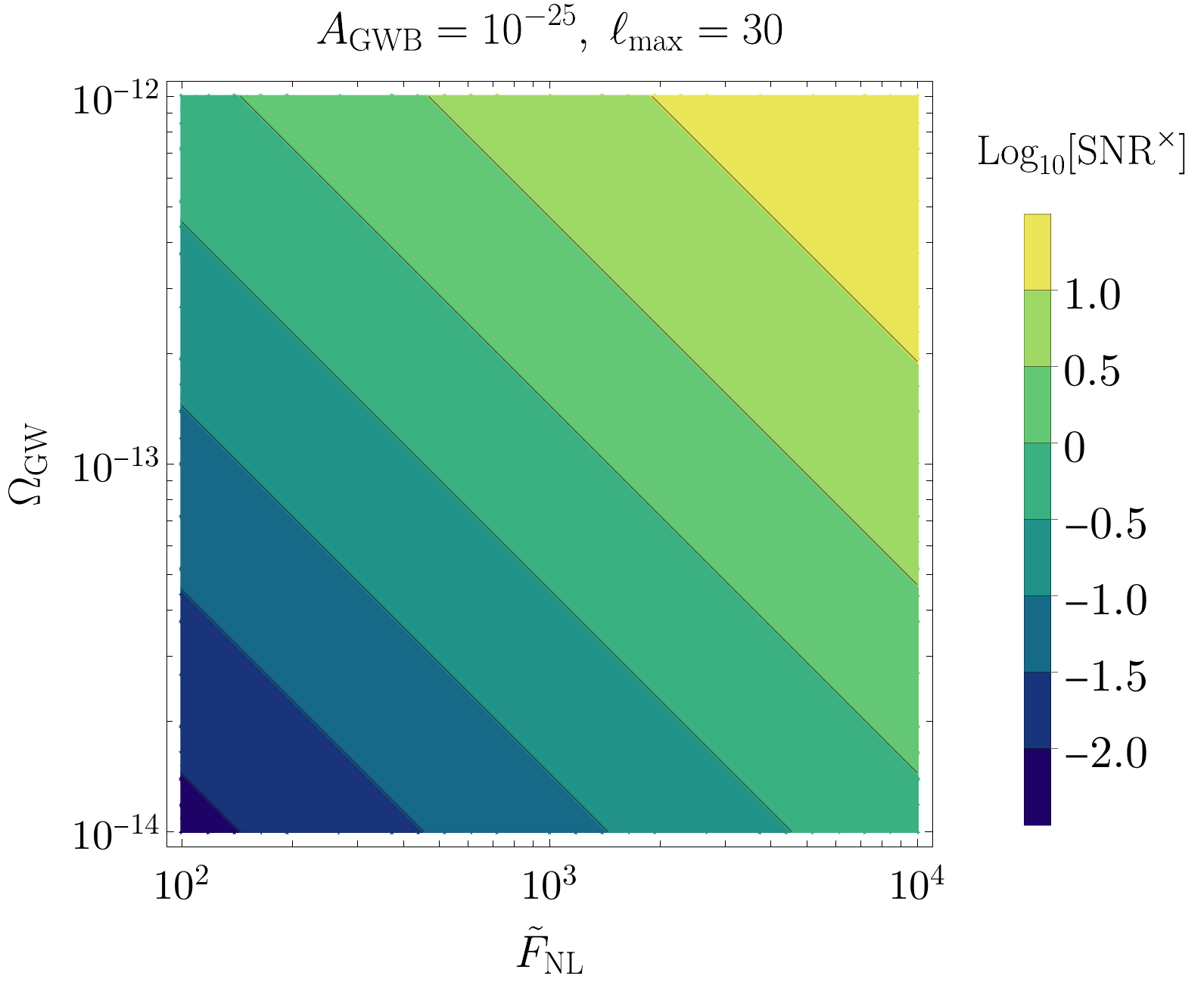}
\end{minipage}    
\begin{minipage}{0.49\textwidth}
    \centering
    \includegraphics[width=0.99\linewidth]{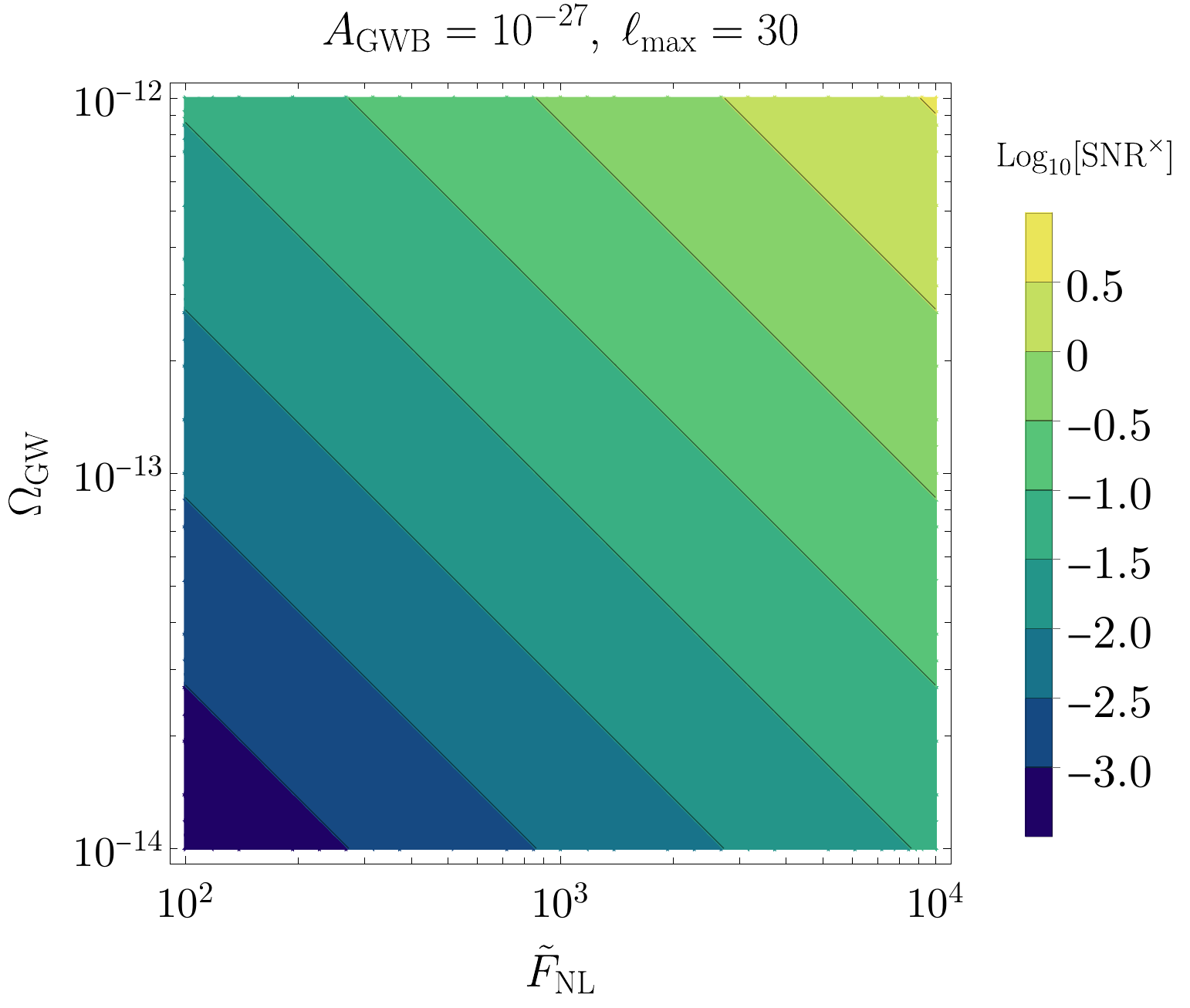}
\end{minipage}
    \caption{Left : SNR for the GW-CMB cross-correlation arising from the monopolar TTS bispectrum plotted as a function of $\Omegagw$ and $|\tlfnls|$, taking $(\ell+1/2)C_{\ell}=A_{\rm GWB}$ for the astrophysical background. Right: SNR for the same cross-correlation but for the quadrupolar TTS bispectrum. }
    \label{fig:snr_cross}
\end{figure}
\subsection{Cross-correlations with anisotropies from \texorpdfstring{$\langle \gamma \gamma \gamma \rangle$}{TTT} bispectrum}
\begin{figure}
    \centering
    \includegraphics[width=0.75\linewidth]{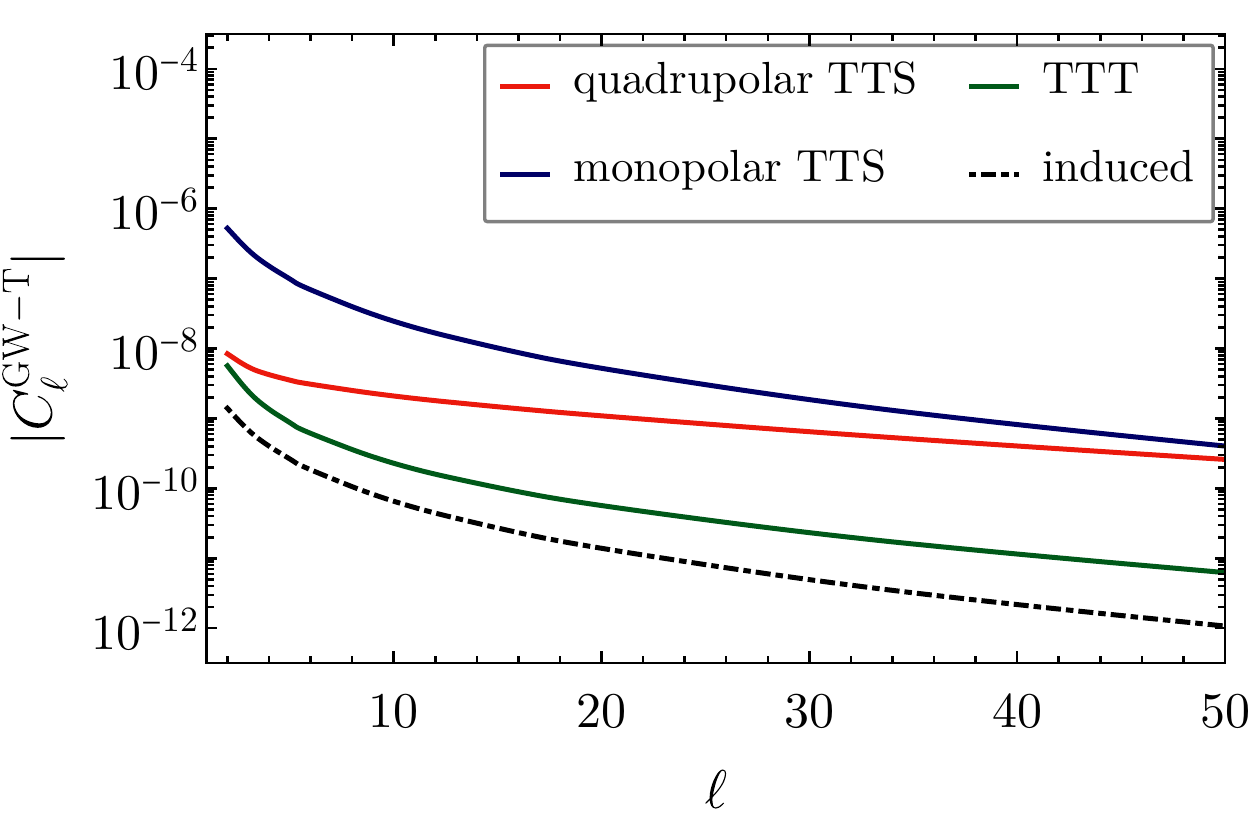}
     \caption{{The cross-correlation of the SGWB anisotropies as a function of $\ell$ plotted for $|\tlfnls|=|\fnlt|=10^3$ and a tensor-to-scalar ratio $r=0.05$.}} 
    \label{fig:cross_model_ind}
\end{figure}

We now calculate the cross-correlation of the intrinsic GW anisotropies from the TTT bispectrum, Eq.~\eqref{eq:deltagw_ttt}, with the CMB temperature anisotropies sourced by the large scale tensor modes. This contribution to the CMB temperature anisotropies can be written as \cite{Challinor_2009},
\begin{align}
    \deltaT(\hat n) = -\frac{1}{2}\sum_{s=\pm 2}\int\,
    d\eta\,\frac{d^3q}{\left(2\pi\right)^3}\frac{\partial\gamma^{s}_{\vq}}{\partial\eta} \epsilon_{ij}^{s}(\hat q)\hat{n}^i \hat{n}^je^{-i\chi(\eta)\hat n\cdot \vq}\,,
\end{align}
where $\chi(\eta)=\eta_0-\eta$ and the integral over the conformal time spans from $\eta_i=\eta_{\rm rec}$ to $\eta_f = \eta_0$. 
The spherical harmonic coefficients can be obtained using Eq.~\eqref{eq:tensor_id},
\begin{align}
    \deltaT_{\ell m} = \pi \, (-i)^\ell\sqrt{\frac{(\ell+2)!}{(\ell-2)!}}
    \sum_{s=\pm 2}\int\,
    d\eta\,\frac{d^3q}{\left(2\pi\right)^3}\frac{\partial\gamma^{ s}_{\vq}}{\partial\eta}\frac{j_{\ell}(q\chi(\eta))}{(q\chi(\eta))^2}{}_{- s}Y_{L M}^*(\hat q) \, .
\end{align}
The resulting cross-correlation is
\begin{align}
    \GTT = \frac{(\ell-1)\ell(\ell+1)(\ell+2)}{4\pi}&\sum_{s=\pm 2}\int_{q\ll k} q^2 dq\,F_{\rm NL}^{ttt}(\vk,\vq) P_{\gamma}^{s}(q)\frac{j_\ell(qd)}{(qd)^2}\nonumber\\
    &\times\int d\eta\,\frac{\partial\mathcal{T}(k,\eta)}{\partial\eta}\frac{j_{\ell}(q\chi(\eta))}{(q\chi(\eta))^2}\,.
\end{align}
For modes that re-enter the horizon after the universe becomes matter dominated ($k<k_{\rm eq}$) we have \cite{Watanabe:2006qe},
\begin{align}
    \mathcal{T}_{\gamma}(k,\eta) = \frac{3j_1(k,\eta)}{k\eta}\,.
\end{align}
Limiting ourselves to these modes with $k<k_{\rm eq}$, we can approximate the cross-correlation as
\begin{align}
    \GTT \simeq \frac{(\ell-1)\ell(\ell+1)(\ell+2)}{4\pi}&\sum_{s=\pm 2}\int_{1/\eta_0}^{k_{\rm eq}} q^2 dq\,\fnlt(\vk,\vq) P_{\gamma}^s(q)\frac{j_\ell(qd)}{(qd)^2}\nonumber\\
    &\times\int_{\eta_{\rm rec}}^{\eta_0} d\eta\,\frac{\partial\mathcal{T}(k,\eta)}{\partial\eta}\frac{j_{\ell}(q\chi(\eta))}{(q\chi(\eta))^2}\,.
    \label{eq:ttt_cross}
\end{align}
The magnitudes of the different contributions to the cross-correlation are plotted in Fig.~\ref{fig:cross_model_ind}. In the next section, we will consider the signatures studied in this section for a specific realisation of inflation. The analysis of the projected constraints on $\fnlt$ is  presented in Sec.~\ref{sec:ttt_error_spin2}.

\section{Constraints on the extra spin-2 setup} \label{sec:model_spin}

\subsection{Description of the model}

 In this section, we consider an effective field theory approach to inflation comprising an extra spin-2 field $\sigma_{ij}$ non-minimally coupled to the inflaton \cite{Bordin:2018pca}. This direct coupling allows $\sigma_{ij}$ to be effectively light compared to the Hubble scale (avoiding the so-called Higuchi bound \cite{Higuchi:1986py}) and, in turn, the bispectrum to have a significant squeezed component (as already mentioned e.g. in \cite{Bordin:2018pca,Dimastrogiovanni:2018gkl}). The action describing this model is given by
\begin{equation}
S = S_{\pi} +S_{\sigma} +S_{\rm int} \, ,
\end{equation} 
where  $S_{\pi}$ denotes the standard  generalised slow-roll dynamics captured by the single-field EFT approach to inflation \cite{Cheung:2007sv},  $S_{\sigma}$ is the 
free action for the spin-2 field $\sigma_{ij}$ and $S_{\rm int}$ contains the quadratic and cubic mixing interactions\footnote{From here onwards $\tau$ denotes  conformal time and $'$ stands for differentiation with respect to conformal time.},
\begin{align}
\label{action}
\mathcal{S_{\rm int}} & = \;\; \, \int d\tau \, d^3x\, a^4 \Big[ -\frac{g}{\sqrt{2 \epsilon} H} a^{-2} \partial_i \partial_j \pi_c \sigma^{ij} +\frac{1}{2} a^{-1} g \,\gamma'_c\,_{ij} \sigma^{ij}  \Big]  \; \nonumber\\
&\;\;+\, \int d\tau \, d^3x\, a^4 \Big[ - \frac{g}{2\epsilon H^2 M_{\rm Pl}}a^{-2} (a^{-1} \partial_i\pi_c \partial_j\pi_c {\sigma'}^{ij}\\ \nonumber&\qquad \qquad \qquad\qquad\qquad+2H \partial_i\pi_c \partial_j\pi_c {\sigma}^{ij} ) - \mu(\sigma_{ij})^3 +\dots \Big] \; .
\end{align}
Canonically normalising the fields, we have $\gamma_c \equiv \gamma M_{\rm Pl}$, where $\gamma_{ij}$ describes  the standard traceless and transverse tensor fluctuations  and $\pi_c\equiv\sqrt{2 \epsilon} H M_{\rm Pl}\, \pi$. The field $\pi$ is the canonically normalised Goldstone boson, linearly related to the curvature fluctuation via $\zeta\simeq-H \pi$ \cite{Cheung:2007sv}.  The quantities $g,\mu$ are coupling constants.  The dots in Eq.~(\ref{action}) stand for higher-order mixing interactions. 

We can decompose the spin-2 field $\sigma^{ij}$ into the helicity states as 
\begin{align}
 {\sigma}_{ij} =  {\sigma}_{ij}^{(0)} + {\sigma}_{ij}^{(1)} + {\sigma}_{ij}^{(2)} \, .
\end{align}
As usual, we may neglect helicity-1 modes as they end up being diluted by the inflationary expansion (see e.g. \cite{Riotto:2002yw}). 
The (traceless and transverse) tensor degrees of freedom in the  theory, $\gamma_{ij}$ and $\sigma^{(2)}_{ij}$, can be expanded in the R/L-handed basis  
\begin{align}
\gamma_{ij} = \int \frac{d^3k}{(2 \pi)^3} e^{i \bk \cdot \bx} \sum_{\lambda = R/L} \epsilon^{\lambda}_{ij}(\hat k) \, \gamma^\lambda_\bk(\tau)\, ,
\end{align}
and 
\begin{align}
\sigma^{(2)}_{ij} = \int \frac{d^3k}{(2 \pi)^3} e^{i \bk \cdot \bx} \sum_{\lambda = R/L} \epsilon^{\lambda}_{ij}(\hat k) \, \sigma^{2,\lambda}_\bk(\tau)\, ,
\end{align}
where we have introduced the basis 
\ba \label{eq:def_pol}
\epsilon_{ij}^{R/L} &= \frac{1}{2} \left(\epsilon_{ij}^+ \pm  i \epsilon_{ij}^\times \right) \, ,
\ea
see App. \ref{app:pol_ten} for more details. The polarisation tensors obey the following normalisation conventions
\begin{align}
&\epsilon^{R/L}_{ij}(\hat k) \cdot \epsilon^{L/R}_{ij}(\hat k) = 1 \, , \\
&\epsilon^{R/L}_{ij}(\hat k) \cdot \epsilon^{R/L}_{ij}(\hat k) = 0 \, , \\
&\epsilon^{R/L *}_{ij}(\hat k) = \epsilon^{L/R}_{ij}(\hat k) = \epsilon^{R/L}_{ij}( - \hat k) \, .
\end{align}
We may also write  $\zeta$ and $\sigma^{(0)}_{ij}$, as 
\begin{align}
\zeta =& \int \frac{d^3k}{(2 \pi)^3} e^{i \bk \cdot \bx} \zeta_\bk(\tau) \, ,\\
\sigma^{(0)}_{ij} =& \int \frac{d^3k}{(2 \pi)^3} e^{i \bk \cdot \bx} \epsilon^{0}_{ij}(\hat k) \sigma^0_\bk(\tau) \, ,
\end{align}
where 
\begin{align}
\epsilon^{0}_{ij}(\hat k) = \sqrt{\frac{3}{2}} \left( \hat k_i \hat k_j - \frac{\delta_{ij}}{3}\right) \, ,
\end{align}
with $\epsilon^{0}_{ij}$  conforming to the following normalisation rule
\begin{equation}
\epsilon^{0}_{ij}(\hat k) \cdot \epsilon^{0}_{ij}(\hat k) = 1 \, .   
\end{equation}
One may proceed to quantise fields $X_\bk(\tau)$ by expanding in terms of annihilation and creation operators,
\begin{align}
X_\bk(\tau) = a_{\bf k} u^{X}_{\bf k}(\tau) + a_{- \bf k}^\dag u^{X,*}_{\bf - k}(\tau) \, .
\end{align}
We report below the mode functions of $\zeta$ and $\gamma$,
\begin{align} \label{eq:uzeta}
u^{\zeta}_{\bf k}(\tau) = \frac{i H}{2 M_{Pl} \sqrt{\epsilon k^3}} (1 + i k \tau) e^{- i k \tau} \, , \\
u^{\gamma}_{\bf k}(\tau) = \frac{2 i H}{M_{Pl} \sqrt{2 k^3}} (1 + i k \tau) e^{- i k \tau} \, , \label{eq:ugamma}
\end{align}
and those for the $\sigma_{ij}$ field,
\begin{align}\label{newwf1}
u^{\sigma^{(2)}}_{\bf k}(\tau) = \sqrt{\frac{\pi}{2}} H (-\tau)^{3/2}\left(\frac{c_{2}(\tau)}{c_{2i}}\right)^{1/2} \mathcal H_\nu^{(1)}(- c_2(\tau) k \tau) \, ,\\\label{newwf2}
u^{\sigma^{(0)}}_{\bf k}(\tau) = \sqrt{\frac{\pi}{2}} H (-\tau)^{3/2} \left(\frac{c_{0}(\tau)}{c_{0i}}\right)^{1/2} \mathcal H_\nu^{(1)}(- c_0(\tau) k \tau) \, .
\end{align}
Here $\nu = \sqrt{9/4 - (m_\sigma^2/H^2)}$, $m_\sigma$ being the mass of the ``extra'' spin-2 field, $\mathcal H_\nu^{(1)}$ is the Hankel function of the first kind, and $c_{0/2}$ denote the sound speeds of the helicity-0/2 components, with $c_{0i}$ and $c_{2i}$ the initial sound speeds. We verified\footnote{See e.g. \cite{Chen:2006nt} for the full derivation of the mode functions in the case of a time-varying sound speed. Our solutions for the wave-functions coincide with   those obtained in \cite{Chen:2006nt} 
in terms of canonically normalised fields.} that the solutions (\ref{newwf1}) and (\ref{newwf2}) reproduce the Bunch-Davies vacuum at early times, and that they reduce to those obtained in the $c_{0,2}=$ \textsl{constant} case when the sound speeds are time-independent \cite{Bordin:2018pca}.
 By combining together the expressions for the different sound speeds, the following independent relation follows \cite{Bordin:2018pca}
\begin{equation} \label{eq:sound_speeds_rel}
c_1^2 = \frac{1}{4} c_2^2 + \frac{3}{4} c_0^2 \, ,
\end{equation}
which connects the sound speeds of the different helicity modes. We shall require that the  expressions above are valid under the assumption of slowly varying sound speeds. Whenever the time dependence of the sound speeds is more sharp one ought to employ different approaches to the solutions (see e.g. \cite{Achucarro:2013cva}).

\subsection{Sound speed(s) scaling} \label{sub:time_sound}

In this work we will adopt scale dependent sound speeds. This can be taken simply as an ansatz but one may show that in terms of cosmological correlators this choice  corresponds, to a good approximation, to employing weakly time-dependent sound speeds. Let us start by considering the following parametrisation for the helicity-$j$ sound speed
\begin{equation}
\label{eq:sound_t}
c_j(t)=c_j^i\, e^{-s_j N} + c_j^f \, , 
\end{equation}
where $N = \int_{t_i}^t H(t') \, dt'$ is the number of e-folds between a given reference time $t_i$ and $t$, with $c_j^i = c_j(t_i)$. Our choice to add an asymptotic value $c_j^f$ is slightly different from the one adopted in \cite{Iacconi:2020yxn}: it ensures that, regardless of the duration of inflation,  there is a lower limit  $c_j^f \ll c_j^i$ in place for the sound speeds. This is very convenient in view of perturbativity bounds on $c_j$ as the latter impose a lower limit on $c_j$ of the order of $10^{-3}$. Each $s_j$ is taken to be a constant positive parameter. Their meaning is most clear in the $c_j^f \ll c_j^i \, e^{-s_j N}$ regime, where they approach the slow-roll parameter $\tilde{s}$ usually defined \cite{Seery:2005wm} as $\tilde s_j = \dot c_j/c_j H $. A weak time dependence for the sound-speeds is then tantamount to  requiring $s_j \ll 1$. In conformal time Eq.~\eqref{eq:sound_t} reads
\begin{equation} \label{eq:sound_tau}
c_j(\tau) = c_j^i \left(\frac{\tau}{\tau_i}\right)^{s_j} + c_j^f \, ,
\end{equation}
where for reference we take $\tau_i=1/k_0$ with $k_0 =  a_0 H_0$. It is well known that, when employing the in-in formalism for cosmological correlators, the main contribution comes from the time when mode-functions are at horizon crossing. This is because mode functions exhibit highly oscillating behaviour deep inside the(ir) horizon.\\
\indent As explicit in our calculations, each $\sigma$ mode function is proportional to functions of the   $\mathcal H_\nu^{(1,2)}[- c_j(\tau) \, k \tau]$ type, so the horizon is at $- c_j(\tau) \, k \tau\sim 1$. Given that $c_j$ values have a rather narrow range between $c_2^i$ and $c_2^f$, one may solve for $\tau$ and verify that (i) at CMB scales (i.e. $k=k_{\rm CMB}$) one finds $c_j(\tau^{\rm CMB})\simeq c_2^i$ and (ii) at e.g. BBO scales (i.e. $k=k_{\rm BBO}$) one finds $c_j(\tau^{\rm BBO})\simeq c_2^f$ to a very good approximation. This is precisely what one finds also in employing scale-dependent sound speeds:
\begin{equation} \label{eq:sound_kk}
c_j(k) = c_j^i \left(\frac{k}{k_0}\right)^{-s_j} + c_j^f \, ,
\end{equation}
which we shall adopt henceforth. The above line of reasoning is based on the notion that the key contributions to correlators come from the horizon of the mode functions at hand. This is easily verified to be the case when all wavefunctions in the integral corresponding to a given vertex share the same horizon. Whenever there are instead mode functions with different arguments (and therefore different horizons) in the same integral, the 
horizon of choice is clear: that corresponding to the mode that exits the horizon last (see e.g. \cite{Maldacena:2002vr,Chen:2009zp}). The reasoning is always the same: the integrand at any earlier time would display a highly oscillating behaviour due to at least one mode function. We will implement this criterion in all our in-in calculations.

\subsection{The \texorpdfstring{$\langle \gamma \gamma \gamma \rangle$ and $\langle \gamma \gamma \zeta \rangle$ bispectra}{TTT and TTS bispectra}}

\begin{figure}
\begin{minipage}{0.49\textwidth}
     \centering\includegraphics[width=\linewidth]{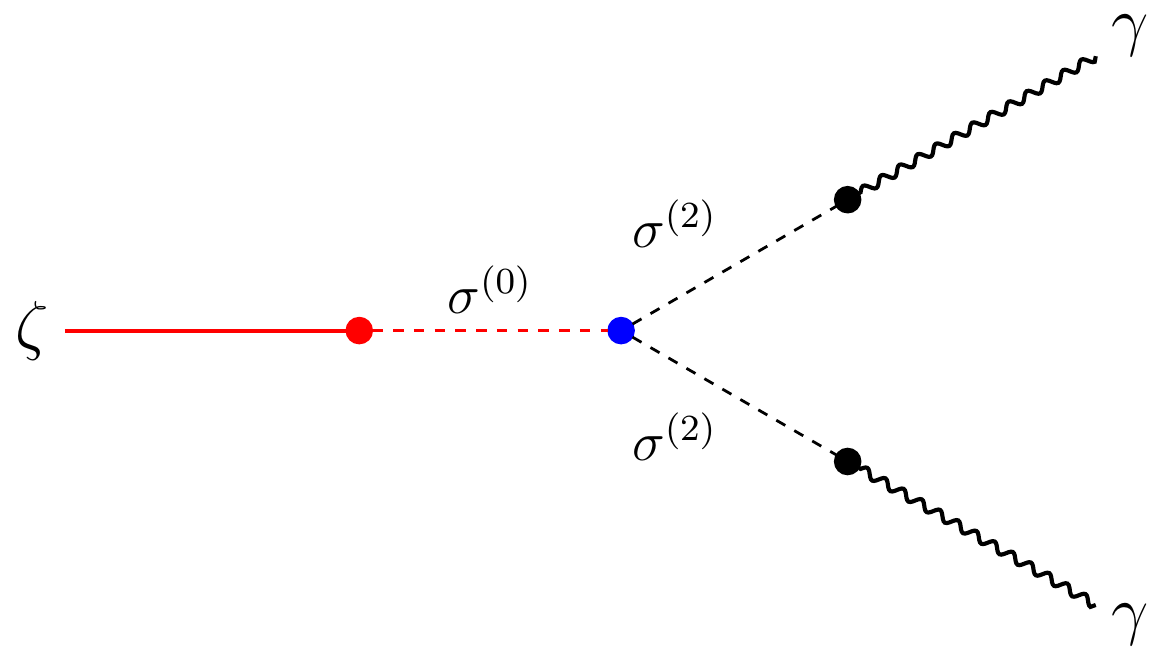}
\end{minipage}
\begin{minipage}{0.49\textwidth}
     \centering\includegraphics[width=\linewidth]{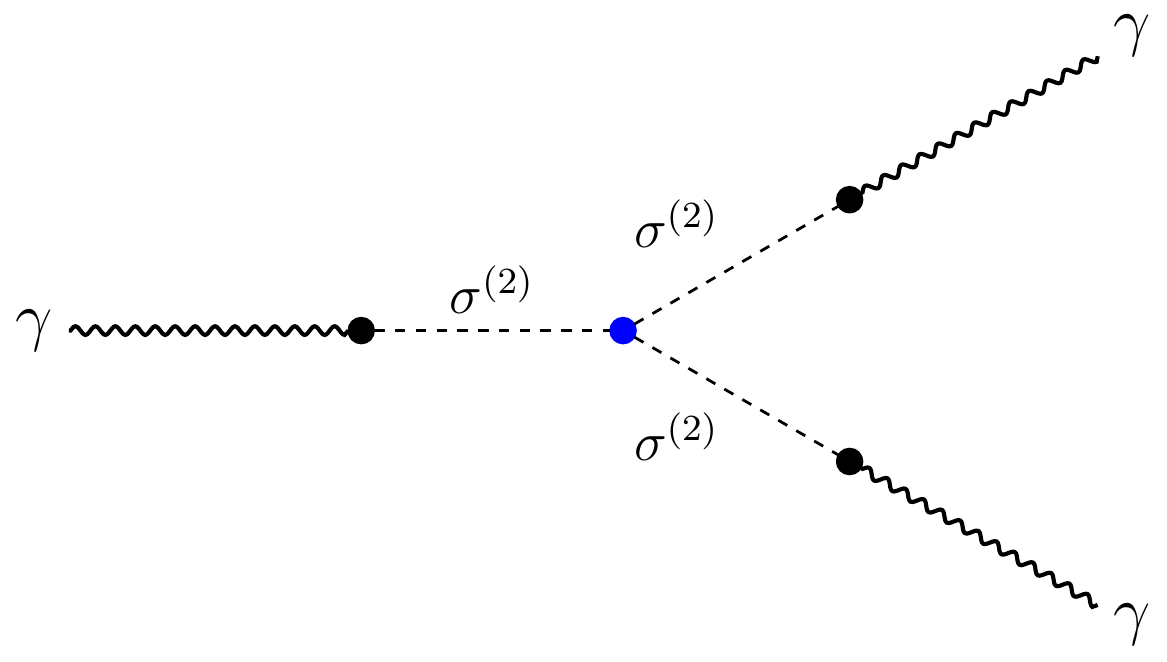}
\end{minipage}
\caption{Leading $\sigma$-mediated contributions to $\langle \gamma \gamma \zeta  \rangle$ and $\langle \gamma \gamma \gamma \rangle$. Straight lines correspond to $\zeta$, wiggly lines correspond to $\gamma$, red (black) dashed lines correspond to $\sigma^{(0)}$ ($\sigma^{(2)}$) fields.}
    \label{fig:feyn_diag}
\end{figure}

The mixing action in Eq.~\eqref{action} can be made more explicit by writing it in terms of helicity-fields
\begin{align}
\label{action2}
\mathcal{S_{\rm int}} & = \;\; \, \int d\tau \, d^3x\, a^4 \Big[ -\frac{g}{\sqrt{2 \epsilon} H} a^{-2} \partial_i \partial_j \pi_c \sigma^{(0),ij} +\frac{1}{2} a^{-1} g \,\gamma'_c\,_{ij} \sigma^{(2),ij}  \Big]  \; \nonumber\\
&\;\;+\, \int d\tau \, d^3x\, a^4 \Big[ - \frac{g}{2\epsilon H^2 M_{\rm Pl}}a^{-2} (a^{-1} \partial_i\pi_c \partial_j\pi_c {\sigma'}^{(0),ij} + a^{-1} \partial_i\pi_c \partial_j\pi_c {\sigma'}^{(2),ij}\\ 
\nonumber&\qquad \qquad \qquad+2H \partial_i\pi_c \partial_j\pi_c {\sigma}^{(0), ij} +2H \partial_i\pi_c \partial_j\pi_c {\sigma}^{(2), ij}) - \mu(\sigma^{(2), ij})^3 - 3 \mu \, \sigma^{(2)}_{ij}\cdot \sigma_{jk}^{(2)} \cdot \sigma_{ki}^{(2)} + \dots \Big] \; .
\end{align}
The interaction vertices of interest for our analysis are the following 
\begin{align}
& H_{\sigma^{(2)} \gamma} = - \int  d^3x \,\frac{g}{2} M_{Pl} \, a^3 \gamma'_{ij} \sigma^{(2), ij} = - \int  \frac{d^3q}{(2 \pi)^3} \,\frac{g}{2} M_{Pl} \, a^3 \sum_{\lambda= R/L} \gamma'^{\lambda}_{\bq} \, \sigma^{(2),  \lambda}_{-\bq} \, ,  \label{Hs2g}\\
& H_{(\sigma^{(2)})^3} = \int  d^3x \, \mu \, a^4  \, \sigma^{(2)}_{ij}\cdot \sigma_{jk}^{(2)} \cdot \sigma_{ki}^{(2)} = \int  \frac{d^3q}{(2 \pi)^3} \int  \frac{d^3q'}{(2 \pi)^3} \int  \frac{d^3q''}{(2 \pi)^3} (2 \pi)^3 \delta^{(3)}(\bq + \bq' + \bq'')  \nonumber \\
&  \qquad  \qquad \qquad \times  \mu \, a^4  \, \sum_{\lambda, \lambda', \lambda^{''}= R/L} \left(  \epsilon^\lambda_{ij}(\hat q) \cdot \epsilon_{jk}^{\lambda'}(\hat q') \cdot \epsilon_{ki}^{\lambda''}(\hat q'') \right)  \, \sigma^{(2), \lambda}_{\bq} \, \sigma^{(2), \lambda'}_{\bq'} \, \sigma^{(2), \lambda''}_{\bq''} \, , \label{Hsss}\\
& H_{(\sigma^{(2)})^2 \sigma^{(0)}} = \int  d^3x \, 3\mu \, a^4  \, \sigma^{(2)}_{ij}\cdot \sigma_{jk}^{(2)} \cdot \sigma_{ki}^{(0)} = \int  \frac{d^3q}{(2 \pi)^3} \int  \frac{d^3q'}{(2 \pi)^3} \int  \frac{d^3q''}{(2 \pi)^3} (2 \pi)^3 \delta^{(3)}(\bq + \bq' + \bq'')  \nonumber \\
&  \qquad  \qquad \qquad \times 3 \sqrt{\frac{3}{2}} \mu \, a^4  \, \sum_{\lambda, \lambda'= R/L} \left( \hat q_i \hat q_l \cdot \epsilon^{\lambda}_{ij}(\hat q') \cdot \epsilon_{jl}^{\lambda'}(\hat q'') - \frac{1}{3} \epsilon^\lambda_{ij}(\hat q') \cdot \epsilon_{ij}^{\lambda'}(\hat q'')\right) \, \sigma^{(0)}_{\bq} \, \sigma^{(2), \lambda}_{\bq'} \, \sigma^{(2), \lambda'}_{\bq''} \, , \label{Hsss0}\\
& H_{\sigma^{(0)} \zeta} =  -\int  d^3x \, \frac{g}{H} M_{Pl} \, a^2 \partial_i \partial_j \zeta \,\sigma^{(0), ij} =   \int \frac{d^3q}{(2 \pi)^3} \, \sqrt{\frac{2}{3}} \, \frac{g}{H} M_{Pl} \, a^2 q^2 \, \zeta_{\bq} \,\sigma^{(0)}_{-\bq} \, . \label{Hs0z}
\end{align}
These lead to $\sigma$-mediated contributions to primordial correlators (Fig.~\ref{fig:feyn_diag}), which can be evaluated with in-in techniques. In this work we are interested in the effects of such contributions to the three point functions $\langle \gamma \gamma \zeta \rangle$ and $\langle \gamma \gamma \gamma \rangle$. In particular, we are interested in those diagrams that end up giving the leading contributions to the three-point correlators above in view of our sampling the $c_2 \ll 1$ region of parameter space. As we show in Secs. \ref{sec:self_corr} and \ref{sec:cross}, this regime is the one for which the effect of primordial correlators on GW anisotropies is the strongest.

It is worth at this stage to comment on the squeezed limit of the STT and TTT correlators and consistency relations. For single-field slow-roll models inflation it is known \cite{Maldacena:2002vr,Hinterbichler:2013dpa} that the leading contribution of these bispectra in the squeezed limit is in fact a gauge artifact. The physical contribution is instead to be found at sub-leading order (i.e. one typically pays the price of a $k_L^2/k_S^2$ suppression). Consistency relations may instead be broken\footnote{We write ``broken'' to conform to standard terminology here. For our purposes it is enough for consistency relation to be modified, they can still be in place. In single field slow-roll the leading contribution to the three-point function has the same effect as a gauge transformation. In multi-field models it is often the case that a \textit{specific linear combination of different contributions}  to a diagram may be described as a gauge transformation, but, crucially, not each contribution taken on its own. There may be, in the multi-field case, more than one field that non-linearly transforms under the diffeomorphism behind the consistency relation and that is why it is a linear combination of contributions to correspond to the gauge transformation. See e.g. \cite{Assassi:2012zq} for interesting examples along these lines.} in multi-field scenarios, for non-Bunch-Davies initial conditions, in the case of non-attractor (followed by an attractor phase) solutions and for models with non-standard symmetry breaking patterns (e.g. solid inflation), to name a few. For the model at hand, it is straightforward to show the breaking of consistency conditions parametrically, in that certain coefficients appear only at cubic order and are not present in the (tree-level) power spectra \cite{Bordin:2018pca,Dimastrogiovanni:2018gkl}. The diagrams under study then give a contribution that is both the leading one and physical. The main $\sigma$-mediated diagram in the $\langle \gamma \gamma \gamma \rangle$ bispectrum is given by 
\begin{align}
    \langle \gamma_{k_1}^{\lambda_1} \gamma_{k_2}^{\lambda_2} \gamma_{k_3}^{\lambda_3}\rangle = (2\pi)^3\delta^{(3)}(\vec{k}_1 + \vec{k}_2+\vec{k}_3) \, \mathcal{A}^{\lambda_1 \lambda_2 \lambda_3}B_{\rm ttt}(k_1,k_2,k_3)\,, 
\end{align}
where the function $B_{\rm ttt}(k_1,k_2,k_3)$ in the squeezed limit $k_L = k_3 \ll k_1 \simeq k_2 = k_S$ reads
\begin{align} \label{eq:model_ttt}
    B_{\rm ttt}(k_S,k_S,k_L) = \frac{24\times 2^{\nu}\pi^2}{k_S^{9/2-\nu}k_L^{3/2+\nu}}\frac{\mu}{H}\left(\frac{g}{M_{\rm Pl}}\right)^3 \left(\frac{c_{2}(k_{S})}{c_{2i}}\right)^{5/2}\left(\frac{c_{2}(k_{L})}{c_{2i}}\right)^{1/2} \mathcal I(c_2, \nu)\, ,
\end{align}
and 
\begin{align}
   \mathcal A^{\lambda_1 \lambda_2 \lambda_3} &=  \epsilon^{\lambda_1}_{ij}(\hat{k}_1)\epsilon^{\lambda_2}_{jl}(\hat{k}_2)\epsilon^{\lambda_1}_{lk}(\hat{k}_3) \, .
\end{align}
In the same limit the quantity $\mathcal{A}$ reads
\begin{align}
   \mathcal A^{\lambda_1 \lambda_2 \lambda_3}|_{\rm sq.} = - \frac{1}{2} \hat k_S^i \hat k_S^j \, \epsilon_{ij}^{\lambda_1}(\hat k_L) \times \begin{cases}
1 \quad \mbox{if} \qquad \lambda_{1} = \lambda_{2} \\
0 \quad \mbox{if} \qquad \lambda_{1} \neq \lambda_{2}
\end{cases}\, .
\end{align}
The function $\mathcal I(c_2, \nu)$ is given by Eq.~\eqref{eq:I} upon replacing $c_0$ with $c_2$, obtaining
\begin{eqnarray}
\mathcal I(c_2,\nu) &=& \frac{\Gamma(\nu)}{c^\nu_2(k_S)} \int_{-\infty}^{0} dx_{1}\int_{-\infty}^{x_{1}} dx_{2}\int_{-\infty}^{x_{2}} dx_{3} \, (-x_{1})^{-1/2} \,\nonumber \\
&&  \times \Big\{(-x_{2})^{1/2-\nu} (-x_{3})^{-1/2} \, \times \sin[- x_{1}] \text{Im}\left[e^{i x_{3}} \mathcal H_\nu^{(1)}(- c_2(k_S) x_2) \mathcal H_\nu^{(2)}(- c_2(k_S) x_3)\right] \nonumber\\
&&\quad\quad\times \text{Im}\left[\mathcal H_\nu^{(1)}(- c_2(k_S) x_1) \mathcal H_\nu^{(2)}(- c_2(k_S) x_2)\right] +\nonumber \\
&&  \quad + (-x_{2})^{-1/2} (-x_{3})^{1/2-\nu} \sin[- x_{1}] \sin[- x_{2}] \, \text{Im}\Big[\mathcal H_\nu^{(1)}(- c_2(k_S) x_3) \mathcal H_\nu^{(1)}(- c_2(k_S) x_3)  \nonumber\\
&& \quad \quad  \quad \times \mathcal H_\nu^{(2)}(- c_2(k_S) x_1) \mathcal H_\nu^{(2)}(- c_2(k_S) x_2) \Big] \Big\} \times \nonumber \\
&&\times \left(\int_{-\infty}^{0} dy\, (- y)^{-1/2} \text{Re}\left[e^{-i y} \mathcal H_\nu^{(1)}(- c_2(k_L) y)\right] \right) \, ,
\end{eqnarray}
and is well-fit by the following power law in $c_2$ 
\begin{align} \label{eq:ssq}
   \mathcal I(c_2, \nu) \simeq \frac{a(\nu)}{c_2(k_L)^{\nu}c_2(k_S)^{3\nu}} \,,
\end{align}
where $a$ is a parameter dependent on $\nu$ (see Tab. \ref{tab:fit_power_nu} for a sample set of possible values). By proceeding in a similar fashion (see App. \ref{app:tss} for technical details), we obtain the following $\sigma$-mediated contribution to $\langle \gamma \gamma \zeta \rangle$
\begin{align}
    \langle \gamma_{k_1}^{\lambda_1} \gamma_{k_2}^{\lambda_2} \zeta_{k_3} \rangle = (2\pi)^3\delta^{(3)}(\vec{k}_1 + \vec{k}_2+\vec{k}_3) \, \mathcal{A}^{\lambda_1 \lambda_2} B_{\rm tts}(k_1,k_2,k_3) \, , 
\end{align}
where the function $B_{\rm tts}(k_1,k_2,k_3)$ in the squeezed limit $k_L = k_3 \ll k_1 \simeq k_2 = k_S$ is given by
\begin{align} \label{eq:model_tts}
    B_{\rm tts}(k_L,k_S,k_S) = &- \, \frac{2 \pi^2}{\epsilon}  \frac{\mu}{H} \left(\frac{g}{M_{Pl}}\right)^3 \, \frac{2^\nu}{k_S^{9/2 - \nu} k_L^{3/2+\nu}} \left(\frac{c_{2}(k_{S})}{c_{2i}}\right)^{2}\left(\frac{c_{0}(k_{S})}{c_{0i}}\right)^{1/2}\left(\frac{c_{0}(k_{L})}{c_{0i}}\right)^{1/2} \, \nonumber\\
    &\times \mathcal I(c_0,c_2,\nu) \, ,
\end{align}
and 
\begin{align}
\mathcal{A}^{\lambda_1 \lambda_2} &=  \left(\epsilon^{\lambda_1}_{ij}(\hat k_1) \cdot \epsilon_{ij}^{\lambda_2}(\hat k_2) - 3 \, \hat k_3^i \hat k_3^l \cdot \epsilon^{\lambda_1}_{ij}(\hat k_1) \cdot \epsilon_{jl}^{\lambda_2}(\hat k_2)\right) \, .
\end{align}
In the squeezed limit $\mathcal{A}$ simplifies to
\begin{equation}
\mathcal{A}^{\lambda_1 \lambda_2}_{\rm sq.} = \frac{4 \pi}{5} \sum_{M} Y_{2 M}(\hat k_L) \, Y^*_{2 M}(\hat k_S)  \times \begin{cases}
1 \quad \mbox{if} \qquad \lambda_{1} = \lambda_{2} \\
0 \quad \mbox{if} \qquad \lambda_{1} \neq \lambda_{2}
\end{cases}  \, .  
\end{equation}
The function $\mathcal I(c_0,c_2,\nu)$ is given by Eq. \eqref{eq:I}, which reads 
\begin{eqnarray}
\mathcal I(c_0,c_2,\nu) &=& \frac{\Gamma(\nu)}{c^\nu_0(k_S)} \int_{-\infty}^{0} dx_{1}\int_{-\infty}^{x_{1}} dx_{2}\int_{-\infty}^{x_{2}} dx_{3} \, (-x_{1})^{-1/2} \,\nonumber \\
&&  \times \Big\{(-x_{2})^{1/2-\nu} (-x_{3})^{-1/2} \, \times \sin[- x_{1}] \text{Im}\left[e^{i x_{3}} \mathcal H_\nu^{(1)}(- c_2(k_S) x_2) \mathcal H_\nu^{(2)}(- c_2(k_S) x_3)\right] \nonumber\\
&&\quad\quad\times \text{Im}\left[\mathcal H_\nu^{(1)}(- c_2(k_S) x_1) \mathcal H_\nu^{(2)}(- c_2(k_S) x_2)\right] +\nonumber \\
&&  \quad + (-x_{2})^{-1/2} (-x_{3})^{1/2-\nu} \sin[- x_{1}] \sin[- x_{2}] \, \text{Im}\Big[\mathcal H_\nu^{(1)}(- c_2(k_S) x_3) \mathcal H_\nu^{(1)}(- c_2(k_S) x_3)  \nonumber\\
&& \quad \quad  \quad \times \mathcal H_\nu^{(2)}(- c_2(k_S) x_1) \mathcal H_\nu^{(2)}(- c_2(k_S) x_2) \Big] \Big\} \times \nonumber \\
&&\times \left(\int_{-\infty}^{0} dy\, (- y)^{-1/2} \text{Re}\left[e^{-i y} \mathcal H_\nu^{(1)}(- c_0(k_L) y)\right] \right) \, ,
\end{eqnarray}
and is well fit by the following power law in $c_0$ and $c_2$, Eq. \eqref{fit_pow},
\begin{align} \label{eq:tsq}
   \mathcal I(c_0, c_2, \nu) \simeq \frac{a(\nu)}{c_0(k_L)^{\nu}c_0(k_S)^{\nu}c_2(k_S)^{2\nu}} \, .
\end{align}
 The key quantities that provide a handle on non-Gaussianities via GW anisotropies are $F^{\rm ttt}_{\rm NL}(\vec k_S, \vec k_L)$ and $F^{\rm tts}_{\rm NL}(\vec k_S, \vec k_L)$, defined respectively in  Eq.~\eqref{eq:fnlttt_def} and
Eq.~\eqref{eq:fnltts_def}.
The full scalar and tensor power spectra of the model under scrutiny are given by (see also  \cite{Bordin:2018pca,Iacconi:2019vgc})
\begin{align}
P_\zeta(k) =& \frac{H^2}{4 M^2_{Pl} \epsilon k^3}  \left[ 1+ \frac{\mathcal C_\zeta(\nu)}{\epsilon c^{2 \nu}_0(k)} \left(\frac{c_{0}(k)}{c_{0i}}\right)\left(\frac{g}{H}\right)^2\right]\, , \\
P_\gamma(k) =& \frac{4 H^2}{M^2_{Pl} k^3}  \left[ 1+ \frac{\mathcal C_\gamma(\nu)}{c^{2 \nu}_2(k)}\left(\frac{c_{2}(k)}{c_{2i}}\right) \left(\frac{g}{H}\right)^2 \right] \, ,
\end{align}
where the analytical form of the $\nu$-dependent functions $\mathcal C_\zeta(\nu)$ and $\mathcal C_\gamma(\nu)$ can be found in Ref. \cite{Bordin:2018pca} in some specific configurations.
Our interest lies in GW anisotropies at intermediate and small scales. The EFT at hand can deliver sufficiently large GW spectrum and anisotropic component at the appropriate frequencies provided we work in the $c_2\ll 1$ regime. A small helicity-2 sound speed at all scales may run afoul of non-Gaussianity bounds from CMB measurements. It is therefore convenient to consider a scale dependent $c_2$, and a blue GW spectrum in particular.\\

As we shall see, choosing a $c_2$ that decreases towards smaller scales and a  $c_0\sim \mathcal{O}(1) $ that exhibits a very limited variation across the frequency range, makes for a very interesting phenomenology. At CMB scales scalar and tensor spectra are close to those of single-field slow-roll inflation, i.e. the sourced contribution is sub-leading. Going towards smaller scales, the scalar spectrum remains largely dominated by vacuum fluctuations whilst the GW signal is due to the sourced contribution
\begin{align}
P_\zeta(k_L) \simeq& \frac{H^2}{4 M^2_{Pl} \epsilon k_L^3}  \, , \\
P_\gamma(k_L) \simeq& \frac{4 H^2}{M^2_{Pl}  k_L^3}  \, , \\
P_\gamma(k_S) \simeq& \frac{4 H^2}{ M^2_{Pl} k_S^3} \frac{\mathcal C_\gamma(\nu)}{c^{2 \nu}_2(k_S)}\left(\frac{c_{2}(k_{S})}{c_{2i}}\right) \left(\frac{g}{H}\right)^2 \label{eq:power_tensor_kS}\, ,
\end{align}
where one should picture $k_L$ at CMB scales and $k_S$ corresponding to e.g. PTA or interferometer frequencies. Given such behaviour for the power spectra, we may proceed to evaluate the non-linear parameters as defined in Eqs. \eqref{eq:fnlttt_def} and \eqref{eq:tts_quad}, to obtain
\begin{align} 
F^{\rm ttt}_{\rm NL}(k_S, k_L, \nu) = & 3 \, \pi^2 \,2^\nu \frac{a(\nu)}{C_\gamma(\nu)} \left(\frac{M_{Pl}}{H}\right) \left(\frac{c_2(k_S)^{\frac{3}{2}-\nu}}{ c_2(k_L)^{\nu-\frac{1}{2}}\,c_{2i}^2}\right)
\left(\frac{g}{H}\right) \left(\frac{\mu}{H}\right) \left(\frac{k_L}{k_S}\right)^{3/2 - \nu} \, ,\label{eq:tilde_fnl_ttt_m} \\
\tilde F^{\rm tts}_{\rm NL}(k_S, k_L, \nu) = & - 4 \, \pi^2 \,2^\nu \frac{a(\nu)}{C_\gamma(\nu)} \left(\frac{M_{Pl}}{H}\right)
\left(\frac{c_2(k_S)}{c_{0}(k_L)^{\nu-\frac{1}{2}}c_{0}(k_S)^{\nu-\frac{1}{2}}\,c_{2i}\,c_{0i}}\right)
\left(\frac{g}{H}\right) \left(\frac{\mu}{H}\right) \left(\frac{k_L}{k_S}\right)^{\frac{3}{2} - \nu} \, . \label{eq:tilde_fnl_tts_m}
    \end{align}
As clear from the common $(k_L/k_S)^{3/2 - \nu}$ scaling, a small exponent (i.e. a light field)  will provide the most striking signatures for both the auto- and cross-correlations we are after. Combining Eq.~\eqref{eq:tilde_fnl_ttt_m} with Eq.~\eqref{eq:tilde_fnl_tts_m} to obtain
\begin{equation}F^{\rm ttt}_{\rm NL}(k_S, k_L, \nu) = - \frac{3}{4}\left(\frac{c_{0i}}{c_{2i}}\right)
\left(\frac{c_{0}(k_{S})c_0(k_L)}{c_{2}(k_{S})c_2(k_L)}\right)^{\nu - \frac{1}{2}}
\tilde F^{\rm tts}_{\rm NL}(k_S, k_L, \nu) \,, 
\end{equation}
shows that whenever $c_2 \ll c_0$, for a sufficiently light field the TTT non-linear parameter will be enhanced with respect to the TTS one.

\subsection{Parameter space}
\label{sec:param}
Let us now discuss the parameter space of the theory. First of all, following the previous subsection, we fix the following parametrisation for the helicity-2 sound speed $c_2$,
\ba 
c_2(k) = c_{2i}\left(\frac{k_0}{k}\right)^{s_2}+c_{2f} \; .
\label{eq:c2k}
\ea
For the helicity-0 component we take the following parametrisation
\ba
c_0(k) = c_{0i}\left(\frac{k_0}{k}\right)^{s_0} \, .
\ea
Note that here we are neglecting $c_{0f}$. The reason is that we are interested in scenarios where the running with scale of the helicity-0 sound speed is almost absent, with $c_0(k) \simeq 1$. 

A number of constraints on the parameter space of our model are already in place. These come from: (i) consistency of the theory (i.e. gradient instabilities, perturbativity); (ii) constraints on the scalar power spectrum amplitude and spectral index (e.g. it is important to keep $c_{0}$ fairly close to 1 at CMB scales so that the scalar power spectrum is dominated by the vacuum and there is no issue with the spectral index being in agreement with observations); (iii) constraints on the tensor power spectrum; (iv) constraints on scalar and tensor non-Gaussianities. Points (i)-(iii) are discussed and summarised in Secs.~3.1, 3.2 and 3.3 of \cite{Iacconi:2019vgc}. Point (iv) is discussed e.g. in \cite{Iacconi:2020yxn}.   

\begin{figure}
    \centering
    \includegraphics[width=0.6\textwidth]{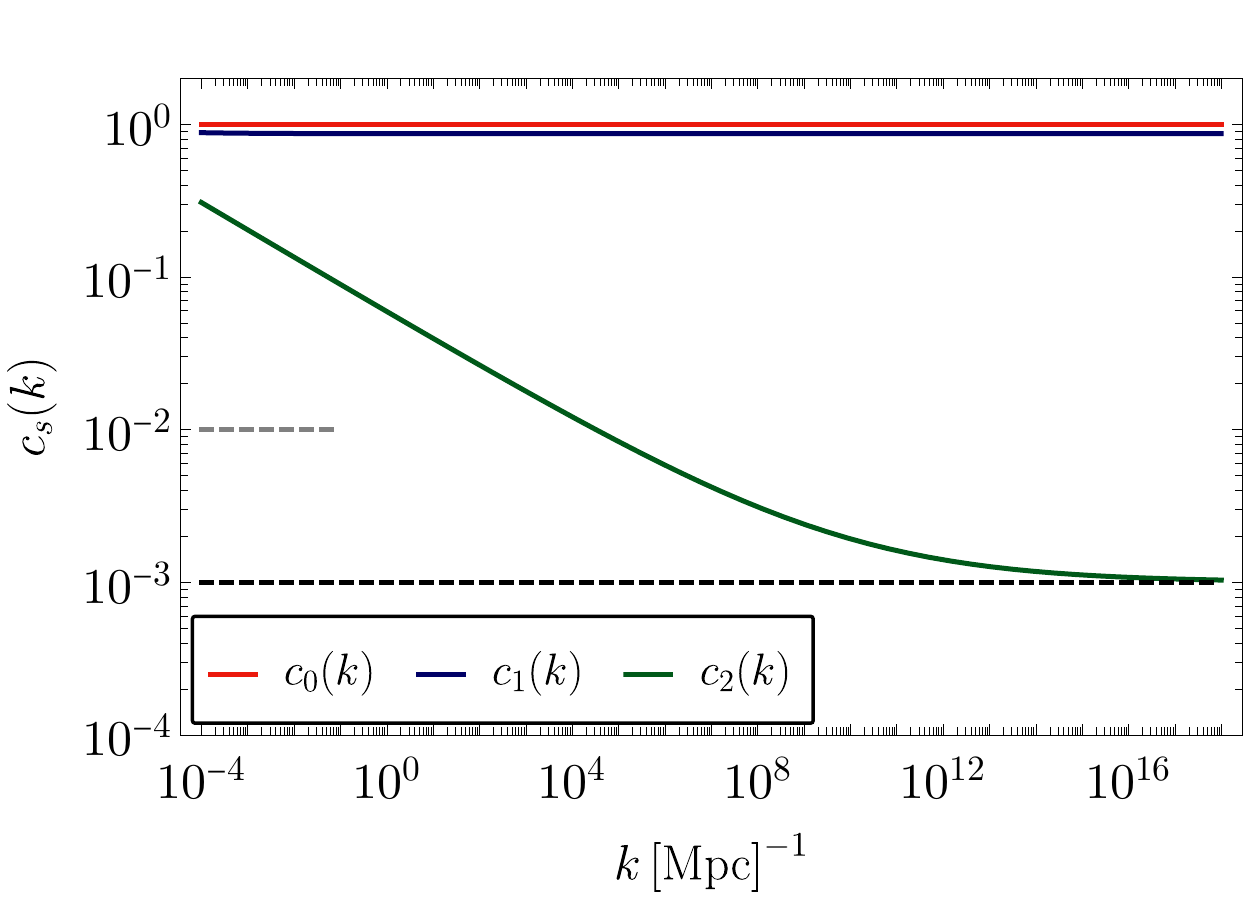}
    \caption{The scale-dependence of the sound speeds $c_s(k)$ taking $c_{2i}=2.5\times 10^{-1},\,c_{2f}=10^{-3},\, s_2=1.8\times 10^{-1},\,c_{0i}=1,\, s_0=0$. The black and grey dashed lines show the bounds from perturbativity and CMB non-Gaussianity respectively.}
    \label{fig:cs_plot}
\end{figure}

The existing constraints on primordial non-Gaussianity at CMB scales require $ c_{2}\gtrsim 10^{-2}$, while one must impose $c_{2f}\gtrsim 10^{-3}$ throughout due to the perturbativity bound. 
Moreover, taking $c_{0i}$ close to $1$ with $s_0\ll s_2$ implies the following relation for the sound speeds, $c_0\simeq c_1 \gg c_2$ \footnote{We are assuming conservatively that the relation in Eq.~\eqref{eq:sound_speeds_rel} holds for scale-dependent sound speeds as well.}. The momentum dependence of the sound speeds is plotted in Fig.~\ref{fig:cs_plot} for a set of parameters that maximises the GW amplitudes at direct detection scales. 
Note that $c_0$ has been chosen as large as possible (close to 1) throughout its scale dependence due to the constraint $g/H\ll \sqrt{\epsilon c_0^2}$ imposed to avoid gradient instabilities \cite{Iacconi:2019vgc}. This allows us to take larger values of  $g/H$, leading to an observable GW spectrum on small scales (the sourced contribution to the tensor power spectrum scales as $(g/H)^2$, Eq.~\eqref{eq:power_tensor_kS}).\\
\indent The same constraint also ensures that the sourced contribution to the scalar power spectrum is subdominant compared to the vacuum one whose amplitude and spectral index are taken to be $A_S=2.09\times 10^{-9}$ and $n_S = 0.9649$ in accordance with the Planck values \cite{Akrami:2018odb}. A detailed analysis of the constraints on the parameter space of this model has previously been carried out in \cite{Iacconi:2019vgc,Iacconi:2020yxn} and we have ensured that our choices for the parameters $\{c_{s,i},\,c_{s,f},s\,,\,\mu/H,\,\rho/H,\,\nu\}$ lie within the allowed region.
\begin{figure}
    \begin{minipage}{0.49\textwidth}
    \centering
    \includegraphics[width=\textwidth]{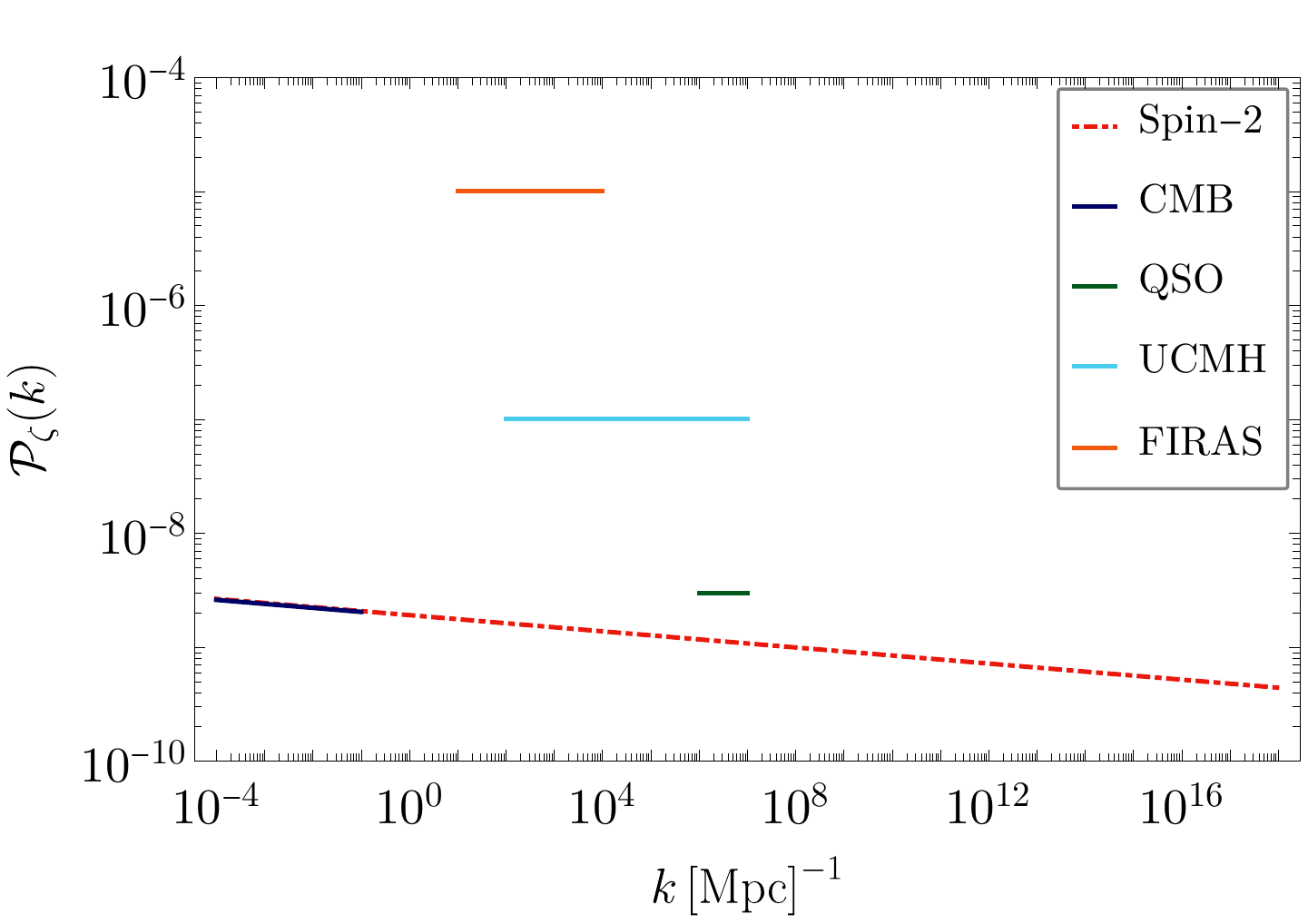}
    \end{minipage}
    \begin{minipage}{0.49\textwidth}
    \centering
    \includegraphics[width=\textwidth]{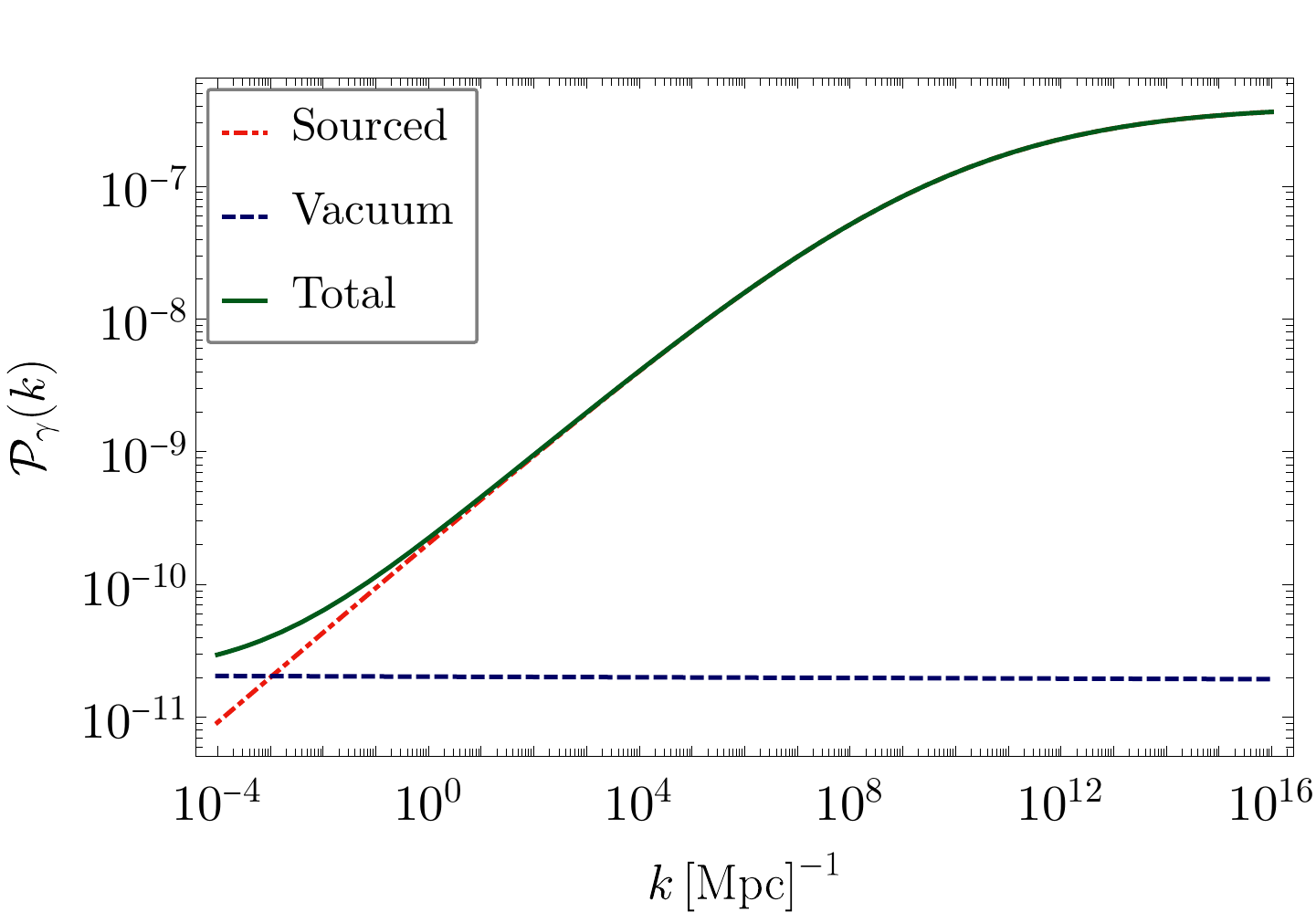}
    \end{minipage}
    \caption{Left : The scalar power spectrum for the spin-2 model along with the measured power spectrum from CMB observations \cite{Akrami:2018odb} and the constraints from various experiments (see \cite{Kalaja:2019uju,Yoshiura:2020soa} for the exact constraints). Right: The different contributions to the tensor power spectrum for the spin-2 model. The choice of parameters for both panels is $c_{2i}=2.5\times 10^{-1},\,c_{2f}=10^{-3},\, s_2=1.8\times 10^{-1},\,c_{0i}=1,\, s_0=0,\,g/H = 4\times 10^{-3},\nu=1.45$.}
    \label{fig:scalar_tensor_plot}
\end{figure}

\begin{figure}
    \centering
    \includegraphics[width=0.75\textwidth]{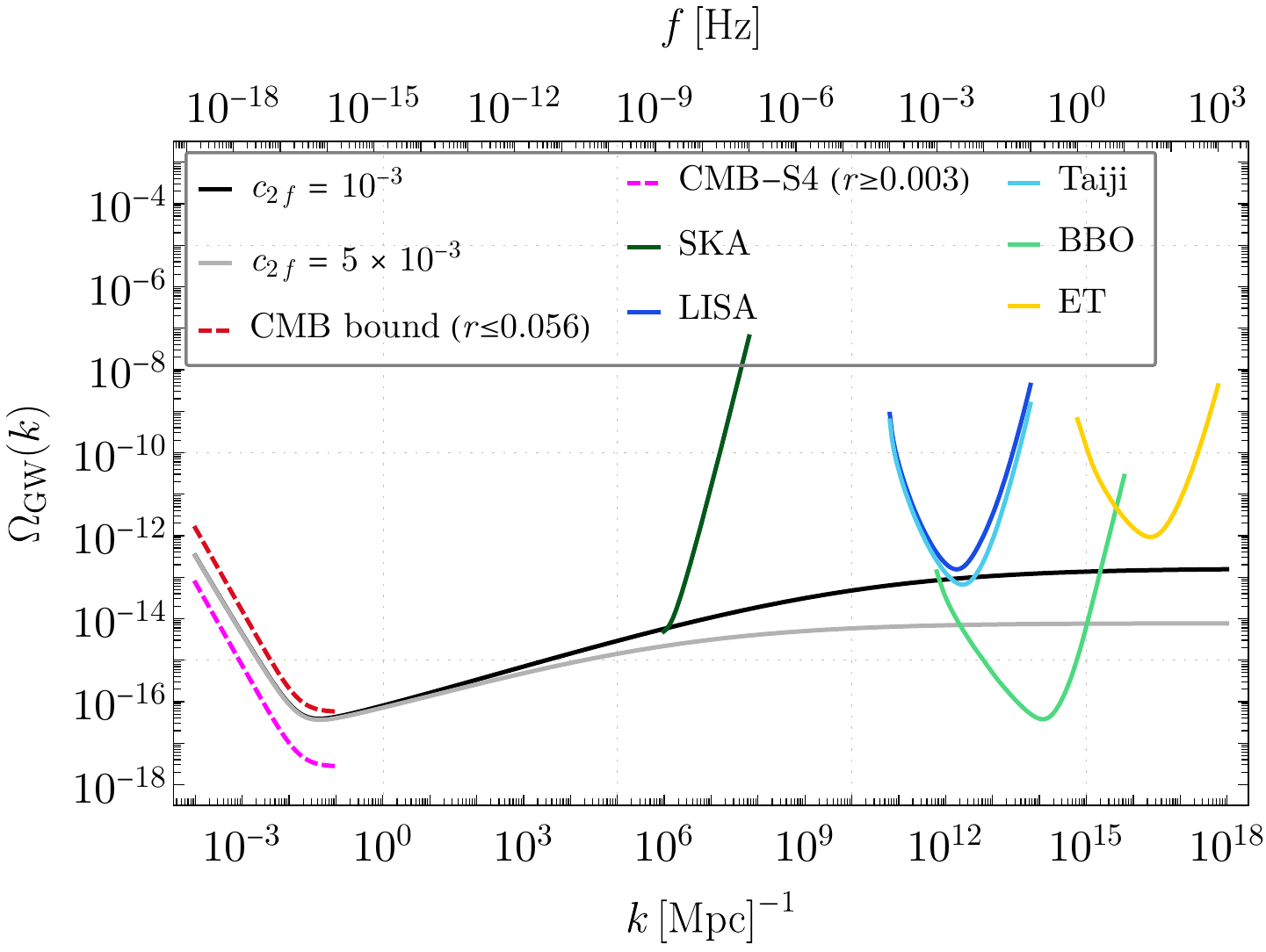}
    \caption{$\Omegagw(k)$ for different values of $c_{2f}$ plotted alongside the power law sensitivity curves for SKA, LISA, Taiji, BBO and ET. We also plot the current bound from the CMB as well as the expected sensitivity of CMB-S4 \cite{CMB-S4:2020lpa}.}
    \label{fig:omegagwplot}
\end{figure}

\begin{figure}
    \includegraphics[width=\textwidth]{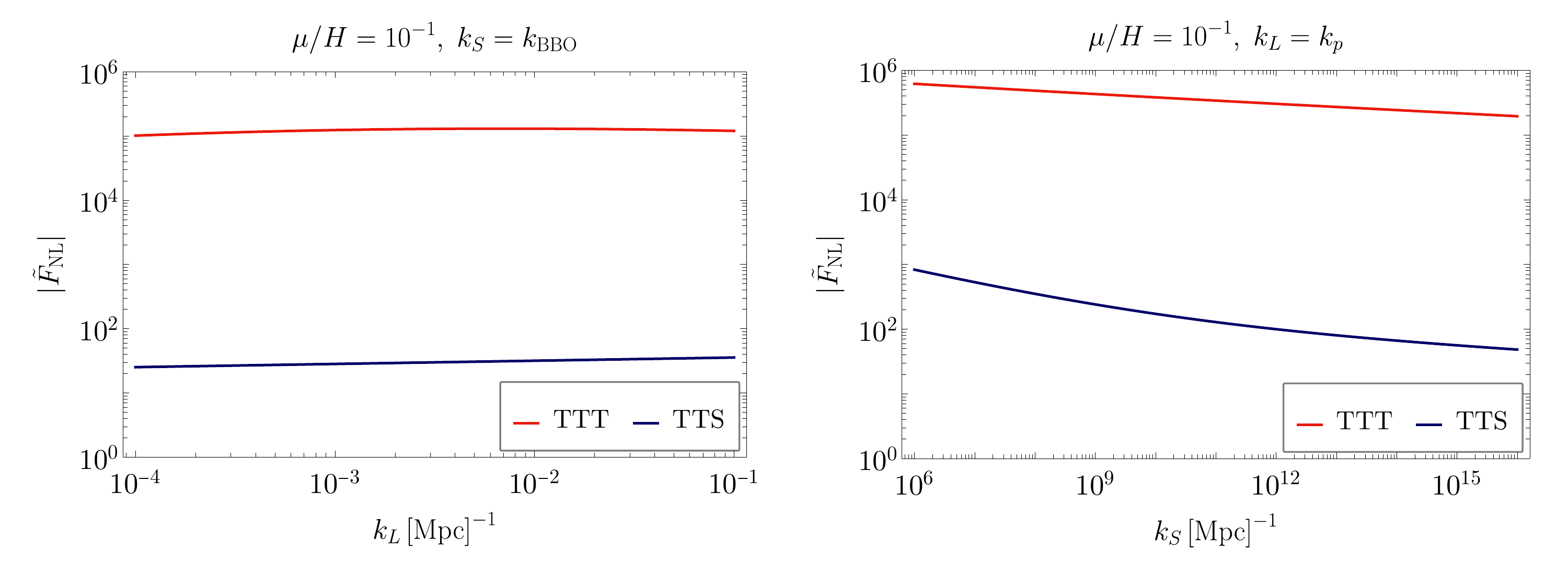}
    \caption{The non-linearity parameters $\tilde F_{\rm NL}(k_S,k_L)$ for the TTS and TTT bispectra plotted as a function of $k_S$ and $k_L$ for $\mu/H = 10^{-1}$.} 
    \label{fig:fnlplot}
\end{figure}

\indent The scalar and tensor power spectra for this model are plotted in Fig.~\ref{fig:scalar_tensor_plot} with the parameter choice $c_{2i}=2.5\times 10^{-1},\,c_{2f}=10^{-3},\, s_2=1.8\times 10^{-1},\,c_{0i}=1,\, s_0=0,\,g/H = 4\times 10^{-3},\nu=1.45,\,H/M_{\rm pl}=10^{-5}$ and the GW spectrum is plotted in Fig.~\ref{fig:omegagwplot} for different values of $c_{2f}$. We see that for $c_{2f}= 10^{-3}$, the GW spectrum falls within the sensitivity range of SKA, Taiji and BBO as well as next generation CMB experiments like CMB-S4. For the same choice of parameters, in Fig.~\ref{fig:fnlplot} we plot the TTS and TTT non-linearity parameters taking $\tilde F_{\rm NL}$ with the expressions given in Eq.~\eqref{eq:tilde_fnl_ttt_m} and Eq.~\eqref{eq:tilde_fnl_tts_m}.
The expression for $\tlfnls$ is exact since the scalar power spectrum is dominated by the vacuum contribution throughout. For $\fnlt$, note that since the tensor power spectrum is dominated by the sourced contribution on smaller scales Eq.~\eqref{eq:tilde_fnl_tts_m} is an approximation that works well only on the largest scales ($k_L<10^{-2}$). While this approximation reproduces the correct scaling as well as the order of magnitude in this range, the exact results are used for all the plots and in calculating the correlations in Sec.~\ref{sec:spin2_cl}. 
The behaviour of both $\fnlt$ and $\tlfnls$ as a function of momentum can be understood as a result of the scale dependence arising from the factor $(k_L/k_S)^{3/2 -\nu}$. There is also a moderate scale dependence due to the running sound speed $c_2$ whereas $c_0$ remains constant throughout. 
\subsection{Angular power spectra of GW anisotropies}
\label{sec:spin2_cl}
\begin{figure}
\begin{minipage}{0.49\textwidth}
    \centering
    \includegraphics[width=0.98\textwidth]{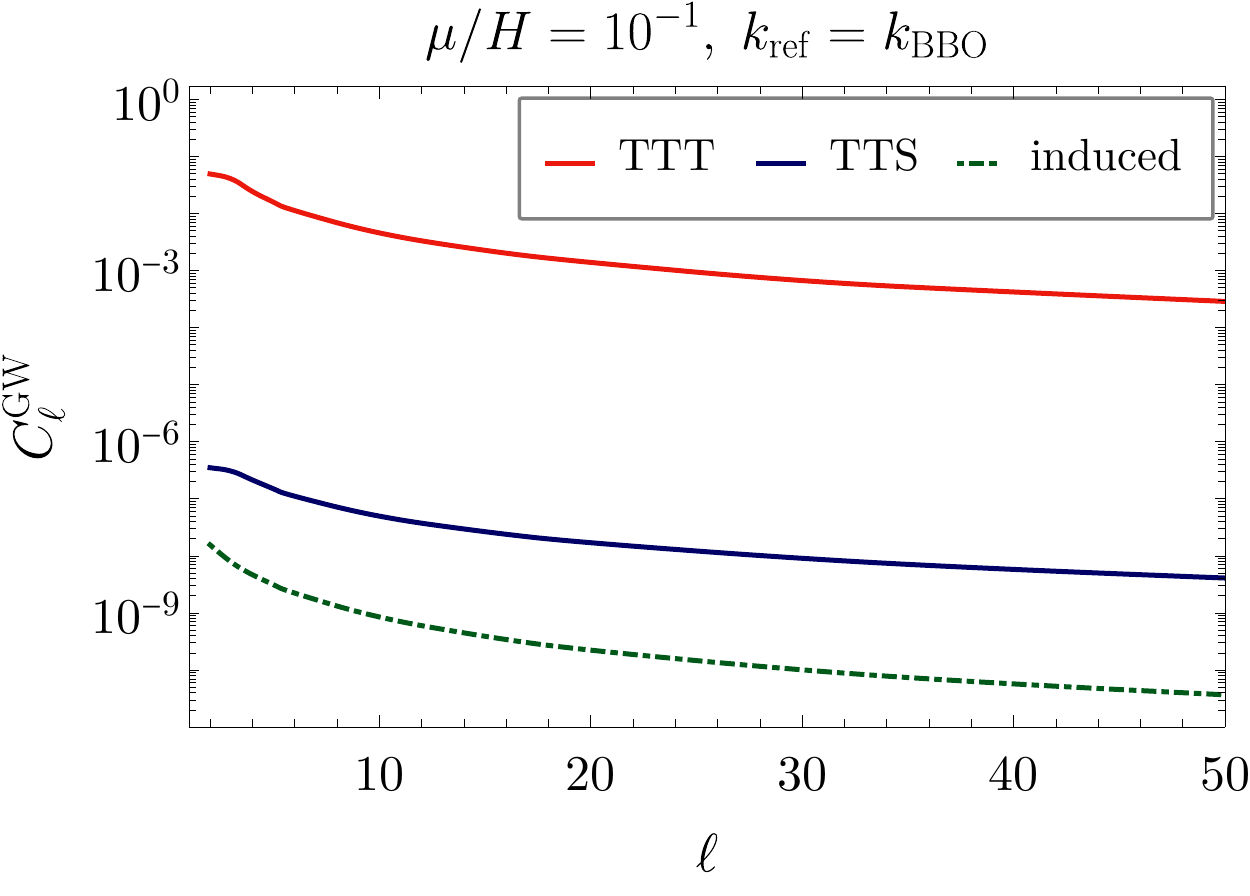}
\end{minipage}
\begin{minipage}{0.49\textwidth}
    \centering
    \includegraphics[width=0.98\textwidth]{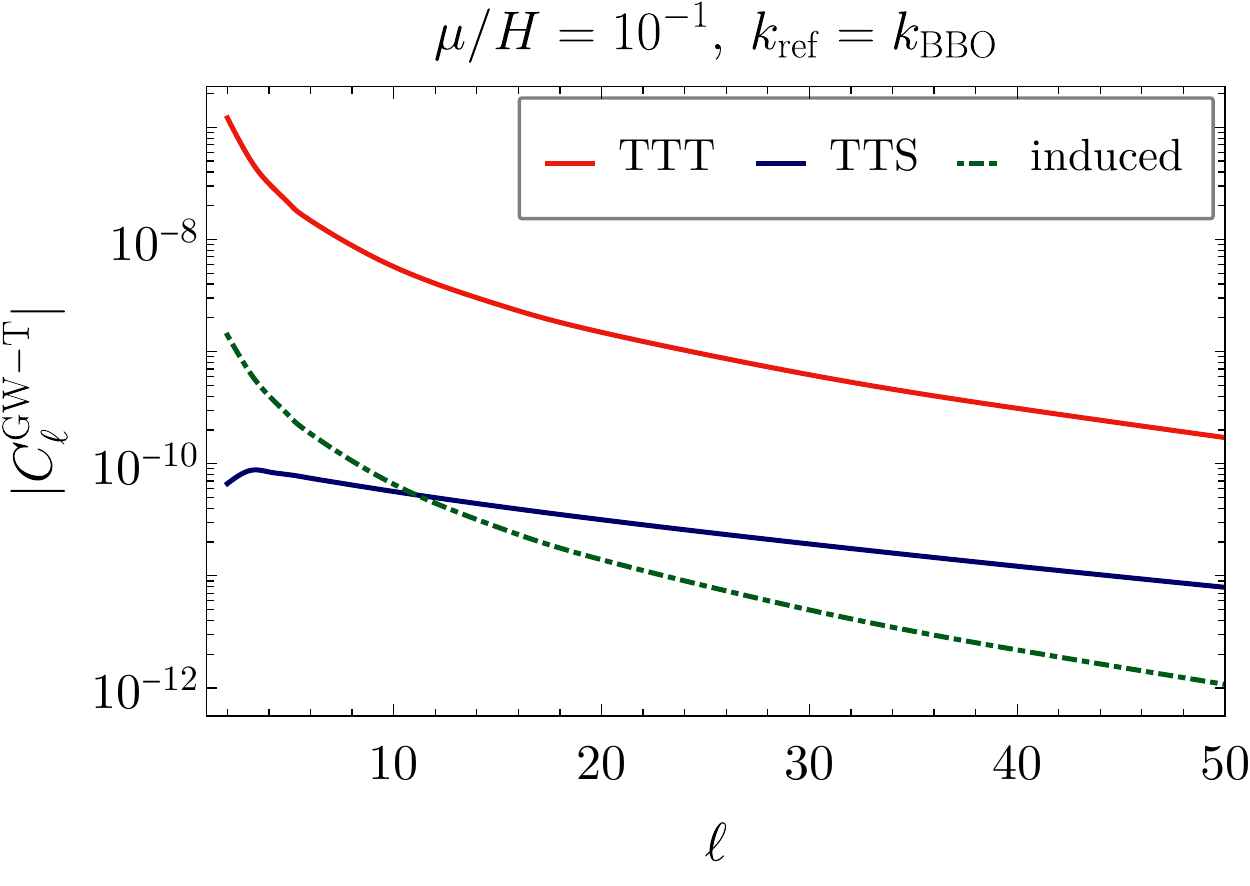}
\end{minipage}
    \caption{The different contributions to the auto-correlation $\GG$ and cross-correlation $\GT$ plotted for the spin-2 model. For comparison we also plot the induced anisotropies from propagation.}
    \label{fig:cl_spin2}
\end{figure}

With the above choice of parameters and armed with the results of Sec.~\ref{sec:clgw} and \ref{sec:cross}, we are now able to calculate the angular power spectra of the GW anisotropies for the spin-2 model. We plot these in Fig.~\ref{fig:cl_spin2} at the frequency scales associated to BBO. We see that both the auto-correlation and the cross-correlation are dominated by the TTT contribution as a result of the small sound speed of the helicity-2 component of the spin-2 field.
\begin{figure}
    \centering
    \includegraphics[width=0.6\linewidth]{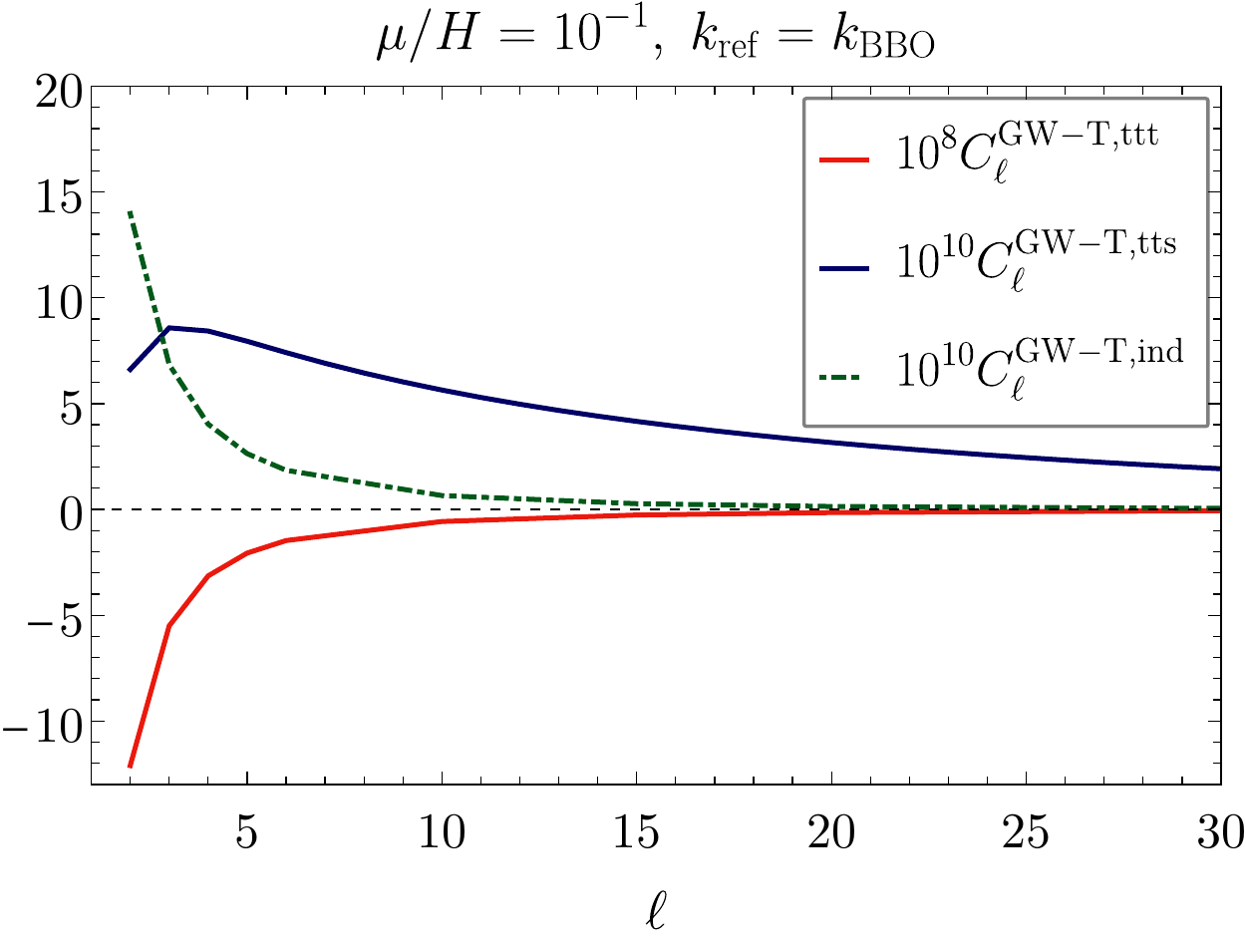}
    \caption{The cross-correlation $\GT$ for the spin-2 model. For comparison we also plot the anisotropies induced by propagation.}
    \label{fig:sign_clgwt_spin2}
\end{figure}

Interestingly, as a result of the $(k_L/k_S)^{3/2-\nu}$ scale dependence, the CMB-GW cross-correlation from the TTS bispectrum is suppressed at the largest angular scales, in particular at $\ell=2$ (Fig.~\ref{fig:sign_clgwt_spin2}), in contrast to the case where $\tlfnls$ is scale independent (Fig.~\ref{fig:cross_model_ind}). While this suppression can be a potential signature of the mass of the spin-2 field (recall that $\nu = \sqrt{9/4-m_{\sigma}^2/H^2}$), the TTS cross-correlation is always smaller than the cross-correlation from the TTT bispectrum and thus this effect is unlikely to be observable. We also see that $\GTT<0$ (Fig.~\ref{fig:sign_clgwt_spin2}) which can be physically understood from the fact that the CMB temperature anisotropies from tensor perturbations arise from the subhorizon decay of the tensor modes. Thus, in Eq.~\eqref{eq:ttt_cross} we have $d\gamma/d\eta < 0$ and so when $\fnlt>0$ we will have $\GTT<0$. Additionally, the TTT and TTS cross-correlations have opposite signs which reduces the cross-correlation signal.

For reference note that the GW energy density and $\tlfnls,\fnlt$ at BBO scales can be approximated as a function of the model parameters $\mu/H,c_{2f}$ as 
\begin{align}\label{eq:param_omegafnl}
     \Omegagw(k_{\rm BBO}) &\simeq 1.3\times 10^{-13}\left(\frac{c_{2f}}{10^{-3}}\right)^{-1.7},\nonumber\\
     \fnlt(k_{\rm BBO},k_p) &\simeq  1.2\times 10^{6}\left(\frac{\mu}{H}\right),\\ \tlfnls(k_{\rm BBO},k_p)&\simeq -3.5 \times 10^2 \left(\frac{\mu}{H}\right)  \,\nonumber .
\end{align}
Note that we have fixed all the other parameters to the values considered in Sec.~\ref{sec:param}. 

As a consequence of the scale dependence of $\tlfnls,\,\fnlt$, the resulting $\GG,\,\GT$ will also be scale dependent\footnote{The additional scale dependence arising from the running of the sound speed $c_2(k)$ on small scales can be safely neglected since this is nearly constant throughout ($c_{2}(k_S)\simeq 10^{-3}$, see Fig.~\ref{fig:cs_plot}). }, i.e.,
\begin{align}
    \GG(k_{\rm ref}) &\simeq \GG(k_{\rm BBO})\times\left(\frac{k_{\rm BBO}}{k_{\rm ref}}\right)^{3-2\nu},\nonumber\\
    \GT(k_{\rm ref}) &\simeq \GT(k_{\rm BBO})\times\left(\frac{k_{\rm BBO}}{k_{\rm ref}}\right)^{3/2-\nu},
\end{align}
for the anisotropies arising from both the TTS and TTT non-Gaussianity. 
\subsection{\texorpdfstring{Projected constraints on $F_{\rm NL}$}{Projected constraints on Fnl}}
\label{sec:ttt_error_spin2}
We can now estimate the error in the measurement of ${F}_{\rm NL}$. For the spin-2 model we have seen that the anisotropies from the TTT bispectrum are dominant compared to those from the TTS bispectrum. Thus, we focus here on the error in the estimation of ${F}_{\rm NL}^{\rm ttt}$ using both auto-correlation and cross-correlation measurements (see Sec.~\ref{sec:tts_error} for the details). 
\begin{figure}
    \centering
    \includegraphics[width=\textwidth]{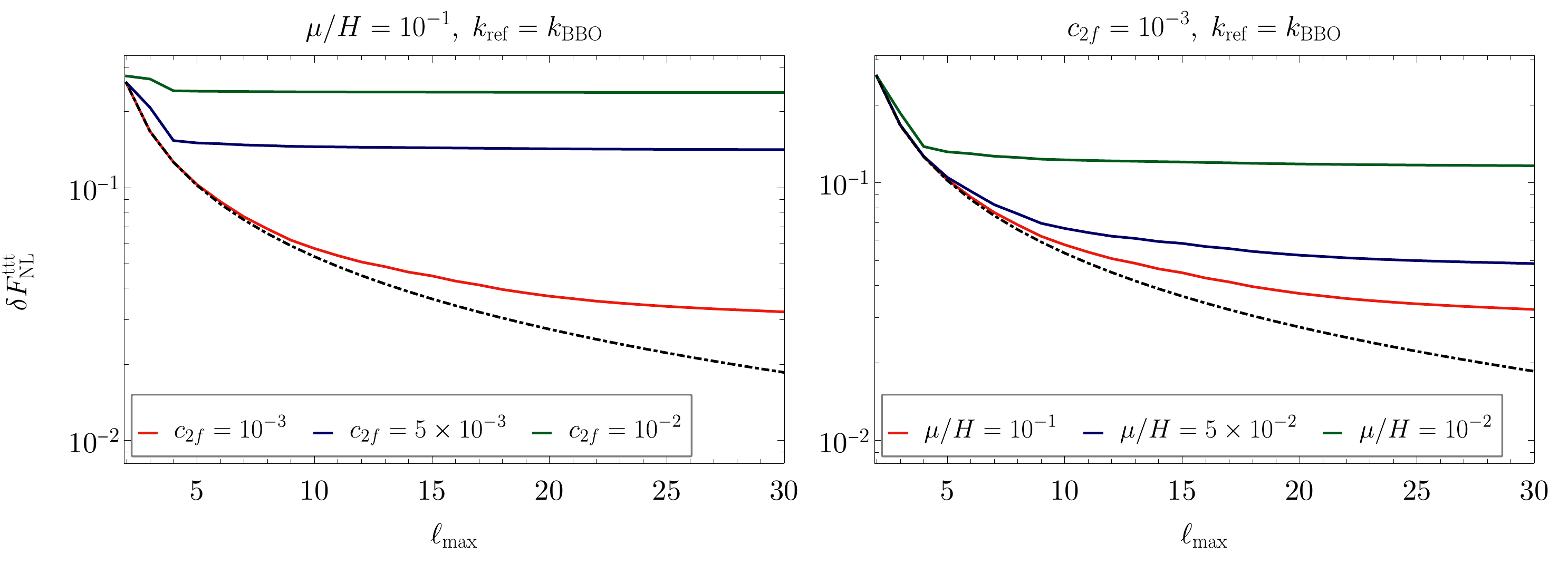}
    \caption{The relative error in the measurement of ${F}_{\rm NL}^{\rm ttt}$ as a function of $\ell_{\rm max}$ with BBO for different values of $\mu,c_{2f}$. The dashed curves show the errors for an idealised, cosmic variance limited measurement.}
    \label{fig:error_ttt}
\end{figure}
In Fig.~\ref{fig:error_ttt} we plot the relative error in the measurement of ${F}_{\rm NL}^{\rm ttt}$ defined as
\begin{align}
    \delta{F}_{\rm NL}^{\rm ttt} \equiv {\Delta {F}_{\rm NL}^{\rm ttt}}/{{F}_{\rm NL}^{\rm ttt}}\,,
\end{align} for different values of $\mu,c_{2f}$. For reference, note that at BBO scales, the GW amplitude and $\fnlt$ can be approximated as a function of $c_{2f},\mu/H$ as given in Eq.~\eqref{eq:param_omegafnl}.
We see that in the case of $c_{2f}= 10^{-3},\mu/H=10^{-1}$ we can achieve a relative error $\delta{F}_{\rm NL}^{\rm ttt}\sim 10^{-2}$; a similar error is achievable for the noiseless cosmic-variance limited case which is understandable from the fact that for this choice of parameters we have $C_{\ell}^{\rm{ GW,ttt}} \gg \NGW$. For smaller values of $\mu$ and larger $c_{2f}$, i.e. smaller ${F}_{\rm NL}^{\rm ttt},\,\Omegagw$, we see that the error saturates extremely quickly, around $\ell_{\rm max}\sim 6\operatorname{--}10$, due to the fact that the detector noise increases rapidly after the first few multipoles (see Fig.~\ref{fig:Nell_Omega}). 
\subsection*{SNR of the CMB-GW cross-correlation}
As in Sec.~\ref{sec:cross}, we also estimate the signal to noise ratio of the CMB-GW cross-correlation for the spin-2 model in the presence of an astrophysical foreground. 
\begin{figure}
\includegraphics[width=0.99\linewidth]{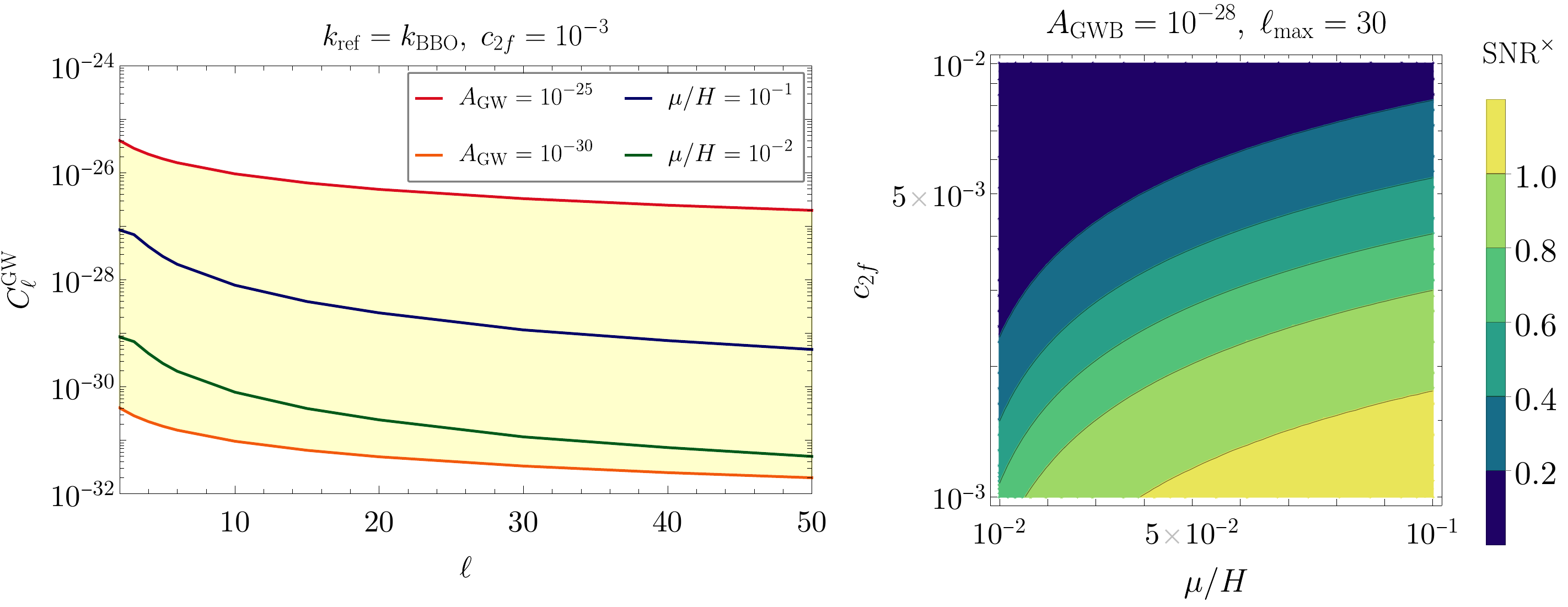}
    \label{fig:astro_spin2}
    \caption{Left: The $\GG$ for the astrophysical background (yellow shaded region) and for the CGWB anisotropies in the spin-2 model for different values of $\mu/H$ taking $c_{2f}=10^{-3}$. Right: Signal to noise of the total cross-correlation for the spin-2 model as a function of $c_{2f},\,\mu/H$.}
\end{figure}
To arrive at a detectable signal one needs to consider a relatively small $A_{\rm GWB}$. This can be understood from an argument similar to the one made in Sec.~\ref{sec:cross_TTS}. In the expression for the SNR Eq.~\eqref{eq:snr_def} one has the following term,
\begin{align}
    \SNRx \simeq \left[\sum_{\ell}(2\ell+1)\frac{\left(C_{\ell}^{\rm GW-T,signal}\right)^2}{C_{\ell}^{\rm GW,total}\TT}\right]^{1/2}.
\end{align}
For the spin-2 model we have seen that the CMB-GW cross-correlation as well as the auto-correlation is dominated by the TTT contribution, however the variance of the CMB temperature anisotropies is dominated by the scalar term rather than the tensor term. Thus, for the spin-2 model anisotropies, we again have $C _{\ell}^{\rm GW-T,signal}\ll ({C_{\ell}^{\rm GW,total}\TT})^{1/2}$ unlike that of the monopolar TTS considered in Sec.~\ref{sec:cross_TTS} where we found $C_{\ell}^{\rm GW-T,signal}\simeq ({C_{\ell}^{\rm GW,total}\TT})^{1/2}$. Furthermore, we also see from Fig.~\ref{fig:sign_clgwt_spin2} that $\GTT$ and $\GTS$ for the spin-2 model have opposite signs which reduces the total signal appearing in the SNR.

\section{Conclusions}
\label{conclusions}
During inflation, scalar and tensor quantum fluctuations are generated around the background solution. These modes eventually re-enter the horizon and ``perturb'' the path of any field that propagates trough structure. One such field is the massless graviton. The effect of GW propagation in a perturbed universe manifests itself in the intermediate/small scale GW power spectrum in the form of an anisotropic component (see e.g. \cite{Alba:2015cms,Contaldi:2016koz,Bartolo:2019oiq,Bartolo:2019yeu}). It is clear that such anisotropies are ultimately of inflationary origin and furthermore that they are universal in the sense that they rely on there being an inflationary phase but require no further model-dependent assumption. Naturally, detection requires the existence of an observable SGWB at the scales  relevant for each given probe. In our nomenclature, these are \textsl{induced} anisotropies and their amplitude is typically  $\delta^{\rm GW}_{\rm ind}\sim \sqrt{A_s}\sim \zeta$. \\
\indent In this work we have studied another source of anisotropies, one that relies on the existence of sufficiently large primordial mixed and/or tensor non-Gaussianities. We have shown that, in order to be the \textit{leading} source of anisotropies, this mechanism requires that the non-linear parameters satisfy:  $F^{\rm tts}_{\rm NL} \gg 1$ and/or $\sqrt{r}\, F^{\rm ttt}_{\rm NL}\gg1$. These requirements limit the models that can be put to the test. On the other hand, by testing anisotropies of this nature we are automatically probing not only inflation but in particular inflationary interactions, which allows us to rule out significant regions of parameter space and possibly model space as well. Furthermore, it is often the case that the same (rich) field content that produces a GW signal within reach of e.g. PTAs or laser interferometers also has widen in the parameter space for those non-Gaussianities that give a leading contribution to GW anisotropies.\\
\indent The first part of this paper took a phenomenological approach. Here we studied auto- and cross-correlations of GW anisotropies and identified the parameter space (in terms of $\Omega_{\rm GW}, F_{\rm NL}$) that is within reach for upcoming GW probes. Our analysis revealed the importance of the angular dependence of the primordial bispectra especially in the case of cross-correlations. For pre-determined angular behaviour in the case of temperature anisotropies, having e.g. a monopolar or quadrupolar TTS signal can make the difference between having an observable cross-correlation \textit{vs} an undetectable one. \\
\indent The focus on a stochastic gravitational wave background of cosmological origin should not distract  from the fact that there most certainly is a background of astrophysical origin. This must be negotiated within our effort to test early universe physics. We have accounted for the AGWB in our analysis, assuming its angular power spectrum amplitude varies between $10^{-30}$ and $10^{-25}$ (in the $10^{-2}$ to $\sim 100$Hz range). We have seen that the use of cross-correlations with the CMB is likely the most powerful tool at our disposal in this case. Indeed, upon using cross-correlations, and depending on the parameter space, the primordial signal can be detected even when the anisotropies angular power spectrum is dominated by the AGWB component.\\
\indent In the last part of this paper, we considered a concrete example of an inflationary model equipped with (i) a blue GW spectrum and (ii) sufficiently large non-Gaussian amplitude to grant a percent level relative error on $F^{\rm ttt}_{\rm NL}$ (for $F^{\rm ttt}_{\rm NL}\sim 10^4$--$10^{5}$) obtained via GW anisotropies measurements. Our specific realisation underscores the fact that, despite the unavoidable $\sim \sqrt{r}$ suppression TTT-sourced anisotropies suffer w.r.t. to their TTS counterpart, it is still possible for TTT to generate the leading anisotropic component of the GW spectrum. It will be very interesting to consider other inflationary realisations for which this is not the case. Another direction worth pursuing is the study of models with a monopolar TTS contribution. We leaves these pursuits to future work.

\section*{Acknowledgements}
M.\,F. would like to acknowledge support from the
``Atracci\'{o}n de Talento'' grant 2019-T1/TIC-15784. G.O. and P.D.M. acknowledge support from the Netherlands organisation for scientific research (NWO) VIDI grant (dossier 639.042.730).

\appendix

\section{Helicity-2 polarisation tensors} \label{app:pol_ten}
In this section we set our conventions for the definition of the helicity-2 polarisation tensors. If the helicity-2 field wave-vector is written in polar coordinates as 
\be
\hat k = (\sin\theta\cos\phi,\sin\theta\sin\phi,\cos\theta)\, ,
\ee
we can define the linear polarisation tensors in terms of two unit vectors perpendicular to $\hat k$ as
\ba
\epsilon_{ij}^{+} &= (u_1)_i (u_1)_j - (u_2)_i (u_2)_j \, ,\\
\epsilon_{ij}^{\times} &= (u_1)_i (u_2)_j + (u_2)_i (u_1)_j \, ,
\ea
where 
\be
u_1 = \left(\sin \phi , - \cos \phi, 0\right)\, ,  \qquad u_2 = \begin{cases} \left(\cos \theta \cos \phi , \cos \theta \sin \phi, - \sin \theta \right) \qquad \mbox{if } \theta < \pi/2 \\
 - \left(\cos \theta \cos \phi , \cos \theta \sin \phi, - \sin \theta \right) \quad \mbox{if } \theta > \pi/2 \, .
\end{cases}
\ee
For a given bispectrum one can always exploit the momentum conservation, $\mathbf{k_1}+\mathbf{k_2}+\mathbf{k_3}=0$, and the invariance under rotations to make the three wave vectors lie on the same plane. If that is chosen to be $(x,y)$ plane, one can parametrise a generic wave-vector $\mathbf{k_i}$ as
\begin{equation}
\mathbf{k_i}=k_i \, (\cos\phi,\sin\phi,0)\, ,
\end{equation}
where $\phi$ is the angle of $\hat k_i$ with respect to the chosen $x$-axis. Employing Eq.~\eqref{eq:def_pol}, the polarisation tensor in the helicity basis reads
\begin{equation}
\epsilon^{(s)}(\mathbf{k_i})=\frac{1}{2}
\begin{pmatrix}
\sin^2\phi & -\sin\phi \cos\phi & -i\lambda_s \sin\phi\\
-\sin\phi \cos\phi & \cos^2\phi & i\lambda_s \cos\phi\\
-i\lambda_s \sin\phi & i\lambda_s \cos\phi & -1\\
\end{pmatrix},
\end{equation}   
where $\lambda_s=\pm 1$ for $s=R$ and $s=L$, respectively.

\section{Computation of \texorpdfstring{$\sigma$-mediated $\langle \gamma \gamma \zeta \rangle$}{sigma mediated TTS bispectrum}} 
\label{app:tss}

In this section we give the detailed computation of the $\sigma$-mediated contribution to $\langle \gamma \gamma \zeta \rangle$ in the model described in Sec. \ref{sec:model_spin}. 

As we are interested in the $c_2 \ll 1$ limit as a way to maximise our primordial bispectra, the dominant in-in contribution being represented in (left panel of) Fig. \ref{fig:feyn_diag}. This turns out to be the tree-level diagram with the highest negative powers of $c_2$. Using the \textsl{in-in} formalism, one can express this diagram in terms of nested commutators (see e.g. \cite{Chen:2009zp})
\begin{align} \label{eq:in-in_formula}
\langle \gamma^{\lambda_1}_{\bk_1}(\tau) \gamma^{\lambda_2}_{\bk_2}(\tau) \zeta_{\bk_3}(\tau)\rangle = &\int_{- \infty}^\tau d\tau_1 \int_{- \infty}^{\tau_1} d\tau_2 \int_{- \infty}^{\tau_2} d\tau_3   \int_{- \infty}^{\tau_3} d\tau_4 \times \nonumber\\
&\times \langle [H_{I}(\tau_4),[H_{I}(\tau_3),[H_{I}(\tau_2), [H_{I}(\tau_1), \gamma^{\lambda_1}_{\bk_1}(\tau) \gamma^{\lambda_2}_{\bk_2}(\tau)\zeta_{\bk_3}(\tau)]]]]\rangle \, ,
\end{align}
where all the integrands are written under a single time-ordered integral. We refer to this form as the \textit{commutator form}. The first step is to sum over all the possible terms one can form by replacing one of the $H_I$ with $H_{(\sigma^{(2)})^2 \sigma^{(0)}}$ i.e. Eq.~\eqref{Hsss0}, another with $H_{\sigma^{(0)} \zeta}$ i.e. Eq.~\eqref{Hs0z}, and the rest with $H_{\sigma^{(2)} \gamma}$, Eq.~\eqref{Hs2g}. By using the mode-functions as specified in Eqs. \eqref{eq:uzeta}, \eqref{eq:ugamma}, \eqref{newwf1} and \eqref{newwf2}, we get
\begin{align} \label{eq:tts}
\langle  \gamma^{\lambda_1}_{\bk_1}(0) \gamma^{\lambda_2}_{\bk_2}(0) \zeta_{\bk_3}(0) \rangle|_{\sigma} = - (2 \pi)^3 \delta^{(3)}(\sum_i \bk_i) \, \frac{\pi^3}{\epsilon}  \frac{\mu}{H} \frac{g^3}{M^3_{Pl}} \frac{1}{k_{1}k_{2}k_{3}}\left[ A + B + C \right] \mathcal A^{\lambda_1 \lambda_2} +\rm{k_1 \leftrightarrow k_2} \, ,
\end{align}
where
\begin{align} \label{eq:Al}
\mathcal A^{\lambda_1 \lambda_2} = \left(\epsilon^{\lambda_1}_{ij}(\hat k_1) \cdot \epsilon_{ij}^{\lambda_2}(\hat k_2) - 3 \, \hat k_3^i \hat k_3^l \cdot \epsilon^{\lambda_1}_{ij}(\hat k_1) \cdot \epsilon_{jl}^{\lambda_2}(\hat k_2)\right) \, ,
\end{align}
and
\begin{align}
A =& \int_{-\infty}^{0} d\tau_{1}\int_{-\infty}^{\tau_{1}} d\tau_{2}\int_{-\infty}^{\tau_{2}} d\tau_{3}\int_{-\infty}^{\tau_{3}} d\tau_{4} \, \sqrt{\frac{\tau_{2}}{\tau_{1}\tau_{3}\tau_{4}}} \, \left(\frac{c_0(\tau_2)}{c_{0i}}\right)^{1/2} \, \left(\frac{c_2(\tau_2)}{c_{2i}}\right) \, \Big\{A_1 + A_2 + A_3 \Big\}\,, \nonumber \\
B =& - \int_{-\infty}^{0} d\tau_{1}\int_{-\infty}^{\tau_{1}} d\tau_{2}\int_{-\infty}^{\tau_{2}} d\tau_{3}\int_{-\infty}^{\tau_{3}} d\tau_{4} \, \sqrt{\frac{\tau_{3}}{\tau_{1}\tau_{2}\tau_{4}}} \, \left(\frac{c_0(\tau_3)}{c_{0i}}\right)^{1/2} \, \left(\frac{c_2(\tau_3)}{c_{2i}}\right) \, \Big\{B_1 + B_2 + B_3 \Big\}\,, \nonumber \\
C =& \int_{-\infty}^{0} d\tau_{1}\int_{-\infty}^{\tau_{1}} d\tau_{2}\int_{-\infty}^{\tau_{2}} d\tau_{3}\int_{-\infty}^{\tau_{3}} d\tau_{4} \, \sqrt{\frac{\tau_{4}}{\tau_{1}\tau_{2}\tau_{3}}} \, \left(\frac{c_0(\tau_4)}{c_{0i}}\right)^{1/2} \, \left(\frac{c_2(\tau_4)}{c_{2i}}\right) \, \Big\{C_1 + C_2 + C_3 \Big\} \, .
\end{align}
Here
\begin{align}\label{eq:A2}
A_1=& \left(\frac{c_0(\tau_1)}{c_{0i}}\right)^{1/2}  \left(\frac{c_2(\tau_3)}{c_{2i}}\right)^{1/2}  \left(\frac{c_2(\tau_4)}{c_{2i}}\right)^{1/2} \,  \times \left(\sin[-k_{3}\tau_{1}] + k_3 \tau_1 \cos[-k_{3}\tau_{1}]\right) \nonumber \\
& \qquad \times \text{Im}\left[e^{i k_{2}\tau_{3}} \mathcal H_\nu^{(1)}(- c_2(\tau_2) k_2 \tau_2) \mathcal H_\nu^{(2)}(- c_2(\tau_3) k_2 \tau_3)\right] \nonumber\\
& \qquad \times \text{Im}\left[\mathcal H_\nu^{(1)}(- c_0(\tau_1) k_3 \tau_1) \mathcal H_\nu^{(2)}(- c_0(\tau_2) k_3 \tau_2)\right]\nonumber\\
& \qquad \times\text{Im}\left[e^{-i k_{1}\tau_{4}}\mathcal H_\nu^{(1)}(- c_2(\tau_4) k_1 \tau_4) \mathcal H_\nu^{(2)}(- c_2(\tau_2) k_1 \tau_2)\right] \,,  \\
A_2=& \left(\frac{c_0(\tau_4)}{c_{0i}}\right)^{1/2}  \left(\frac{c_2(\tau_1)}{c_{2i}}\right)^{1/2}  \left(\frac{c_2(\tau_3)}{c_{2i}}\right)^{1/2} \,  \times  \sin[-k_{1}\tau_{1}] \nonumber \\
& \qquad \times \text{Im}\left[e^{i k_{2}\tau_{3}} \mathcal H_\nu^{(1)}(- c_2(\tau_2) k_2 \tau_2) \mathcal H_\nu^{(2)}(- c_2(\tau_3) k_2 \tau_3)\right] \nonumber \\ 
& \qquad \times \text{Im}\left[\mathcal H_\nu^{(1)}(- c_2(\tau_1) k_1 \tau_1) \mathcal H_\nu^{(2)}(- c_2(\tau_2) k_1 \tau_2)\right] \nonumber \\
& \qquad \times \text{Im}\left[e^{-i k_{3}\tau_{4}}(1+i k_{3}\tau_{4}) \mathcal H_\nu^{(1)}(- c_0(\tau_4) k_3 \tau_4) \mathcal H_\nu^{(2)}(- c_0(\tau_2) k_3 \tau_2) \right] \,, \\
A_3 =& \left(\frac{c_0(\tau_3)}{c_{0i}}\right)^{1/2}  \left(\frac{c_2(\tau_1)}{c_{2i}}\right)^{1/2}  \left(\frac{c_2(\tau_4)}{c_{2i}}\right)^{1/2} \, \times  \sin[-k_{1}\tau_{1}] \nonumber \\
& \qquad \times \text{Im}\left[e^{i k_{3}\tau_{3}}(1-i k_{3}\tau_{3}) \mathcal H_\nu^{(1)}(- c_0(\tau_2) k_3 \tau_2) \mathcal H_\nu^{(2)}(- c_0(\tau_3) k_3 \tau_3)\right] \nonumber\\
&\qquad \times \text{Im}\left[\mathcal H_\nu^{(1)}(- c_2(\tau_1) k_1 \tau_1) \mathcal H_\nu^{(2)}(- c_2(\tau_2) k_1 \tau_2) \right] \nonumber \\
&\qquad \times \text{Im}\left[e^{-i k_{2}\tau_{4}} \mathcal H_\nu^{(1)}(- c_2(\tau_4) k_2 \tau_4) \mathcal H_\nu^{(2)}(- c_2(\tau_2) k_2 \tau_2)\right] \,, \\
\label{eq:B2}
B_1 = & \left(\frac{c_0(\tau_1)}{c_{0i}}\right)^{1/2}  \left(\frac{c_2(\tau_2)}{c_{2i}}\right)^{1/2}  \left(\frac{c_2(\tau_4)}{c_{2i}}\right)^{1/2} \, \times  \left( \sin[-k_{3}\tau_{1}] + k_3 \tau_1 \cos[-k_{3}\tau_{1}]\right)  \nonumber \\
& \qquad \times \text{Im}\left[\mathcal H_\nu^{(1)}(- c_0(\tau_3) k_3 \tau_3) \mathcal H_\nu^{(1)}(- c_2(\tau_3) k_2 \tau_3) \mathcal H_\nu^{(2)}(- c_0(\tau_1) k_3 \tau_1) \mathcal H_\nu^{(2)}(- c_2(\tau_2) k_2 \tau_2)\right] \nonumber \\
& \qquad \times \text{Im}\left[e^{i k_{1}\tau_{4}} \mathcal H_\nu^{(1)}(- c_2(\tau_3) k_1 \tau_3) \mathcal H_\nu^{(2)}(- c_2(\tau_4) k_1 \tau_4)\right] \times \sin[-k_{2}\tau_{2}] \,,  \\
B_2 = & \left(\frac{c_0(\tau_4)}{c_{0i}}\right)^{1/2}  \left(\frac{c_2(\tau_1)}{c_{2i}}\right)^{1/2}  \left(\frac{c_2(\tau_2)}{c_{2i}}\right)^{1/2} \, \times  \sin[-k_{1}\tau_{1}] \nonumber \\
&\times \text{Im}\left[\mathcal H_\nu^{(1)}(- c_2(\tau_3) k_1 \tau_3) \mathcal H_\nu^{(1)}(- c_2(\tau_3) k_2 \tau_3) \mathcal H_\nu^{(2)}(- c_2(\tau_1) k_1 \tau_1) \mathcal H_\nu^{(2)}(- c_2(\tau_2) k_2 \tau_2) \right] \, \nonumber \\
& \qquad \times \text{Im}\left[e^{i k_{3}\tau_{4}}(1-i k_3 \tau_4) \mathcal H_\nu^{(1)}(- c_0(\tau_3) k_3 \tau_3) \mathcal H_\nu^{(2)}(- c_0(\tau_4) k_3 \tau_4) \right] \times \sin[-k_{2}\tau_{2}] \,, \\
B_3 = & \left(\frac{c_0(\tau_2)}{c_{0i}}\right)^{1/2}  \left(\frac{c_2(\tau_1)}{c_{2i}}\right)^{1/2}  \left(\frac{c_2(\tau_4)}{c_{2i}}\right)^{1/2} \, \times  \sin[-k_{1}\tau_{1}] \nonumber \\
& \times \text{Im}\left[\mathcal H_\nu^{(1)}(- c_2(\tau_3) k_1 \tau_3) \mathcal H_\nu^{(1)}(- c_0(\tau_3) k_3 \tau_3) \mathcal H_\nu^{(2)}(- c_2(\tau_1) k_1 \tau_1) \mathcal H_\nu^{(2)}(- c_0(\tau_2) k_3 \tau_2)\right] \, \nonumber \\
& \qquad \times \text{Im}\left[e^{i k_{2}\tau_{4}} \mathcal H_\nu^{(1)}(- c_2(\tau_3) k_2 \tau_3) \mathcal H_\nu^{(2)}(- c_2(\tau_4) k_2 \tau_4) \right] \nonumber \\
& \qquad \times \sin[-k_{2}\tau_{2}]  \left( \sin[-k_{3}\tau_{2}] + k_3 \tau_2 \cos[-k_{3}\tau_{2}]\right)  \,,  \\
\label{eq:C2}
C_1 =& \left(\frac{c_0(\tau_1)}{c_{0i}}\right)^{1/2}  \left(\frac{c_2(\tau_2)}{c_{2i}}\right)^{1/2}  \left(\frac{c_2(\tau_3)}{c_{2i}}\right)^{1/2} \, \times  \left( \sin[-k_{3}\tau_{1}] + k_3 \tau_1 \cos[-k_{3}\tau_{1}]\right)  \sin[-k_{2}\tau_{2}] \sin[-k_{1}\tau_{3}]  \nonumber \\
& \qquad \times \text{Im}\Big[\mathcal H_\nu^{(1)}(- c_0(\tau_4) k_3 \tau_4) \mathcal H_\nu^{(1)}(- c_2(\tau_4) k_2 \tau_4)\mathcal H_\nu^{(1)}(- c_2(\tau_4) k_1 \tau_4)  \nonumber \\ 
& \qquad \qquad \qquad \times \mathcal H_\nu^{(2)}(- c_0(\tau_1) k_3 \tau_1) \mathcal H_\nu^{(2)}(- c_2(\tau_2) k_2 \tau_2) \mathcal H_\nu^{(2)}(- c_2(\tau_3) k_1 \tau_3) \Big]   \, , \\
C_2 =& \left(\frac{c_0(\tau_3)}{c_{0i}}\right)^{1/2}  \left(\frac{c_2(\tau_1)}{c_{2i}}\right)^{1/2}  \left(\frac{c_2(\tau_2)}{c_{2i}}\right)^{1/2} \, \times  \left(\sin[-k_{3}\tau_{3}] + k_3 \tau_3 \cos[-k_{3}\tau_{3}]\right)  \sin[-k_{2}\tau_{2}] \sin[-k_{1}\tau_{1}]  \nonumber\\
& \qquad \times \text{Im}\Big[\mathcal H_\nu^{(1)}(- c_0(\tau_4) k_3 \tau_4) \mathcal H_\nu^{(1)}(- c_2(\tau_4) k_2 \tau_4)\mathcal H_\nu^{(1)}(- c_2(\tau_4) k_1 \tau_4)  \nonumber \\ 
& \qquad \qquad \qquad \times \mathcal H_\nu^{(2)}(- c_0(\tau_3) k_3 \tau_3) \mathcal H_\nu^{(2)}(- c_2(\tau_2) k_2 \tau_2) \mathcal H_\nu^{(2)}(- c_2(\tau_1) k_1 \tau_1) \Big]   \, , \\
C_3 =& \left(\frac{c_0(\tau_2)}{c_{0i}}\right)^{1/2}  \left(\frac{c_2(\tau_1)}{c_{2i}}\right)^{1/2}  \left(\frac{c_2(\tau_3)}{c_{2i}}\right)^{1/2} \, \times \left( \sin[-k_{3}\tau_{2}] + k_3 \tau_2 \cos[-k_{3}\tau_{2}]\right)  \sin[-k_{2}\tau_{3}] \sin[-k_{1}\tau_{1}]  \nonumber \\
& \qquad \times \text{Im}\Big[\mathcal H_\nu^{(1)}(- c_0(\tau_4) k_3 \tau_4) \mathcal H_\nu^{(1)}(- c_2(\tau_4) k_2 \tau_4)\mathcal H_\nu^{(1)}(- c_2(\tau_4) k_1 \tau_4)  \nonumber \\ 
& \qquad \qquad \qquad \times \mathcal H_\nu^{(2)}(- c_0(\tau_2) k_3 \tau_2) \mathcal H_\nu^{(2)}(- c_2(\tau_3) k_2 \tau_3) \mathcal H_\nu^{(2)}(- c_2(\tau_1) k_1 \tau_1) \Big]   \, .
\end{align}
One may want to rewrite the integrals by  introducing the dimensionless time $x_i = k_1 \tau_i$, finding
\begin{align}
A =& \frac{1}{k_1^3} \int_{-\infty}^{0} dx_{1}\int_{-\infty}^{x_{1}} dx_{2}\int_{-\infty}^{x_{2}} dx_{3}\int_{-\infty}^{x_{3}} dx_{4} \, \sqrt{\frac{x_{2}}{x_{1}x_{3}x_{4}}}  \, \left(\frac{c_0(\frac{x_2}{k_1})}{c_{0i}}\right)^{1/2} \, \left(\frac{c_2(\frac{x_2}{k_1})}{c_{2i}}\right) \, \Big\{A_1 + A_2 + A_3 \Big\} \,,\nonumber \\
B =& - \frac{1}{k_1^3}  \int_{-\infty}^{0} dx_{1}\int_{-\infty}^{x_{1}} dx_{2}\int_{-\infty}^{x_{2}} dx_{3}\int_{-\infty}^{x_{3}} dx_{4} \, \sqrt{\frac{x_{3}}{x_{1}x_{2}x_{4}}} \, \left(\frac{c_0(\frac{x_3}{k_1})}{c_{0i}}\right)^{1/2} \, \left(\frac{c_2(\frac{x_3}{k_1})}{c_{2i}}\right) \,  \Big\{B_1 + B_2 + B_3 \Big\} \,,\nonumber \\
C =& \frac{1}{k_1^3}  \int_{-\infty}^{0} dx_{1}\int_{-\infty}^{x_{1}} dx_{2}\int_{-\infty}^{x_{2}} dx_{3}\int_{-\infty}^{x_{3}} dx_{4} \, \sqrt{\frac{x_{4}}{x_{1}x_{2}x_{3}}} \, \left(\frac{c_0(\frac{x_4}{k_1})}{c_{0i}}\right)^{1/2} \, \left(\frac{c_2(\frac{x_4}{k_1})}{c_{2i}}\right) \, \Big\{C_1 + C_2 + C_3 \Big\} \, ,
\end{align}
where
\begin{align}\label{eq:A3}
A_1 =& \left(\frac{c_0(\frac{x_1}{k_1})}{c_{0i}}\right)^{1/2}  \left(\frac{c_2(\frac{x_3}{k_1})}{c_{2i}}\right)^{1/2}  \left(\frac{c_2(\frac{x_4}{k_1})}{c_{2i}}\right)^{1/2} \, \times \left( \sin[-\frac{k_3}{k_1}x_{1}] + \frac{k_3}{k_1} x_1 \cos[- \frac{k_3}{k_1} x_{1}]\right) \times \nonumber\\
&\times \text{Im}\left[e^{i \frac{k_{2}}{k_{1}}x_{3}} \mathcal H_\nu^{(1)}(- c_2(\frac{x_2}{k_1}) \frac{k_2}{k_1} x_2) \mathcal H_\nu^{(2)}(- c_2(\frac{x_3}{k_1}) \frac{k_2}{k_1} x_3)\right] \nonumber \\
&\times \text{Im}\left[\mathcal H_\nu^{(1)}(- c_0(\frac{x_1}{k_1}) \frac{k_3}{k_1} x_1) \mathcal H_\nu^{(2)}(- c_0(\frac{x_2}{k_1}) \frac{k_3}{k_1} x_2)\right] \nonumber\\
&\times\text{Im}\left[e^{-i x_{4}}\mathcal H_\nu^{(1)}(- c_2(\frac{x_4}{k_1}) x_4) \mathcal H_\nu^{(2)}(- c_2(\frac{x_2}{k_1}) x_2)\right]  \,, \\
A_2=& \left(\frac{c_0(\frac{x_4}{k_1})}{c_{0i}}\right)^{1/2}  \left(\frac{c_2(\frac{x_1}{k_1})}{c_{2i}}\right)^{1/2}  \left(\frac{c_2(\frac{x_3}{k_1})}{c_{2i}}\right)^{1/2} \, \times  \sin[- x_{1}] \nonumber \\
& \times \text{Im}\left[e^{i \frac{k_{2}}{k_{1}} x_{3}} \mathcal H_\nu^{(1)}(- c_2(\frac{x_2}{k_1}) \frac{k_{2}}{k_{1}} x_2) \mathcal H_\nu^{(2)}(- c_2(\frac{x_3}{k_1}) \frac{k_{2}}{k_{1}} x_3)\right] \nonumber \\ 
& \,\,\times \text{Im}\left[\mathcal H_\nu^{(1)}(- c_2(\frac{x_1}{k_1}) x_1) \mathcal H_\nu^{(2)}(- c_2(\frac{x_2}{k_1}) x_2)\right] \nonumber \\
& \,\,\times \text{Im}\left[e^{-i \frac{k_{3}}{k_{1}} x_{4}}(1+i \frac{k_{3}}{k_{1}} x_{4}) \mathcal H_\nu^{(1)}(- c_0(\frac{x_4}{k_1}) \frac{k_{3}}{k_{1}} x_4) \mathcal H_\nu^{(2)}(- c_0(\frac{x_2}{k_1}) \frac{k_{3}}{k_{1}} x_2) \right] \,, \label{eq:exampleA2}\\
A_3 =& \left(\frac{c_0(\frac{x_3}{k_1})}{c_{0i}}\right)^{1/2}  \left(\frac{c_2(\frac{x_1}{k_1})}{c_{2i}}\right)^{1/2}  \left(\frac{c_2(\frac{x_4}{k_1})}{c_{2i}}\right)^{1/2} \, \times \sin[- x_{1}] \nonumber \\
& \times \text{Im}\left[e^{i \frac{k_{3}}{k_{1}} x_{3}}(1-i \frac{k_{3}}{k_{1}} x_{3}) \mathcal H_\nu^{(1)}(- c_0(\frac{x_2}{k_1}) \frac{k_{3}}{k_{1}} x_2) \mathcal H_\nu^{(2)}(- c_0(\frac{x_3}{k_1}) \frac{k_{3}}{k_{1}} x_3)\right] \nonumber\\
&\,\,\times \text{Im}\left[\mathcal H_\nu^{(1)}(- c_2(\frac{x_1}{k_1}) x_1) \mathcal H_\nu^{(2)}(- c_2(\frac{x_2}{k_1}) x_2) \right] \nonumber \\
& \times \text{Im}\left[e^{-i \frac{k_{2}}{k_{1}} x_{4}} \mathcal H_\nu^{(1)}(- c_2(\frac{x_4}{k_1}) \frac{k_{2}}{k_{1}} x_4) \mathcal H_\nu^{(2)}(- c_2(\frac{x_2}{k_1}) \frac{k_{2}}{k_{1}} x_2)\right]\,, \\
\label{eq:B3}
B_1 =& \left(\frac{c_0(\frac{x_1}{k_1})}{c_{0i}}\right)^{1/2}  \left(\frac{c_2(\frac{x_2}{k_1})}{c_{2i}}\right)^{1/2}  \left(\frac{c_2(\frac{x_4}{k_1})}{c_{2i}}\right)^{1/2} \, \times \left( \sin[-\frac{k_3}{k_1} x_1] + \frac{k_3}{k_1} x_1 \cos[-\frac{k_3}{k_1} x_1]\right) \times \nonumber \\
&\times \text{Im}\left[\mathcal H_\nu^{(1)}(- c_0(\frac{x_3}{k_1}) \frac{k_3}{k_1} x_3) \mathcal H_\nu^{(1)}(- c_2(\frac{x_3}{k_1}) \frac{k_2}{k_1} x_3) \mathcal H_\nu^{(2)}(- c_0(\frac{x_1}{k_1}) \frac{k_3}{k_1} x_1) \mathcal H_\nu^{(2)}(- c_2(\frac{x_2}{k_1}) \frac{k_2}{k_1} x_2)\right] 
\nonumber\\
&\times \text{Im}\left[e^{i x_{4}} \mathcal H_\nu^{(1)}(- c_2(\frac{x_3}{k_1}) x_3) \mathcal H_\nu^{(2)}(- c_2(\frac{x_4}{k_1}) x_4)\right]  \sin[-\frac{k_2}{k_1} x_2]  \,, \\
B_2=& \left(\frac{c_0(\frac{x_4}{k_1})}{c_{0i}}\right)^{1/2}  \left(\frac{c_2(\frac{x_1}{k_1})}{c_{2i}}\right)^{1/2}  \left(\frac{c_2(\frac{x_2}{k_1})}{c_{2i}}\right)^{1/2} \, \times \sin[- x_{1}] \nonumber\\
& \times \text{Im}\left[\mathcal H_\nu^{(1)}(- c_2(\frac{x_3}{k_1}) x_3) \mathcal H_\nu^{(1)}(- c_2(\frac{x_3}{k_1}) \frac{k_2}{k_1} x_3) \mathcal H_\nu^{(2)}(- c_2(\frac{x_1}{k_1}) x_1) \mathcal H_\nu^{(2)}(- c_2(\frac{x_2}{k_1}) \frac{k_2}{k_1} x_2) \right] \, \nonumber \\
& \,\,\times \text{Im}\left[e^{i \frac{k_3}{k_1} x_{4}}(1-i \frac{k_3}{k_1} x_4) \mathcal H_\nu^{(1)}(- c_0(\frac{x_3}{k_1}) \frac{k_3}{k_1} x_3) \mathcal H_\nu^{(2)}(- c_0(\frac{x_4}{k_1}) \frac{k_3}{k_1} x_4) \right] \times \sin[-\frac{k_2}{k_1} x_{2}] \,, \\
B_3=& \left(\frac{c_0(\frac{x_2}{k_1})}{c_{0i}}\right)^{1/2}  \left(\frac{c_2(\frac{x_1}{k_1})}{c_{2i}}\right)^{1/2}  \left(\frac{c_2(\frac{x_4}{k_1})}{c_{2i}}\right)^{1/2} \, \times \sin[- x_{1}] \nonumber \\
& \times \text{Im}\left[\mathcal H_\nu^{(1)}(- c_2(\frac{x_3}{k_1}) x_3) \mathcal H_\nu^{(1)}(- c_0(\frac{x_3}{k_1}) \frac{k_3}{k_1} x_3) \mathcal H_\nu^{(2)}(- c_2(\frac{x_1}{k_1}) x_1) \mathcal H_\nu^{(2)}(- c_0(\frac{x_2}{k_1}) \frac{k_3}{k_1} x_2)\right] \, \nonumber \\
& \,\,\times \text{Im}\left[e^{i \frac{k_2}{k_1} x_{4}} \mathcal H_\nu^{(1)}(- c_2(\frac{x_3}{k_1}) \frac{k_2}{k_1} x_3) \mathcal H_\nu^{(2)}(- c_2(\frac{x_4}{k_1}) \frac{k_2}{k_1} x_4) \right] \nonumber\\
& \,\, \times \sin[- \frac{k_2}{k_1} x_{2}]  \left( \sin[- \frac{k_3}{k_1} x_{2}] + \frac{k_3}{k_1} x_2 \cos[- \frac{k_3}{k_1} x_{2}]\right)  \,, \\
\label{eq:C3}
C_1 =& \left(\frac{c_0(\frac{x_1}{k_1})}{c_{0i}}\right)^{1/2}  \left(\frac{c_2(\frac{x_2}{k_1})}{c_{2i}}\right)^{1/2}  \left(\frac{c_2(\frac{x_3}{k_1})}{c_{2i}}\right)^{1/2} \, \times \left( \sin[-\frac{k_3}{k_1} x_1] + \frac{k_3}{k_1} x_1 \cos[-\frac{k_3}{k_1} x_1]\right) \times \nonumber\\
& \times \sin[-\frac{k_2}{k_1} x_2]\sin[-x_{3}] \times \text{Im}\Big[\mathcal H_\nu^{(1)}(- c_0(\frac{x_4}{k_1}) \frac{k_3}{k_1} x_4) \mathcal H_\nu^{(1)}(- c_2(\frac{x_4}{k_1}) \frac{k_2}{k_1} x_4)\mathcal H_\nu^{(1)}(- c_2(\frac{x_4}{k_1}) x_4) \times \nonumber \\
& \qquad \qquad \qquad \qquad\qquad \qquad \times \mathcal H_\nu^{(2)}(- c_0(\frac{x_1}{k_1}) \frac{k_3}{k_1} x_1) \mathcal H_\nu^{(2)}(- c_2(\frac{x_2}{k_1}) \frac{k_2}{k_1} x_2) \mathcal H_\nu^{(2)}(- c_2(\frac{x_3}{k_1}) x_3) \Big] \,, \\
C_2 =& \left(\frac{c_0(\frac{x_3}{k_1})}{c_{0i}}\right)^{1/2}  \left(\frac{c_2(\frac{x_1}{k_1})}{c_{2i}}\right)^{1/2}  \left(\frac{c_2(\frac{x_2}{k_1})}{c_{2i}}\right)^{1/2} \, \times \left( \sin[-\frac{k_3}{k_1} x_3] + \frac{k_3}{k_1} x_3 \cos[-\frac{k_3}{k_1} x_3]\right) \times \nonumber\\
& \times \sin[-\frac{k_2}{k_1} x_2]\sin[-x_{1}]\times 
\text{Im}\Big[\mathcal H_\nu^{(1)}(- c_0(\frac{x_4}{k_1}) \frac{k_3}{k_1} x_4) \mathcal H_\nu^{(1)}(- c_2(\frac{x_4}{k_1}) \frac{k_2}{k_1} x_4)\mathcal H_\nu^{(1)}(- c_2(\frac{x_4}{k_1}) x_4) \times \nonumber \\ 
& \qquad \qquad \qquad \qquad\qquad \qquad \times \mathcal H_\nu^{(2)}(- c_0(\frac{x_3}{k_1}) \frac{k_3}{k_1} x_3) \mathcal H_\nu^{(2)}(- c_2(\frac{x_2}{k_1}) \frac{k_2}{k_1} x_2) \mathcal H_\nu^{(2)}(- c_2(\frac{x_1}{k_1}) x_1) \Big]  \,, \\
C_3 =& \left(\frac{c_0(\frac{x_2}{k_1})}{c_{0i}}\right)^{1/2}  \left(\frac{c_2(\frac{x_1}{k_1})}{c_{2i}}\right)^{1/2}  \left(\frac{c_2(\frac{x_3}{k_1})}{c_{2i}}\right)^{1/2} \, \times  \left( \sin[-\frac{k_3}{k_1} x_2] + \frac{k_3}{k_1} x_2 \cos[-\frac{k_3}{k_1} x_2]\right) \nonumber \\
& \times \sin[-\frac{k_2}{k_1} x_3] \sin[-x_{1}] \times \text{Im}\Big[\mathcal H_\nu^{(1)}(- c_0(\frac{x_4}{k_1}) \frac{k_3}{k_1} x_4) \mathcal H_\nu^{(1)}(- c_2(\frac{x_4}{k_1}) \frac{k_2}{k_1} x_4)\mathcal H_\nu^{(1)}(- c_2(\frac{x_4}{k_1}) x_4) \times \nonumber \\ 
& \qquad \qquad \qquad \qquad \qquad \qquad \times \mathcal H_\nu^{(2)}(- c_0(\frac{x_2}{k_1}) \frac{k_3}{k_1} x_2) \mathcal H_\nu^{(2)}(- c_2(\frac{x_3}{k_1}) \frac{k_2}{k_1} x_3) \mathcal H_\nu^{(2)}(- c_2(\frac{x_1}{k_1}) x_1) \Big]  \,.
\end{align}
At this stage it is convenient to employ the approximations discussed in Sec.~\ref{sub:time_sound} whereby we use a scale-dependent sound speed in lieu of a time-dependent one. This leads to the next step in our calculation, which is performed in the squeezed limit\footnote{Note here that working in the squeezed configuration amounts to enforcing a hierarchy among the momenta that is instrumental in determining the appropriate horizon discussed in Sec.~\ref{sub:time_sound}.}.
\vspace{0.2cm}

\centerline{\bf Squeezed limit}

\vspace{0.2cm}
\noindent In this subsection we evaluate the squeezed limit of the previous result when the scalar mode is much smaller than the tensor modes, $k_L = k_3 \ll k_1 \simeq k_2 = k_S$. By using the same arguments of \cite{Iacconi:2020yxn}, we note that the terms $A_2$ and $B_2$ are of lowest order in the ratio $k_L/k_S \rightarrow 0$. Thus, they represent the dominant contribution in the squeezed limit. In this limit one obtains
\begin{align}
A =& \frac{1}{k_S^3} \left(\frac{c_{2}(k_{S})}{c_{2 i}}\right)^{2}\left(\frac{c_{0}(k_{S})}{c_{0 i}}\right)^{1/2}  \left(\frac{c_{0}(k_{L})}{c_{0 i}}\right)^{1/2} \int_{-\infty}^{0} dx_{1}\int_{-\infty}^{x_{1}} dx_{2}\int_{-\infty}^{x_{2}} dx_{3}\int_{-\infty}^{x_{3}} dx_{4} \, \sqrt{\frac{x_{2}}{x_{1}x_{3}x_{4}}}  \, A_2 \, , \\
B =& - \frac{1}{k_S^3} \left(\frac{c_{2}(k_{S})}{c_{2 i}}\right)^{2}\left(\frac{c_{0}(k_{S})}{c_{0 i}}\right)^{1/2}  \left(\frac{c_{0}(k_{L})}{c_{0 i}}\right)^{1/2} \int_{-\infty}^{0} dx_{1}\int_{-\infty}^{x_{1}} dx_{2}\int_{-\infty}^{x_{2}} dx_{3}\int_{-\infty}^{x_{3}} dx_{4} \, \sqrt{\frac{x_{3}}{x_{1}x_{2}x_{4}}}  \, B_2 \, ,
\end{align}
where
\begin{eqnarray}\label{eq:A5}
A_2&=&  \text{Im}\left[e^{i x_{3}} \mathcal H_\nu^{(1)}(- c_2(k_S) x_2) \mathcal H_\nu^{(2)}(- c_2(k_S) x_3)\right] \nonumber \\ 
&& \,\,\times \text{Im}\left[\mathcal H_\nu^{(1)}(- c_2(k_S) x_1) \mathcal H_\nu^{(2)}(- c_2(k_S) x_2)\right] \nonumber \\
&& \,\,\times \text{Im}\left[e^{-i \frac{k_{L}}{k_{S}} x_{4}} \mathcal H_\nu^{(1)}(- c_0(k_L) \frac{k_{L}}{k_{S}} x_4) \mathcal H_\nu^{(2)}(- c_0(k_S) \frac{k_{L}}{k_{S}} x_2) \right] \times \sin[- x_{1}] \,,
\end{eqnarray}
and
\begin{eqnarray}
\label{eq:B5}
B_2&=& \text{Im}\left[\mathcal H_\nu^{(1)}(- c_2(k_S) x_3) \mathcal H_\nu^{(1)}(- c_2(k_S) x_3) \mathcal H_\nu^{(2)}(- c_2(k_S) x_1) \mathcal H_\nu^{(2)}(- c_2(k_S) x_2) \right] \, \nonumber \\
&& \,\,\times \text{Im}\left[e^{i \frac{k_L}{k_S} x_{4}} \mathcal H_\nu^{(1)}(- c_0(k_S) \frac{k_L}{k_S} x_3) \mathcal H_\nu^{(2)}(- c_0(k_L) \frac{k_L}{k_S} x_4) \right] \times  \sin[- x_{1}] \sin[- x_{2}] \, .\;
\end{eqnarray}
By virtue of the change of variable $k_L/k_S\, x_4 = y$, these can be rewritten as
\begin{eqnarray}\label{eq:A6}
A&=& k_S^{-3} \, \left(\frac{k_L}{k_S}\right)^{-1/2} \left(\frac{c_{2}(k_{S})}{c_{2 i}}\right)^{2}\left(\frac{c_{0}(k_{S})}{c_{0 i}}\right)^{1/2}  \left(\frac{c_{0}(k_{L})}{c_{0 i}}\right)^{1/2} \nonumber \\
&& \times \int_{-\infty}^{0} dx_{1}\int_{-\infty}^{x_{1}} dx_{2}\int_{-\infty}^{x_{2}} dx_{3}\int_{-\infty}^{k_L/k_S x_{3}} dy \, \sqrt{\frac{x_{2}}{x_{1}x_{3}y}} \, \sin[- x_{1}] \nonumber\\
&& \, \times \text{Im}\left[e^{i x_{3}} \mathcal H_\nu^{(1)}(- c_2(k_S) x_2) \mathcal H_\nu^{(2)}(- c_2(k_S) x_3)\right] \nonumber \\ 
&& \,\,\times \text{Im}\left[\mathcal H_\nu^{(1)}(- c_2(k_S) x_1) \mathcal H_\nu^{(2)}(- c_2(k_S) x_2)\right] \nonumber \\
&& \,\,\times \text{Im}\left[e^{-i y} \mathcal H_\nu^{(1)}(- c_0(k_L) y) \mathcal H_\nu^{(2)}(- c_0(k_S) \frac{k_{L}}{k_{S}} x_2) \right]  \, ,
\end{eqnarray}
and
\begin{eqnarray}
\label{eq:B6}
B&=& - k_S^{-3} \,\left(\frac{c_{2}(k_{S})}{c_{2 i}}\right)^{2}\left(\frac{c_{0}(k_{S})}{c_{0 i}}\right)^{1/2}  \left(\frac{c_{0}(k_{L})}{c_{0 i}}\right)^{1/2} \nonumber \\
&& \times \int_{-\infty}^{0} dx_{1}\int_{-\infty}^{x_{1}} dx_{2}\int_{-\infty}^{x_{2}} dx_{3}\int_{-\infty}^{k_L/k_S x_{3}} dy \, \sqrt{\frac{x_{3}}{x_{1}x_{2}y}} \,  \sin[- x_{1}] \sin[- x_{2}] \nonumber \\
&&  \times \text{Im}\left[\mathcal H_\nu^{(1)}(- c_2(k_S) x_3) \mathcal H_\nu^{(1)}(- c_2(k_S) x_3) \mathcal H_\nu^{(2)}(- c_2(k_S) x_1) \mathcal H_\nu^{(2)}(- c_2(k_S) x_2) \right] \, \nonumber \\
&& \,\,\times \text{Im}\left[e^{i y} \mathcal H_\nu^{(1)}(- c_0(k_S) \frac{k_L}{k_S} x_3) \mathcal H_\nu^{(2)}(- c_0(k_L) y) \right] \,.
\end{eqnarray}
This last equation can be rewritten as 
\begin{eqnarray}
B&=& k_S^{-3} \, \left(\frac{c_{2}(k_{S})}{c_{2 i}}\right)^{2}\left(\frac{c_{0}(k_{S})}{c_{0 i}}\right)^{1/2}  \left(\frac{c_{0}(k_{L})}{c_{0 i}}\right)^{1/2} \nonumber \\
&& \times \int_{-\infty}^{0} dx_{1}\int_{-\infty}^{x_{1}} dx_{2}\int_{-\infty}^{x_{2}} dx_{3}\int_{-\infty}^{k_L/k_S x_{3}} dy \, \sqrt{\frac{x_{3}}{x_{1}x_{2}y}} \, \sin[- x_{1}] \sin[- x_{2}] \nonumber\\
&&  \times \text{Im}\left[\mathcal H_\nu^{(1)}(- c_2(k_S) x_3) \mathcal H_\nu^{(1)}(- c_2(k_S) x_3) \mathcal H_\nu^{(2)}(- c_2(k_S) x_1) \mathcal H_\nu^{(2)}(- c_2(k_S) x_2) \right] \, \nonumber \\
&& \,\,\times \text{Im}\left[e^{-i y} \mathcal H_\nu^{(2)}(- c_0(k_S) \frac{k_L}{k_S} x_3) \mathcal H_\nu^{(1)}(- c_0(k_L) y) \right]  \,.
\end{eqnarray}
The Hankel functions of the kind $\mathcal H_\nu^{(2)}(- c_2(k_S) x_3)$, oscillate and, as a result, suppress the integral for $-c_2(k_S) x_3 \gg 1$ \footnote{In this regards, a clarification should be made. In principle, in the in-in integrals we should introduce the $i\epsilon$ prescription at the far past, which projects the interacting vacuum of the full theory into the vacuum of the free theory (see e.g. \cite{Baumann:2018muz}). Therefore, by giving an imaginary component to the integration contour in the asymptotic past, the oscillatory behaviour at $-\infty$ of the Hankel functions, sines and cosines turn into an exponential decay, suppressing the integration in the large argument limit.}. On small scales we want $c_2(k_S)$ small up to $10^{-3}$, therefore only values $|x_3| \lesssim 10^3$ are relevant for the integral computation. As a consequence, the upper limit of the integral in $y_4$ is effectively zero provided that $k_L/k_S \ll 10^{-3}$. Since we want to consider the scenario where $k_S$ is a small scale ($k_S> 10^{5} \mbox{ Mpc}^{-1}$) and $k_L$ is a CMB scale ($k_L \lesssim 10^{-2} \mbox{ Mpc}^{-1}$), this requirement is easily satisfied. This allows one to factorise the $y$ integration.

Looking at the Hankel functions $\mathcal H_\nu^{(2)}(- c_2(k_S) x_j)$, one can further infer that only values of $x_j$ for which $- c_2(k_S) x_j \lesssim 1$ (so $|x_j| \lesssim c^{-1}_2(k_S)$)  contribute to the integral. So, the argument of the Hankel functions $H_\nu^{(2)}(- c_0(k_S) k_L/k_S \,x_j)$ is smaller than $10^{-7}\times c_0(k_S)/c_2(k_S) < 10^{-4}$, as we want $c_0(k_S)$ to be of order $1$. This suggests that the Hankel function $\mathcal H_\nu^{(2)}(- c_0(k_S) k_L/k_S \,x_j)$ can be approximated in the small argument limit as (see e.g. \cite{NIST:DLMF})
\begin{equation}
\mathcal H_\nu^{(2)}(- c_0(k_S) k_L/k_S \,x_j) \simeq  i \frac{ 2^\nu \, \Gamma(\nu) (-x_j)^{-\nu}}{\pi c^\nu_0(k_S)} \left(\frac{k_S}{k_L}\right)^{\nu}\, . 
\end{equation}
As a result of these approximations, the previous integrals reduce into
\begin{eqnarray}\label{eq:A7}
A&=& \frac{2^\nu \, \Gamma(\nu)}{\pi c^\nu_0(k_S)} \, \left(\frac{c_{2}(k_{S})}{c_{2 i}}\right)^{2}\left(\frac{c_{0}(k_{S})}{c_{0 i}}\right)^{1/2}  \left(\frac{c_{0}(k_{L})}{c_{0 i}}\right)^{1/2} \,  k_L^{-1/2-\nu} k_S^{-5/2 + \nu} \nonumber \\
&& \times \int_{-\infty}^{0} dx_{1}\int_{-\infty}^{x_{1}} dx_{2}\int_{-\infty}^{x_{2}} dx_{3} \,  (-x_{1})^{-1/2} (-x_{2})^{1/2-\nu} (-x_{3})^{-1/2}  \, \sin[- x_{1}] \nonumber \\
&&  \times \, \text{Im}\left[e^{i x_{3}} \mathcal H_\nu^{(1)}(- c_2(k_S) x_2) \mathcal H_\nu^{(2)}(- c_2(k_S) x_3)\right] \text{Im}\left[\mathcal H_\nu^{(1)}(- c_2(k_S) x_1) \mathcal H_\nu^{(2)}(- c_2(k_S) x_2)\right] \nonumber \\
&&\times \left(\int_{-\infty}^{0} dy\, (- y)^{-1/2} \text{Re}\left[e^{-i y} \mathcal H_\nu^{(1)}(- c_0(k_L) y)\right] \right) \, ,
\end{eqnarray}
and
\begin{eqnarray}
\label{eq:B7}
B&=& \frac{2^\nu \, \Gamma(\nu)}{\pi c^\nu_0(k_S)} \, \left(\frac{c_{2}(k_{S})}{c_{2 i}}\right)^{2}\left(\frac{c_{0}(k_{S})}{c_{0 i}}\right)^{1/2}  \left(\frac{c_{0}(k_{L})}{c_{0 i}}\right)^{1/2} \, k_L^{-1/2-\nu} k_S^{-5/2 + \nu}  \,  \nonumber \\
&& \times \int_{-\infty}^{0} dx_{1}\int_{-\infty}^{x_{1}} dx_{2}\int_{-\infty}^{x_{2}} dx_{3} \, (-x_{1})^{-1/2} (-x_{2})^{-1/2} (-x_{3})^{1/2-\nu} \, \sin[- x_{1}] \sin[- x_{2}] \nonumber \\
&&\times  \, \text{Im}\left[\mathcal H_\nu^{(1)}(- c_2(k_S) x_3) \mathcal H_\nu^{(1)}(- c_2(k_S) x_3) \mathcal H_\nu^{(2)}(- c_2(k_S) x_1) \mathcal H_\nu^{(2)}(- c_2(k_S) x_2) \right] \, \nonumber \\
&& \times \left(\int_{-\infty}^{0} dy\, (- y)^{-1/2} \text{Re}\left[e^{-i y} \mathcal H_\nu^{(1)}(- c_0(k_L) y)\right] \right) \, .
\end{eqnarray}
With this in mind, the bispectrum in Eq.~\eqref{eq:tts} to leading order in the squeezed limit becomes
\begin{align}
\langle  \gamma^{\lambda_1}_{k_S}(0) \gamma^{\lambda_2}_{k_S}(0) \zeta_{k_L}(0) \rangle|^{\rm squeez}_{\sigma} = & - (2 \pi)^3 \delta^{(3)}(\sum_i \bk_i) \, \frac{2 \pi^2}{\epsilon}  \frac{\mu}{H} \frac{g^3}{M^3_{Pl}} \, 2^\nu \, k_L^{-3/2-\nu} k_S^{-9/2 + \nu} \times \nonumber \\
& \times  \,\left(\frac{c_{2}(k_{S})}{c_{2 i}}\right)^{2}\left(\frac{c_{0}(k_{S})}{c_{0 i}}\right)^{1/2}  \left(\frac{c_{0}(k_{L})}{c_{0 i}}\right)^{1/2} \times \nonumber\\
& \times \, \mathcal I(c_{0}, c_{2}, \nu)
\times  \frac{4 \pi}{5} \sum_{M} Y_{2 M}(\hat k_L) \, Y^*_{2 M}(\hat k_S) \times \begin{cases}
1 \quad \mbox{if} \qquad \lambda_{1} = \lambda_{2} \\
0 \quad \mbox{if} \qquad \lambda_{1} \neq \lambda_{2}
\end{cases}\, ,
\end{align}
where
\begin{eqnarray}
\label{eq:I}
\mathcal I(c_{0}, c_{2}, \nu)&=& \frac{\Gamma(\nu)}{c^\nu_0(k_S)} \int_{-\infty}^{0} dx_{1}\int_{-\infty}^{x_{1}} dx_{2}\int_{-\infty}^{x_{2}} dx_{3} \, (-x_{1})^{-1/2} \,\nonumber \\
&&  \times \Big\{(-x_{2})^{1/2-\nu} (-x_{3})^{-1/2} \, \times \sin[- x_{1}] \text{Im}\left[e^{i x_{3}} \mathcal H_\nu^{(1)}(- c_2(k_S) x_2) \mathcal H_\nu^{(2)}(- c_2(k_S) x_3)\right] \nonumber\\
&&\quad\quad\times \text{Im}\left[\mathcal H_\nu^{(1)}(- c_2(k_S) x_1) \mathcal H_\nu^{(2)}(- c_2(k_S) x_2)\right] +\nonumber \\
&&  \quad + (-x_{2})^{-1/2} (-x_{3})^{1/2-\nu} \sin[- x_{1}] \sin[- x_{2}] \, \text{Im}\Big[\mathcal H_\nu^{(1)}(- c_2(k_S) x_3) \mathcal H_\nu^{(1)}(- c_2(k_S) x_3)  \nonumber\\
&& \quad \quad  \quad \times \mathcal H_\nu^{(2)}(- c_2(k_S) x_1) \mathcal H_\nu^{(2)}(- c_2(k_S) x_2) \Big] \Big\} \times \nonumber \\
&&\times \left(\int_{-\infty}^{0} dy\, (- y)^{-1/2} \text{Re}\left[e^{-i y} \mathcal H_\nu^{(1)}(- c_0(k_L) y)\right] \right) \, .
\end{eqnarray}
Notice that this is the same quantity as Eq.~(3.12) of \cite{Iacconi:2020yxn} apart for the coefficient exchange $c_2^\nu(k_L) \rightarrow c_0^\nu(k_S)$ outside the integrals, and the exchange $c_2(k_L) \rightarrow c_0(k_L)$ in the $y$ integration. This suggests that it can be fit by the following power law
\begin{align} \label{fit_pow_nu}
\mathcal I(c_0, c_2, \nu) = \frac{a}{c^{\nu}_0(k_L) c^{\nu}_0(k_S) c^{2 \nu}_2(k_S)} \, .
\end{align}
As in our case $c_0(k_L)  \simeq c_0(k_S) = c_0$, in Tab. \ref{tab:fit_power} we present the fit of Eq.~\eqref{eq:I} with the following power law
\begin{align} \label{fit_pow}
\mathcal I(c_0, c_2, \nu) = \frac{a}{c^{b}_0 \, c^{c}_2(k_S)} \, .
\end{align}
The results of the fit suggest that the power law Eq.~\eqref{fit_pow_nu} can be taken as a very good fit of Eq.~\eqref{eq:I} as well. In Tab. \ref{tab:fit_power_nu} we present the fit of Eq.~\eqref{eq:I} with the power law Eq.~\eqref{fit_pow_nu}.
\begin{table}[h!]
\begin{center}
\begin{tabular}{ |c|c|c|c|c| } 
\hline
$\nu$ & $a$ & $b$ & $c$  \\ 
\hline
\hline
0.7 & 0.92 & 1.392 & 1.409  \\ 
\hline
1.1 & 2.91 & 2.199 & 2.205   \\ 
\hline
1.4 &  503.3 &  2.80 & 2.801  \\ 
\hline
1.45 &  7861.2 &  2.899 & 2.901  \\ 
\hline
1.48 & 238569. & 2.960 & 2.960  \\
\hline
\end{tabular}
\end{center}
\caption{Results of the power law fit in Eq.~\eqref{fit_pow} obtained for different mass values, labeled by $\nu = (9/4 - (m_\sigma^2/H^2))^{1/2}$. The results agree with the fit in Eq.~\eqref{fit_pow_nu}.} \label{tab:fit_power}
\end{table} 
\begin{table}[h!]
\begin{center}
\begin{tabular}{ |c|c|c|c| } 
\hline
$\nu$ & $a$  \\ 
\hline
\hline
0.7 & 0.96  \\ 
\hline
1.1 & 2.99  \\ 
\hline
1.4 &  506.2  \\ 
\hline
1.45 &  7879.3  \\ 
\hline
1.48 & 238800.  \\
\hline
\end{tabular}
\end{center}
\caption{Same as Tab. \ref{tab:fit_power} for the power law in Eq.~\eqref{fit_pow_nu}.} \label{tab:fit_power_nu}
\end{table}

We end this section by noting that the computation of the $\sigma$-mediated contribution to the TTT bispectrum resembles the one shown here for TTS, and we end up with Eq.~\eqref{eq:model_ttt}. The main difference in the final result is the replacement $c_0\rightarrow c_2$.

\bibliography{references.bib}
\end{document}